\documentclass[11pt]{article}
\pdfoutput=1
\usepackage{feynmp}
\usepackage{hyperref}
\usepackage{longtable}
\usepackage{graphicx,rotate}

\usepackage[title]{appendix}

\usepackage{float}
\usepackage{amsmath,amssymb,enumerate}
\usepackage{slashed}
\usepackage{amsfonts}
\usepackage{geometry}
\usepackage{color}
\usepackage{appendix}
\usepackage{resizegather}
\usepackage{bm}
\usepackage{subfig}
\usepackage{cite}
\definecolor{nicered}{rgb}{0.7,0.1,0.1}
\definecolor{nicegreen}{rgb}{0.1,0.5,0.1}

\newcommand{\nn}{\nonumber}

\usepackage{array}
\newcolumntype{L}[1]{>{\raggedright\let\newline\\\arraybackslash\hspace{0pt}}m{#1}}
\newcolumntype{C}[1]{>{\centering\let\newline\\\arraybackslash\hspace{0pt}}m{#1}}
\newcolumntype{R}[1]{>{\raggedleft\let\newline\\\arraybackslash\hspace{0pt}}m{#1}}

\usepackage[normalem]{ulem}
\usepackage{setspace}
\usepackage{multirow}

\def\({\left(}
\def\){\right)}
\def\[{\left[}
\def\]{\right]}

\newcommand{\ba}{\begin{array}}
\newcommand{\ea}{\end{array}}
\newcommand{\bd}{\begin{displaymath}}
\newcommand{\ed}{\end{displaymath}}
\newcommand{\be}{\begin{equation}}
\makeatletter
\newcommand*{\rom}[1]{\expandafter\@slowromancap\romannumeral #1@}
\makeatother
\newcommand{\ee}{\end{equation}}
\def\bt{\begin{table}}
\def\et{\end{table}}
\def\bc{\begin{center}}
\def\ec{\end{center}}
\def\bi{\begin{itemize}}
\def\ei{\end{itemize}}
\def\bw{\begin{widetext}}
\def\ew{\end{widetext}}

\def\bea{\begin{eqnarray}}
\def\eea{\end{eqnarray}}
\def\beas{\begin{eqnarray*}}
\def\eeas{\end{eqnarray*}}

\def\e{\ell}

\def\N0{\widetilde{\chi}^0}

\def\d{\delta}
\def\e{\epsilon}

\def\D{\Delta}

\def\ov {\overline}

\def\l{\lambda}

\allowdisplaybreaks

\title{\bf Anatomy of Heavy Gauge Bosons in a Left-Right Supersymmetric Model}

\date{}

\begin{document}

\maketitle
\vspace{-2.00cm}
\begin{center}
\author{Biplob Bhattacherjee$^{*}$\footnote{biplob@iisc.ac.in}, Najimuddin Khan$^{*}$\footnote{najimuddink@iisc.ac.in} and  Ayon Patra$^{\dagger}$\footnote{ayon@okstate.edu} \vspace{0.50cm}}

{$^{*}$Centre for High Energy Physics, Indian Institute of Science, Bengaluru - 560012, India}\\
{$^{\dagger}$Physical Research Laboratory, Navrangpura, Ahmedabad - 380009, India\vspace{1.80cm}}

\end{center}

\begin{abstract}
We perform a detailed study of the various decay channels of the heavy charged and neutral gauge bosons ($W_R$ and $Z_R$ respectively) in a left-right supersymmetric (LRSUSY) framework. The decay branching ratios of the $W_R$ and $Z_R$ bosons depend significantly on the particle spectrum and composition of the SUSY states. We show several combinations of mass spectrum for the SUSY particles to facilitate the decay of theses heavy gauge bosons into various combinations of final states. Finally, we choose two benchmark points and perform detailed collider simulations for these heavy gauge bosons in the context of the high energy and high luminosity run of the large hadron collider. We analyze two SUSY cascade decay channels -- mono-$W$ + $\slashed{E}_T$ and mono-$Z$ + $\slashed{E}_T$ along with the standard dilepton and dijet final states. Our results show that the existence of these heavy gauge bosons can be ascertained in the direct decay channels of dilepton and dijet whereas the other two channels will be required to establish the supersymmetric nature of this model.

{\color{red} }

\end{abstract}

%\singlespacing
\onehalfspacing
%\doublespacing

\renewcommand{\thefootnote}{\fnsymbol{footnote}}

\thispagestyle{empty}

\newpage

\section{Introduction}

The recent discovery of the Higgs-like scalar boson at the Large Hadron Collider (LHC)~\cite{Chatrchyan12} has essentially completed the particle spectrum of the Standard Model (SM). The measured properties of this Higgs-like scalar are consistent with the minimal choice of the scalar sector as in the SM with small uncertainties while they still comfortably allow red for an extended red scalar sector. 
The Higgs boson, responsible for giving mass to all the SM particles \footnote{Neutrinos are massless in the SM framework. The generation of neutrino masses requires an extended framework beyond the SM (BSM) and is an important motivation for BSM scenarios.}, itself has a mass which is very finely tuned in the SM framework. Thus one needs to further extend the SM with new particles or additional symmetries in order to understand the large cancellations required for the observed Higgs boson mass.
There are also a number of other experimental observations which lead us to believe that the SM is only an effective low energy theory with new physics coming in at higher energies. Numerous new physics models have thus been suggested to address the shortcomings of the SM.

Left-right supersymmetric (LRSUSY) models \cite{susylr} are a class of well motivated extensions of the SM as it can provide answers to a number of unresolved issues in the SM. These are actually the supersymmetric (SUSY) versions of the left-right symmetric (LRS) models \cite{lr} where the SM gauge group is extended to $G_{3221}=SU(3)_c \times SU(2)_L \times SU(2)_R \times U(1)_{B-L}$. This extended gauge symmetry facilitates the preservation of parity symmetry at high scales. The observed parity asymmetry in the SM is generated as the LR symmetry is spontaneously broken at some scale $v_R$ much above the SM symmetry breaking scale. Parity being a good symmetry in these models can potentially solve the strong CP problem \cite{strongcp} without introducing an additional Peccei-Quinn symmetry \cite{pq}. The gauge structure in LR models also naturally requires the presence of a right-handed neutrino which can help generate a light neutrino mass through the seesaw mechanisms \cite{csaw}. SUSY models, on the other hand, provide an elegant solution to the hierarchy problem. On top of that, if R-parity is conserved, as will be the case in this paper, the lightest SUSY particle (LSP) becomes stable and can be a good dark matter (DM) candidate. Combining the merits of both LRS and SUSY models, one gets a very attractive LRSUSY framework which warrants a careful examination as will be discussed in details in this paper. 

A variety of LRSUSY models have been discussed in literature with different scalar sectors for the symmetry breaking mechanism \cite{bhm, Babu:2014vba}. The one we consider here is the minimal LRSUSY model with automatic R-partiy conservation \cite{bm0} where the right-handed symmetry is broken by scalar triplets fields as they acquire non-zero vacuum expectation values (VEVs). This VEV is also responsible for generating the Majorana masses for the right-handed neutrinos which eventually will generate light neutrino masses. The spontaneous breaking of the right-handed symmetry gives rise to an additional charged gauge boson $W_R$ and a neutral gauge boson $Z_R$ whose masses are also at the same scale. The discovery of these gauge bosons could be one of the strongest indication towards the existence of left-right symmetry. Experimental searches for these heavy gauge bosons have been performed by the ATLAS and CMS collaborations in various final states with leptons, jets and/or missing transverse energy ($\slashed{E}_T$) which have helped put bounds on their masses. The ATLAS search using 36.1 ${\rm fb^{-1}}$ of proton-proton collision data collected at a centre of mass energy of 13 TeV, sets the exclusion limit for a heavy neutral gauge boson mass $M_{Z'} > 4.5$ TeV~\cite{Aaboud:2017buh} in the dilepton channel for the sequential standard model (SSM). The most stringent mass limit on a heavy charged gauge boson ($W'$), on the other hand, comes from the CMS collaboration for an integrated luminosity of 2.6 $\rm fb^{-1}$ at $\sqrt{s}=13$ TeV energy and is given as $M_{W'} > 4.1$ TeV \cite{Khachatryan:2016jww} in the lepton + $\slashed{E}_T$ final state for SSM. This limit, however, does not hold for our analysis as we have chosen the masses of the right-handed neutrinos to be heavier than the $W_R$ boson. Thus the lepton + $\slashed{E}_T$ cross-section in the final state for $W_R$ decay will be extremely small in our case resulting in no significant bound from this channel. Another recent analysis from the ATLAS collaboration \cite{Aaboud:2017yvp} gives a mass bound of $M_{W'} > 3.6$ TeV using 37 ${\rm fb^{-1}}$ integrated luminosity at 13 TeV com energy in the dijet ($q q'$) final state. The mass limit of 3.6 TeV was obtained by assuming a $W' \rightarrow q q'$ branching ratio (BR) of 75\%. For our analyses, on the other hand, this BR could vary from around $50\%-90\%$ for different benchmark points (BPs), which significantly affects the results. The  heavy right-handed $W_R$ bosons decaying into a top and a bottom quark searches at $\sqrt{s}=13$ TeV by the CMS collaboration also provides a mass limit of $M_{W_R}>2.6$ TeV~\cite{Sirunyan:2017ukk}. This limit however is much weaker compared to the dijet decay channel. On top of this, all these experimental searches have been performed for the charged and neutral heavy gauge bosons decaying directly into SM particles. For LRSUSY models this is not true at all, since these particles can also decay into SUSY particles which eventually decay into SM particles and LSP. These new decay channels can alter their mass limits and allow for new possibilities to discover SUSY particles at the LHC. 

In this paper we perform a detailed study of the heavy charged and neutral gauge bosons in a LRSUSY framework with several possible decay channels. We observe that the decay BR of the $W_R$ and $Z_R$ bosons depend significantly on the particle spectrum and composition of the SUSY particles (mainly charginos and neutralinos). We thus choose our BPs to encompass all possible compositions for charginos and neutralinos with and without mixing among the fields in the gauge basis. Several combinations of mass spectrum for the SUSY particles are also chosen to facilitate the decay of the heavy gauge bosons into various combinations of final states for a more comprehensive study of their properties. This gives us a good understanding of each interaction and how it affects the final decay BR of the heavy gauge bosons. Firstly we consider an almost pure one component LSP keeping all the squarks and sleptons to be heavier than $Z_R$ \footnote{Since $Z_R$ is always heavier than $W_R$, the squarks and sleptons are also heavier than $W_R$.}. This allows us to explore SUSY final states with only charginos and neutralinos. Next we consider the case where the LSP is composed of a significantly mixed state of gauginos and higgsinos. Finally we allow the squarks and sleptons to be light as well, maximizing the SUSY decay BR for the heavy gauge bosons. The experimental bounds on the heavy gauge boson masses will thus change as new (SUSY) decay channels have now opened up.

The discovery prospect of SUSY particles at the LHC is severely constrained by the direct production cross-section of these particles. The production cross-section falls rapidly as the mass of the particles increase, and hence this translates into upper limits on the masses of SUSY particles that may be discovered at the LHC. The presence of heavy gauge bosons can help mitigate this problem as they can decay into final states with SUSY particles which would otherwise evade detection at the LHC. We analyze four different final states in the context of high luminosity LHC (HL-LHC) and high energy LHC (HE-LHC) experiments. Two of these channels are the standard search channels used for heavy resonance searches -- dilepton + $\slashed{E}_T$ final state and dijet + $\slashed{E}_T$ final state. These channels can show large significances in specific benchmark regions in the context of the LHC at $14$ and/or $27$ TeV energies with integrated luminosity of $3000~{\rm fb^{-1}}$. The other two final states arise from the cascade SUSY decays of the heavy gauge bosons resulting in final states with mono $W$ + $\slashed{E}_T$ and mono $Z$ + $\slashed{E}_T$. These signals have already been studied at the LHC for DM searches to constrain them but they have so far not been considered as a search channel for heavy gauge bosons. Our analysis of these final states though give promising results as a discovery channel for a $W_R$ or $Z_R$ boson in LRSUSY framework. We find a large number of events with significance greater than $5\sigma$ in these new mono-$X$ ($X=W,Z$) plus $\slashed{E}_T$ channels through one-step cascade decays, with $X$ decaying into leptonic final states only.

The rest of the paper is organized as follows. In Sec.~\ref{model} we present the details of the model and derive expressions for the masses and relevant interactions for all the particles. Sec.~\ref{BP} gives all the different cases that are important for the heavy gauge bosons decays. Here we consider various possible mixings in the neutralino and chargino sectors which affect the $W_R$ and $Z_R$ BR. We also change the masses of the squarks and sleptons to be heavier or lighter than the gauge bosons so as to study the variation of their decay BR in each case. The possible collider signals arising from the SUSY decays of the heavy gauge bosons is analyzed in Sec.~\ref{coll}. Here we first study the familiar dilepton and dijet final states for the heavy gauge boson decays. We then analyze a mono $W$ + $\slashed{E}_T$ and a mono $Z$ + $\slashed{E}_T$ final states and how they can be used to search for these heavy particles. Finally we conclude in Sec.~\ref{conc} with a discussion of our results. All the relevant interactions of the $W_R$ and $Z_R$ bosons leading to their decays are provided in the Appendices.

\section{Minimal LRSUSY Model with Automatic R-parity} {\label{model}}
Left--right symmetric models have an extended gauge symmetry which is $SU(3)_C \times SU(2)_L \times SU(2)_R \times U(1)_{B-L}$. The chiral fermion sector consist of three families of quark and lepton superfields given as
\begin{eqnarray}
\!\!Q\!\!&=&\!\!\begin{pmatrix}
u\\ d \end{pmatrix}  \sim \left (3,2, 1, \frac13 \right ),~~~~~~~~
Q^c\!=\!\begin{pmatrix}
d^c\\-u^c
\end{pmatrix} \sim \left ( 3^{\ast},1, 2, -\frac13
\right ),\nonumber \\
L&=&\begin{pmatrix}
\nu\\ e \end{pmatrix} \sim\left ( 1,2, 1, -1 \right ),~~~~~~~~
L^c=\begin{pmatrix}
e^c \\ -\nu^c \end{pmatrix} \sim \left ( 1,1, 2, 1 \right ),
\label{eq:matter}
\end{eqnarray}
where  $c$ stands for the charge conjugation and the numbers in brackets are their
$SU(3)_C$, $SU(2)_L$, $SU(2)_R$, $U(1)_{B-L}$ gauge quantum numbers respectively.

The minimal Higgs sector required for a consistent symmetry breaking mechanism, generation of quark and lepton masses and mixings and preservation of an unbroken $R$-parity symmetry is given as 
\begin{eqnarray}
\Delta(1,3,1,2)&=&\begin{pmatrix}
\frac{\delta^{+}}{\sqrt{2}} & \delta^{++}\\ \delta^{0} & -\frac{\delta^{+}}{\sqrt{2}}  \end{pmatrix},~~~~
\overline{\Delta}(1,3,1,-2)=\begin{pmatrix}
\frac{\overline{\delta}^{-}}{\sqrt{2}} & \overline{\delta}^{0}\\ \overline{\delta}^{--} & -\frac{\overline{\delta}^{-}}{\sqrt{2}}  \end{pmatrix},\nonumber \\
\Delta^{c}(1,1,3,-2)&=&\begin{pmatrix}
\frac{\delta^{c^{-}}}{\sqrt{2}} & \delta^{c^{0}}\\ \delta^{c^{--}} & -\frac{\delta^{c^{-}}}{\sqrt{2}}  \end{pmatrix},~~~~
\overline{\Delta}^{c}(1,1,3,2)=\begin{pmatrix}
\frac{\overline{\delta}^{c^{+}}}{\sqrt{2}} & \overline{\delta}^{c^{++}}\\ \overline{\delta}^{c^{0}} & -\frac{\overline{\delta}^{c^{+}}}{\sqrt{2}}  \end{pmatrix},\nonumber \\
\Phi_i(1,2,2,0)&=&{\begin{pmatrix}
\phi^{+}_1 & \phi^{0}_{2} \\ \phi^{0}_{1} & \phi^{-}_{2} \end{pmatrix}_i}~~(i=1,2),~~~~ S(1,1,1,0).
\label{eq:triphig}
\end{eqnarray}

The $SU(2)_R$ triplet Higgs field $\D^c(1,1,3,-2)$ is responsible for breaking the $SU(2)_R \times U(1)_{B-L}$ symmetry into $U(1)_Y$ as its neutral component acquires a non--zero VEV. The coupling of this triplet field with the right-handed neutrinos generates their Majorana masses as well. 
Two bidoublet fields $\Phi_a(1,2,2,0)$ are required to generate the quark and lepton masses and mixings through Yukawa interactions. The simpler case of one bidoublet field, as will be considered in our analysis, cannot produce the Cabibbo-Kobayashi-Maskawa (CKM) mixing angles. In such a scenario, the CKM mixing angles could arise from soft SUSY breaking terms\cite{bdm1}. For a SUSY model, an extra $SU(2)_R$ triplet field $\ov \D^c(1,1,3,+2)$ is also required for anomaly cancellation, and two $SU(2)_L$ triplet fields $\D(1,3,1,2)$ and $\ov \D(1,3,1,-2)$ are needed for parity conservation. The singlet field $S(1,1,1,0)$ is required to decouple the SUSY breaking scale from the right-handed symmetry breaking scale. In absence of the singlet, the SUSY breaking scale and the right-handed symmetry breaking scale becomes equal to each other, hence the singlet $S$ is needed in order to decouple the two scales allowing the right-handed symmetry breaking scale to be higher than the SUSY breaking scale.

The non--zero VEVs of various fields are denoted as
\begin{equation}
\left< \d^{c^0} \right> = v_R,~~~~ \left< \ov \d^{c^0} \right> = \ov v_R,~~~~ \left< \phi_{1_i}^0 \right> = v_{u_i},~~~~ \left< \phi^0_{2_i} \right> = v_{d_i}~,~~~~ \left< S \right> = v_s,
\label{vev}
\end{equation}
with the hierarchy among them chosen as $v_R, \ov v_R > v_s >> v_u, v_d$. The right-handed symmetry breaking scale is chosen larger than the SUSY breaking scale as we want the superpartner masses to be lighter than the heavy $W_R$ and $Z_R$ gauge bosons. For simplicity, the left-handed $\d^0$ and $\ov \d^0$ fields do not get any VEVs and hence the neutrino masses do not get any contribution from Type II seesaw. This choice is a consistent one as it can also be shown that the left-handed triplet fields do not get any induced VEV in this model.

The Yukawa couplings in the model are given by the superpotential
\begin{eqnarray}
W_Y &=& \sum_{j=1}^2 \left( Y_q^{(j)} Q^T \tau_2 \Phi_j \tau_2 Q^c + Y_l^{(j)} L^T \tau_2 \Phi_j \tau_2 L^c \right) + i \frac{f}{2} L^T \tau_2 \D L + i \frac{f^c}{2} {L^c}^T \tau_2 \D^c L^c.
\end{eqnarray}
Here $Y_l^j$ and $Y_q^j$ are the lepton and quark Yukawa coupling matrices respectively while $f$ is the Majorana  Yukawa coupling matrix responsible for  generating large Majorana masses for right-handed neutrinos. The transformation of various fields under parity symmetry is given as $\Phi \rightarrow \Phi^{\dagger}, \D \rightarrow \D^{c^*}, \ov \D \rightarrow \ov \D^{c^*}, S \rightarrow S^*,Q \rightarrow Q^{c^*}, L \rightarrow L^{c^*}, \theta \rightarrow \ov{\theta}$, along with $W^\pm \rightarrow W_R^{\pm *}$. Additionally the Yukawa superpotential is invariant under parity if the Yukawa coupling matrices $Y_q^j$ and $Y_l^j$ are Hermitian and $f^c=f$. The up quarks, down quarks, charged leptons, neutrino Dirac and right-handed Majorana neutrino masses are given as
 \begin{eqnarray}
 {\boldsymbol{M_u}} &=& Y_q^{(1)} v_{u_1} + Y_q^{(2)} v_{u_2}, ~~~ {\boldsymbol {M_d}} = Y_q^{(1)} v_{d_1} + Y_q^{(2)} v_{d_2},\nonumber \\
 {\boldsymbol {M_\ell}} &=& Y_l^{(1)} v_{d_1} + Y_l^{(2)} v_{d_2},~~~ {\boldsymbol {M_\nu^D}} = Y_l^{(1)} v_{u_1} + Y_l^{(2)} v_{u_2},~~~ {\boldsymbol {M_R}} = f v_R
 \end{eqnarray}
respectively. Thus it is easy to see that two bidoublet Higgs fields are needed to generate the CKM mixings as otherwise the up- and down-type quark mass matrices would become proportional to each other. A realistic model would therefore need two bidoublet fields but for simplicity we will only consider a version with one bidoublet in the scalar spectrum. As the main focus of this paper is on the heavy gauge boson properties, having only one bidoublet does not change our results significantly.

The gauge sector of the model has an extra charged $W_R$ and a neutral $Z_R$ gauge boson. The mass-squared matrices for the neutral gauge bosons {\boldsymbol{$M_Z^2$}} in the basis $(B,W_{3L},W_{3R})$ and the charged gauge boson {\boldsymbol{$M_W^2$}} in the basis $(W_L,W_R)$ are given as
%==============
\begin{equation}
{\boldsymbol{M_Z^2}} = \frac{1}{2}\begin{bmatrix}
4g_V^2 (v_R^2+\ov v_R^2)&0& -4g_R g_V (v_R^2+\ov v_R^2)\\ 0&g_2^2 v^2&g_2 g_R v^2 \\-4g_R g_V (v_R^2+\ov v_R^2)&g_2 g_R v^2&g_R^2\left(4v_R^2+4\ov v_R^2+v^2\right)
\end{bmatrix},~~
{\boldsymbol{M_W^2}} = \frac{1}{2}\begin{bmatrix}
g_2^2 v^2&g_2 g_R v_1 v_2 \\ g_2 g_R v_1 v_2&g_R^2\left( 2v_R^2+2\ov v_R^2+ v^2 \right),
\end{bmatrix}.
\end{equation}
where  $v^2 = v_u^2+v_d^2=174.1$ GeV while $g_V$, $g_2$ and $g_R$ are the gauge coupling constants 
corresponding to the $U(1)_{B-L}$, $SU(2)_L$ and $ SU(2)_R $ gauge groups respectively. The mass eigenstates of the heavy gauge bosons can be obtained as
\begin{equation}
M^2_{W_R} \simeq \frac{1}{2} g_R^2 (2 v_R^2+2\ov v_R^2+v^2),~~~~
M^2_{Z_R} \simeq \frac{1}{2}(g_R^2+g_V^2) \left[ 4 (v_R^2+\ov v_R^2)+v^2\cos^4 \theta_R\right],
\label{MassWZ}
\end{equation}
where $\cos^2\theta_R = {g_R^2}/(g_R^2+g_V^2)$. In getting these masses we have neglected the mixing between the left and right-handed charged gauge bosons and neglected terms with $v^4/v_R^4$ or higher. The SM gauge bosons have their usual expressions with the effective $U(1)_Y$ hypercharge coupling identified as $g_Y^2 = g_R^2 g_V^2/(g_R^2+g_V^2)$. The ratio of the heavy gauge boson masses can be approximately written as~\cite{susylr} 
\begin{equation}
\frac{M_{Z_R}}{M_{W_R}} \simeq \frac{\sqrt{2} g_R/g_2}{\sqrt{\left(g_R/g_2\right)^2 - \tan^2 \theta_W}},
\end{equation}
where $\theta_W$ is the Weinberg angle. This relation shows that the ratio $g_R/g_2$ should always be larger than $\tan \theta_W$.

The most general superpotential for the Higgs sector is given as
\begin{eqnarray}
W&=& S\left[{\text{Tr}}(\lambda \Delta \overline{\Delta})+{\text{Tr}}(\lambda ^c \Delta^{c}\overline{\Delta}^{c}) +\dfrac{\lambda^{\prime}}{2} {\text{Tr}}(\Phi^{T}\tau_{2}\Phi\tau_{2})-M^2 \right] \notag \\
&+& \text{Tr} \left[ \mu_1 \D \ov \D+\mu_2 \D^c \ov \D^c+\frac{\mu}{2} \left( \Phi^T \tau_2 \Phi \tau_2 \right) \right] + \frac{\mu_S}{2} S^2+\frac{\kappa}{3} S^3,
\end{eqnarray}
where $\l^c = \l^*$, $\mu_1=\mu_2^*$ while $\l_0$, $M_2$, $\mu$ and $\mu_S$ are all real from the conservation of parity symmetry. The Higgs potential derived from this superpotential will consist of $F$-terms, $D$-terms and soft supersymmetry-breaking terms. So we have 
\begin{equation}
V_{Higgs} = V_F+V_D+V_{Soft},
\end{equation}
with each of the terms being
\begin{eqnarray}
\label{eq:1df}
V_F&=&  \left| {\text{Tr}}(\lambda \Delta \overline{\Delta})+{\text{Tr}}(\lambda^* \Delta^{c}\overline{\Delta}^{c}) +\frac{\lambda^\prime}{2}{\text{Tr}}(\Phi^{T}\tau_{2}\Phi\tau_{2})-M^2+\mu_S S+\kappa S^2 \right|^2 \nonumber \\
&+&	\text{Tr} \left[ \left| \mu_1 \D + \lambda S \D \right| ^2 + \left| \mu_1 \ov{\D} + \lambda S \ov{\D} \right| ^2 + \left| \mu_1^* \D^c + \lambda^* S \D^c \right| ^2 \right. \notag \\
&+& \left. \left| \mu_1^* \ov \D^c + \lambda^* S \ov \D^c \right| ^2  \right] +\text{Tr} \left| \mu \Phi + \lambda ^ \prime S \Phi \right|^2, \\
\label{eq:two}
V_D&=&\frac{g_2^2}{8}\sum \limits_{a=1}^3 \left|{\text{Tr}}(2\D^\dagger \tau_a \D+2\ov\D^\dagger \tau_a \ov\D+\Phi^\dagger \tau_a \Phi)\right|^2\nonumber \\
&+&\frac{g_R^2}{8}\sum \limits_{a=1}^3  \left|{\text{Tr}}(2{\D^c}^\dagger \tau_a \D^c+2\ov{\D^c}^\dagger \tau_a \ov\D^c+\Phi^* \tau_a \Phi^T)\right|^2 \nonumber \\
&+& \frac{g_V^2}{2}\left|{\text{Tr}}(\D^\dagger \D- \ov\D^\dagger \ov\D-{\D^c}^\dagger \D^c+ {\ov \D^c}^\dagger \ov\D^c)\right|^2, \\
V_{Soft}&=&m_1^2{\text{Tr}}({\D^c}^\dagger\D^c)+m_2^2{\text{Tr}}({\ov{\D}^c}^\dagger\ov\D^c)+m_3^2{\text{Tr}}(\D^\dagger\D)+m_4^2{\text{Tr}}(\ov\D^\dagger\ov\D)\nonumber \\
&+&m_S^2 |S|^2+ m_5^2 {\text{Tr}}(\Phi^\dagger \Phi)
+\left[\lambda A_{\lambda} S {\text{Tr}}(\D \ov\D +\D^c \ov{\D}^c)+h.c.\right]\nonumber \\&+&[\lambda^\prime A_{\lambda^\prime} S {\text{Tr}}(\Phi^{T}\tau_{2}\Phi\tau_{2})+h.c.]+(\lambda C_{\lambda} M^2 S +h.c.) + \left( \mu_S B_S S^2 + h.c. \right) \notag \\
&+& \left[ \mu_1 B_1 \text{Tr} \left( \D \ov \D \right)+\mu_1^* B_2 \text{Tr} \left( \D^c \ov \D^c \right)+\mu B \text{Tr} \left( \Phi^T \tau_2 \Phi \tau_2 \right) +\kappa A_\kappa S^3+ h.c. \right].
\label{eq:pot}
\end{eqnarray}

The minimization of the scalar potential proceeds as
\begin{equation}
\Big|\frac{\partial V_{Higgs}}{\partial \phi_i}\Big|_{\phi_i=v_i} = 0
\end{equation}
where $v_i$ are the VEVs of the neutral CP-even scalar fields ($\phi^{0}_{1},\phi^{0}_{2},\delta^{c^{0}},\overline{\delta}^{c^{0}},S$). The minimization equations of the scalar potential for this model are thus given as
%%%%%%%%%%%%%%%%%%%%%%%%%%%%%%%%%%%%%%%%%%%%%%%%%%%%%%%%%%%%%%%%%%%%%%
%%%%%%%%%%%%%%%%%%%%%%%%%%%%%%%%%%%%%%%%%%%%%%%%%%%%%%%%%%%%%%%%%%%%%%
{{\allowdisplaybreaks
\begin{align}
\label{pot:min1}
\frac{\partial V}{\partial \phi^{0}_{1}} &= \frac{1}{2\sqrt{2}} v_u \Big(g_{2}^{2} (- v_{d}^{2}  + v_{u}^{2}) + g_{R}^{2} (2 \bar{v}_{R}^{2}  -2 v_{R}^{2}  - v_{d}^{2}  + v_{u}^{2})\Big)+\sqrt{2} \Big(  \lambda'^2 (v_{d}^{2} + v_{s}^{2}) + \mu^2) + m_5^2\Big)v_u \nonumber \\ 
 &- \lambda' \Big(v_d ( \lambda^c  v_R \bar{v}_R  - M^2) + v_s (-2 {{\mu}} v_u  + v_d ( \mu_s + {k} v_s )\Big)-2 v_d (v_s {\lambda'A_{\lambda'}}  + { \mu B }) = 0,\\ 
 %%%%%%%%%%%%%%%%%%%%
 %%%%%%%%%%%%%%%%%%%%
\frac{\partial V}{\partial \phi^{0}_{2}} &= \frac{1}{2\sqrt{2}} v_d \Big(g_{2}^{2} (v_d^2 - v_u^2) + g_{R}^{2} (-2 \bar{v}_{R}^{2}  + 2 v_{R}^{2}  - v_{u}^{2}  + v_{d}^{2})\Big)+\sqrt{2} \Big(v_d ( {\mu}^{2}  + 2 \lambda' {~ \mu ~} v_s  + m_5^2)\nonumber \\ 
 &- \lambda'  \Big(  v_u ( \mu_s v_s  + {k} v_{s}^{2}  +  \lambda^c  v_R \bar{v}_R  - M^2 )-\lambda' v_d (v_{s}^{2} + v_{u}^{2})\Big) -2 v_u (v_s {\lambda'A_{\lambda'}}  + {\mu B}) = 0,\\ 
 %%%%%%%%%%%%%%%%%%%%
 %%%%%%%%%%%%%%%%%%%%
\frac{\partial V}{\partial \delta^{c^{0}}} &= \frac{1}{\sqrt{2}} \bar{v}_R \Big(2  g_V ^{2} (\bar{v}_{R}^{2}- v_{R}^{2}) + g_{R}^{2} (2 \bar{v}_{R}^{2}  -2 v_{R}^{2}  - v_{d}^{2}  + v_{u}^{2})\Big)+\sqrt{2} \bar{v}_R (2  \lambda^c   \mu_2   v_s  +  m_1^2  +  \mu_2 ^{2})\nonumber \\ 
 &+ \lambda^{c\,2}  \bar{v}_R (v_{R}^{2} + v_{s}^{2})+ \lambda^c v_R (-\lambda' v_d v_u  + \mu_s  v_s  + {k} v_{s}^{2}  - M^2)+v_R (v_s { \lambda A_{\lambda} }  + {\mu_1^*B_2})\Big) = 0, \\
 %%%%%%%%%%%%%%%%%%%%
 %%%%%%%%%%%%%%%%%%%%
\frac{\partial V}{\partial \overline{\delta}^{c^{0}} } &= \frac{1}{\sqrt{2}} v_R \Big(2  g_V ^{2}(v_R^2 -\bar{v}_R^2) + g_{R}^{2} (-2 \bar{v}_{R}^{2}  + 2 v_{R}^{2}  - v_{u}^{2}  + v_{d}^{2})\Big)+\sqrt{2} v_R (2  \lambda^c   \mu_2   v_s  +  m_2^2  +  \mu_2 ^{2})\nonumber \\ 
 & + \lambda^c \bar{v}_R (-\lambda' v_d v_u  + \mu_s  v_s  + k v_{s}^{2}  - M^2) +  \lambda^{c2}  v_R (\bar{v}_{R}^{2} + v_{s}^{2})+\bar{v}_R (v_s { \lambda A_{\lambda} }  + {\mu_1^*B_2}) \Big) = 0,\\ 
 %%%%%%%%%%%%%%%%%%%%
 %%%%%%%%%%%%%%%%%%%%
\frac{\partial V}{\partial S} &= \sqrt{2} \Big(  \lambda^c  v_R \bar{v}_R (2 {k} v_s  +\mu_s) + \lambda^c  \mu_2   (v_{R}^{2} + \bar{v}_{R}^{2})+ \mu \lambda' (v_{d}^{2} + v_{u}^{2}) - \lambda' (2 {k} v_s  +\mu_s) v_d v_u \Big)\nonumber \\ 
 &-(2 {k} v_s  + \mu_s)M^2 +{~ \lambda C_{\lambda} ~}+v_R \bar{v}_R { \lambda A_{\lambda} } + v_s ( (2 {k} v_s  + \mu_s)(\mu_s  + {k} v_s) + 2  {\mu_sB_s} + 3 v_s k A_k \nonumber \\ 
 &+ \lambda'^2 (v_{d}^{2} + v_{u}^{2}) + \lambda^{c2} (v_{R}^{2} + \bar{v}_{R}^{2}) + m_S^2 )-2 v_d v_u {\lambda'A_{\lambda'}} \Big) = 0.
 \label{pot:min}
\end{align} 
}
%%%%%%%%%%%%%%%%%%%%%%%%%%%%%%%%%%%%%%%%%%%%%%%%%%%%%%%%%%%%%%%%%%%%%%
%%%%%%%%%%%%%%%%%%%%%%%%%%%%%%%%%%%%%%%%%%%%%%%%%%%%%%%%%%%%%%%%%%%%%%
\subsection{Particle masses}

In this section we calculate the masses of the particles in various sectors of our LRSUSY model.  

\subsubsection{Higgs sector}

The mass-squared matrices for the charged and neutral Higgs bosons can be obtained from the scalar potential. The minimization conditions in Eqns.~\ref{pot:min1}-\ref{pot:min} will provide further constraints on the parameter in the model. The singlet ($S$), the bidoublet ($\Phi$) and the right-handed triplets ($\Delta^c$ and $\ov \Delta^c$) can mix with each other while the left-handed triplets ($\Delta$ and $\ov \Delta$) get decoupled since they do not acquire any VEVs. Here we will only consider the sector consisting of the right-handed triplets, the bidoublet and the singlet as these will be important for our analysis of the heavy right-handed gauge bosons. 

After electroweak symmetry breaking, the mass--squared matrix for the singly--charged Higgs fields can be expressed as a $4 \times 4$ matrix in the basis $( {\delta^{c^{\pm}}}, \overline{\delta}^{c^{\pm}}, {\phi^{\pm}_2}, \phi^{\pm}_1)$ as
%%%%%%%%%%%%%%%%%%%%%%%%%%%%%%%%%%%%%%%%%%%%%%%%%%%%%%%%%%%%%%%%%%%%%%
%%%%%%%%%%%%%%%%%%%%%%%%%%%%%%%%%%%%%%%%%%%%%%%%%%%%%%%%%%%%%%%%%%%%%%
\begin{equation} 
{\boldsymbol {m^2_{Hm}}} = \left( 
\begin{array}{cccc}
m_{{\delta^{c^{-}}}\delta^{c^{+}}} &m^*_{\overline{\delta}^{c^{-}}\delta^{c^{+}}} &- \frac{1}{\sqrt{2}} g_{R}^{2} \bar{v}_R v_u  &- \frac{1}{\sqrt{2}} g_{R}^{2} v_d \bar{v}_R \\ 
m_{{\delta^{c^{-}}}{\overline{\delta}^{c^{+}}}} &m_{\overline{\delta}^{c^{-}}{\overline{\delta}^{c^{+}}}} &\frac{1}{\sqrt{2}} g_{R}^{2} v_R v_u  &\frac{1}{\sqrt{2}} g_{R}^{2} v_d v_R \\ 
- \frac{1}{\sqrt{2}} g_{R}^{2} \bar{v}_R v_u  &\frac{1}{\sqrt{2}} g_{R}^{2} v_R v_u  &m_{{\phi^{-}_2}\phi^{+}_2} &m^*_{\phi^{-}_1\phi^{+}_2}\\ 
- \frac{1}{\sqrt{2}} g_{R}^{2} v_d \bar{v}_R  &\frac{1}{\sqrt{2}} g_{R}^{2} v_d v_R  &m_{{\phi^{-}_2}{\phi^{+}_1}} &m_{\phi^{-}_1{\phi^{+}_1}}\end{array} 
\right),
\label{eq:chmass}
 \end{equation} 
%%%%%%%%%%%%%%%%%%%%%%%%%%%%%%%%%%%%%%%%%%%%%%%%%%%%%%%%%%%%%%%%%%%%%%
%%%%%%%%%%%%%%%%%%%%%%%%%%%%%%%%%%%%%%%%%%%%%%%%%%%%%%%%%%%%%%%%%%%%%%
where
%%%%%%%%%%%%%%%%%%%%%%%%%%%%%%%%%%%%%%%%%%%%%%%%%%%%%%%%%%%%%%%%%%%%%%
%%%%%%%%%%%%%%%%%%%%%%%%%%%%%%%%%%%%%%%%%%%%%%%%%%%%%%%%%%%%%%%%%%%%%%
{\allowdisplaybreaks
\begin{align}
m_{{\delta^{c^{-}}}\delta^{c^{+}}} &= ( g_V ^{2} + g_{R}^{2})\bar{v}_{R}^{2}  -  g_V ^{2} v_{R}^{2}  +  m_1^2  + ( \lambda^c  v_s  +  \mu_2 )^{2},\nonumber\\
%%%%%%%%%%%%%%%%%%%%%%%%
%%%%%%%%%%%%%%%%%%%%%%% 
m_{{\delta^{c^{-}}}{\overline{\delta}^{c^{+}}}} &= - g_{R}^{2} v_R \bar{v}_R  +  \lambda^c (-\lambda' v_d v_u  + \mu_s  v_s  + {k} v_{s}^{2}  +  \lambda^c  v_R \bar{v}_R  - M^2)+ v_s { \lambda A_{\lambda} }  + {\mu_1^*B_2},\nonumber\\ 
%%%%%%%%%%%%%%%%%%%%%%%%
%%%%%%%%%%%%%%%%%%%%%%% 
m_{\overline{\delta}^{c^{-}}{\overline{\delta}^{c^{+}}}} &=  g_V ^{2} (- \bar{v}_{R}^{2}  + v_{R}^{2} ) 
+ g_{R}^{2} v_{R}^{2} 
 +  m_2^2  + (  \lambda^c  v_s  +  \mu_2 )^{2},\nonumber\\ 
%%%%%%%%%%%%%%%%%%%%%%%%
%%%%%%%%%%%%%%%%%%%%%%% 
m_{{\phi^{-}_2}\phi^{+}_2} &= (\lambda' v_s  + \mu)^{2}  + \frac{1}{4} \Big(g_{2}^{2} (v_{d}^{2} + v_{u}^{2}) + g_{R}^{2} (-2 \bar{v}_{R}^{2}  + 2 v_{R}^{2}  + v_{d}^{2} + v_{u}^{2})\Big) + m_5^2,\nonumber\\ 
%%%%%%%%%%%%%%%%%%%%%%%%
%%%%%%%%%%%%%%%%%%%%%%% 
m_{{\phi^{-}_2}{\phi^{+}_1}} &= \lambda'(\mu_s  v_s  -\lambda' v_d v_u  + {k} v_{s}^{2}  +  \lambda^c  v_R \bar{v}_R  - M^2) + 2 v_s {\lambda'A_{\lambda'}}  + 2 { \mu B } + \frac{(g_{2}^{2} + g_{R}^{2})}{2}v_d v_u, \nonumber\\
%%%%%%%%%%%%%%%%%%%%%%%%
%%%%%%%%%%%%%%%%%%%%%%%  
m_{\phi^{-}_1{\phi^{+}_1}} &= (\lambda' v_s  + \mu)^{2}  + \frac{1}{4} \Big(g_{2}^{2}(v_{d}^{2} + v_{u}^{2}) + g_{R}^{2} (2 \bar{v}_{R}^{2}  -2 v_{R}^{2}  + v_{d}^{2} + v_{u}^{2})\Big) + m_5^2.\nonumber
\end{align}
} 
%%%%%%%%%%%%%%%%%%%%%%%%%%%%%%%%%%%%%%%%%%%%%%%%%%%%%%%%%%%%%%%%%%%%%%
%%%%%%%%%%%%%%%%%%%%%%%%%%%%%%%%%%%%%%%%%%%%%%%%%%%%%%%%%%%%%%%%%%%%%%
The singly-charged fields in the left-handed triplets get decoupled and is not important for our analysis. As a result they have not been included here. The $4 \times 4$ mass-squared matrix in Eqn.~\ref{eq:chmass} can be diagonalized by the transformation $ U^{Hm} {\bf m^2_{Hm} } U^{Hm^\dagger} = m^2_{Hm,diag}$, where $U^{Hm}$ stands for the rotation matrix for the singly--charged Higgs fields. This gives two physical mass eigenstates ($H_1^\pm, H_2^\pm$) for the charged scalar fields  while the remaining two states ($G_1^\pm, G_2^\pm$) become the massless Goldstone bosons. These massless degrees of freedom are eaten up by their corresponding gauge bosons $W^\pm$ and  $W_R$ respectively to give them mass.

The mass-squared matrix for the doubly--charged scalar fields is given by a $2 \times 2$ matrix in the basis $({\delta^{c^{\pm\pm}}}, \overline{\delta}^{c^{\pm\pm}})$. This can be written as
%%%%%%%%%%%%%%%%%%%%%%%%%%%%%%%%%%%%%%%%%%%%%%%%%%%%%%%%%%%%%%%%%%%%%%
%%%%%%%%%%%%%%%%%%%%%%%%%%%%%%%%%%%%%%%%%%%%%%%%%%%%%%%%%%%%%%%%%%%%%%
\begin{equation} 
{\boldsymbol {m^2_{Hmm}}} = \left( 
\begin{array}{cc}
m_{{\delta^{c^{--}}}\delta^{c^{++}}} &m^*_{\overline{\delta}^{c^{--}}\delta^{c^{++}}}\\ 
m_{{\delta^{c^{--}}}{\overline{\delta}^{c^{++}}}} &m_{\overline{\delta}^{c^{--}}{\overline{\delta}^{c^{++}}}}\end{array} 
\right),
\label{d-charged}
\end{equation} 
%%%%%%%%%%%%%%%%%%%%%%%%%%%%%%%%%%%%%%%%%%%%%%%%%%%%%%%%%%%%%%%%%%%%%%
%%%%%%%%%%%%%%%%%%%%%%%%%%%%%%%%%%%%%%%%%%%%%%%%%%%%%%%%%%%%%%%%%%%%%%
where
{\allowdisplaybreaks
\begin{align} 
m_{{\delta^{c^{--}}}\delta^{c^{++}}} &= \frac{g_{R}^{2}}{2} (-2 \bar{v}_{R}^{2}  + 2 v_{R}^{2}  - v_{u}^{2}  + v_{d}^{2}) +  g_V ^{2} (\bar{v}_{R}^{2}- v_{R}^{2}) +  m_1^2+( \lambda^c  v_s  +  \mu_2 )^{2}, \nonumber\\ 
%%%%%%%%%%%%%%%%%%%%%%%%
%%%%%%%%%%%%%%%%%%%%%%%
m_{{\delta^{c^{--}}}{\overline{\delta}^{c^{++}}}} &=  \lambda^c ( \mu_s  v_s  -\lambda' v_d v_u  + {k} v_{s}^{2}  +  \lambda^c  v_R \bar{v}_R  - M^2) + v_s { \lambda A_{\lambda} }  + {\mu_1^*B_2},\nonumber\\ 
%%%%%%%%%%%%%%%%%%%%%%%%
%%%%%%%%%%%%%%%%%%%%%%%
m_{\overline{\delta}^{c^{--}}{\overline{\delta}^{c^{++}}}} &= \frac{g_{R}^{2}}{2} (2 \bar{v}_{R}^{2}  -2 v_{R}^{2}  - v_{d}^{2}  + v_{u}^{2}) +  g_V ^{2} ( v_{R}^{2} - \bar{v}_{R}^{2}) +  m_2^2 +( \lambda^c  v_s  +  \mu_2)^{2}.\nonumber
\end{align}
} 
%%%%%%%%%%%%%%%%%%%%%%%%%%%%%%%%%%%%%%%%%%%%%%%%%%%%%%%%%%%%%%%%%%%%%%
%%%%%%%%%%%%%%%%%%%%%%%%%%%%%%%%%%%%%%%%%%%%%%%%%%%%%%%%%%%%%%%%%%%%%%
It can be shown that this doubly-charged Higgs mass-squared matrix, upon diagonalization, admits a negative eigenvalue which is unphysical as it gives rise to a tachyonic state. This problem can however be solved by including the radiative corrections to the doubly--charged Higgs boson mass which makes it positive~\cite{bm0,Babu:2014vba}.

The CP-even neutral scalars consist of the real part of the neutral Higgs fields. The mass--squared matrix for these fields in the basis $(Re[{ \delta^{c^{0}} }], Re[ { \overline{\delta}^{c^{0}} }], Re[{\phi^{0}_{2}}], Re[{\phi^{0}_{1}}], Re[{ S }] )$ is given as 
%%%%%%%%%%%%%%%%%%%%%%%%%%%%%%%%%%%%%%%%%%%%%%%%%%%%%%%%%%%%%%%%%%%%%%
%%%%%%%%%%%%%%%%%%%%%%%%%%%%%%%%%%%%%%%%%%%%%%%%%%%%%%%%%%%%%%%%%%%%%%
\begin{equation} 
{\boldsymbol {m^2_{H}}} = \left( 
\begin{array}{ccccc}
m_{{ \delta^{c^{0}} }{ \delta^{c^{0}} }} &m_{{ \overline{\delta}^{c^{0}} }{ \delta^{c^{0}} }} &m_{{\phi^{0}_{2}}{ \delta^{c^{0}} }} &m_{{\phi^{0}_{1}}{ \delta^{c^{0}} }} &m_{{ S }{ \delta^{c^{0}} }}\\ 
m_{{ \delta^{c^{0}} }{ \overline{\delta}^{c^{0}} }} &m_{{ \overline{\delta}^{c^{0}} }{ \overline{\delta}^{c^{0}} }} &m_{{\phi^{0}_{2}}{ \overline{\delta}^{c^{0}} }} &m_{{\phi^{0}_{1}}{ \overline{\delta}^{c^{0}} }} &m_{{ S }{ \overline{\delta}^{c^{0}} }}\\ 
m_{{ \delta^{c^{0}} }{\phi^{0}_{2}}} &m_{{ \overline{\delta}^{c^{0}} }{\phi^{0}_{2}}} &m_{{\phi^{0}_{2}}{\phi^{0}_{2}}} &m_{{\phi^{0}_{1}}{\phi^{0}_{2}}} &m_{{ S }{\phi^{0}_{2}}}\\ 
m_{{ \delta^{c^{0}} }{\phi^{0}_{1}}} &m_{{ \overline{\delta}^{c^{0}} }{\phi^{0}_{1}}} &m_{{\phi^{0}_{2}}{\phi^{0}_{1}}} &m_{{\phi^{0}_{1}}{\phi^{0}_{1}}} &m_{{ S }{\phi^{0}_{1}}}\\ 
m_{{ \delta^{c^{0}} }{ S }} &m_{{ \overline{\delta}^{c^{0}} }{ S }} &m_{{\phi^{0}_{2}}{ S }} &m_{{\phi^{0}_{1}}{ S }} &m_{{ S }{ S }}\end{array} 
\right).
 \end{equation}
%%%%%%%%%%%%%%%%%%%%%%%%%%%%%%%%%%%%%%%%%%%%%%%%%%%%%%%%%%%%%%%%%%%%%%
%%%%%%%%%%%%%%%%%%%%%%%%%%%%%%%%%%%%%%%%%%%%%%%%%%%%%%%%%%%%%%%%%%%%%%
The matrix elements are defined as 
%%%%%%%%%%%%%%%%%%%%%%%%%%%%%%%%%%%%%%%%%%%%%%%%%%%%%%%%%%%%%%%%%%%%%%
%%%%%%%%%%%%%%%%%%%%%%%%%%%%%%%%%%%%%%%%%%%%%%%%%%%%%%%%%%%%%%%%%%%%%%
 {\allowdisplaybreaks
\begin{align} 
m_{{ \delta^{c^{0}} }{ \delta^{c^{0}} }} &= \frac{ g_{R}^{2}}{2} ( 6 \bar{v}_{R}^{2}  - v_{d}^{2} -2 v_{R}^{2}  + v_{u}^{2})  -  g_V ^{2}(-3 \bar{v}_{R}^{2}  + v_{R}^{2})  + \lambda^{c2} v_{R}^{2}  +  m_1^2  + ( \lambda^c  v_s  +  \mu_2 )^{2},\nn\\ 
%%%%%%%%%%%%%%%%%%%%%%%%
%%%%%%%%%%%%%%%%%%%%%%%
m_{{ \delta^{c^{0}} }{ \overline{\delta}^{c^{0}} }} &=  \lambda^c  ((2 \lambda^c  v_R \bar{v}_R  + \mu_s v_s) -\lambda' v_d v_u  + {k} v_{s}^{2}  - M^2) -2 ( g_V ^{2} + g_{R}^{2})v_R \bar{v}_R  +  v_s { \lambda A_{\lambda} }  + {\mu_1^*B_2},\nn\\ 
%%%%%%%%%%%%%%%%%%%%%%%%
%%%%%%%%%%%%%%%%%%%%%%%
m_{{ \overline{\delta}^{c^{0}} }{ \overline{\delta}^{c^{0}} }} &=   g_V ^{2} (3 v_{R}^{2}  - \bar{v}_{R}^{2} ) + \frac{g_{R}^{2}}{2}(-2 \bar{v}_{R}^{2}  + 6 v_{R}^{2}  - v_{u}^{2}  + v_{d}^{2}) +  \lambda^{c2}  \bar{v}_{R}^{2}  +  m_2^2 + ( \lambda^c  v_s  +  \mu_2  )^{2},\nn\\ 
%%%%%%%%%%%%%%%%%%%%%%%%
%%%%%%%%%%%%%%%%%%%%%%%
m_{{ \delta^{c^{0}} }{\phi^{0}_{2}}} &= - \lambda^c~ \lambda' v_R v_u  - g_{R}^{2} v_d \bar{v}_R,~~m_{{ \overline{\delta}^{c^{0}} }{\phi^{0}_{2}}} = -  \lambda^c  \lambda' \bar{v}_R v_u  + g_{R}^{2} v_d v_R \nn,\\ 
%%%%%%%%%%%%%%%%%%%%%%%%
%%%%%%%%%%%%%%%%%%%%%%%
m_{{\phi^{0}_{2}}{\phi^{0}_{2}}} &=  \lambda'^2 v_{u}^{2}  + (\lambda' v_s  + {\mu })^{2} + \frac{g_{2}^{2}}{4} (3 v_{d}^{2}  - v_{u}^{2} ) + g_{R}^{2} (-2 \bar{v}_{R}^{2}  + 2 v_{R}^{2}  + 3 v_{d}^{2}  - v_{u}^{2} ) + m_5^2,\nn\\ 
%%%%%%%%%%%%%%%%%%%%%%%%
%%%%%%%%%%%%%%%%%%%%%%%
m_{{ \delta^{c^{0}} }{\phi^{0}_{1}}} &= -\lambda^c~ \lambda' v_d v_R  + g_{R}^{2} \bar{v}_R v_u, ~~~m_{{ \overline{\delta}^{c^{0}} }{\phi^{0}_{1}}} = -\lambda^c~ \lambda' v_d \bar{v}_R  - g_{R}^{2} v_R v_u, \nn\\ 
%%%%%%%%%%%%%%%%%%%%%%%%
%%%%%%%%%%%%%%%%%%%%%%%
m_{{\phi^{0}_{2}}{\phi^{0}_{1}}} &= -(\lambda' (\mu_s v_s  + {k} v_{s}^{2}  -2\lambda' v_d v_u  +  \lambda^c  v_R \bar{v}_R  - M^2) -2 v_s {\lambda'A_{\lambda'}}  -2 { \mu B }-\frac{(g_{2}^{2} + g_{R}^{2})}{2} v_d v_u, \nn\\ 
%%%%%%%%%%%%%%%%%%%%%%%%
%%%%%%%%%%%%%%%%%%%%%%%
m_{{\phi^{0}_{1}}{\phi^{0}_{1}}} &= ( \lambda'^2 v_{d}^{2}  + (\lambda' v_s  + \mu)^{2}) -\frac{ g_{2}^{2}}{4} (-3 v_{u}^{2}  + v_{d}^{2}) - \frac{g_{R}^{2}}{4} (-2 \bar{v}_{R}^{2}  + 2 v_{R}^{2}  -3 v_{u}^{2}  + v_{d}^{2}) + m_5^2,\nn\\ 
%%%%%%%%%%%%%%%%%%%%%%%%
%%%%%%%%%%%%%%%%%%%%%%%
m_{{ \delta^{c^{0}} }{ S }} &=   2  \lambda^c \bar{v}_R ( \lambda^c  v_s  +  \mu_2 ) +  \lambda^c v_R (2{k} v_s  +\mu_s) + v_R { \lambda A_{\lambda} },\nn \\ 
%%%%%%%%%%%%%%%%%%%%%%%%
%%%%%%%%%%%%%%%%%%%%%%%
m_{{ \overline{\delta}^{c^{0}} }{ S }} &=   \lambda^c  \bar{v}_R (2{k} v_s  + \mu_s ) + 2  \lambda^c  v_R ( \lambda^c  v_s  +  \mu_2 ) + \bar{v}_R { \lambda A_{\lambda} }, \nn\\ 
%%%%%%%%%%%%%%%%%%%%%%%%
%%%%%%%%%%%%%%%%%%%%%%%
m_{{\phi^{0}_{2}}{ S }} &= -\lambda'(2{k} v_s  +  \mu_s)v_u  + 4 \lambda' v_d (\lambda' v_s  + \mu) + v_u {\lambda'A_{\lambda'}},\nn\\ 
%%%%%%%%%%%%%%%%%%%%%%%%
%%%%%%%%%%%%%%%%%%%%%%%
m_{{\phi^{0}_{1}}{ S }} &= -\lambda'  (2{k} v_s  +  \mu_s)v_d -4 \lambda'(\lambda' v_s  + \mu)v_u  + v_d {\lambda'A_{\lambda'}}, \nn\\ 
%%%%%%%%%%%%%%%%%%%%%%%%
%%%%%%%%%%%%%%%%%%%%%%%
m_{{ S }{ S }} &= m_S^2+ \lambda^c  (2 k v_R \bar{v}_R  +  \lambda^c  (v_{R}^{2} + \bar{v}_{R}^{2}))+54 k_{s}^{2} v_{s}^{2} +\lambda' (-2{k} v_d v_u  +\lambda' (v_{d}^{2} + v_{u}^{2})) \nonumber \\ 
 &+ \mu_s^{2}+k (18 \mu_s v_s  - 6  M^2)+2 {\mu_sB_s} +6 v_s {k A_k}. \nn
\end{align}
} 
%%%%%%%%%%%%%%%%%%%%%%%%%%%%%%%%%%%%%%%%%%%%%%%%%%%%%%%%%%%%%%%%%%%%%%
%%%%%%%%%%%%%%%%%%%%%%%%%%%%%%%%%%%%%%%%%%%%%%%%%%%%%%%%%%%%%%%%%%%%%%
This scalar matrix can be diagonalized by the rotation matrix $Z^H$ as $Z^{H} {\bf m^2_{H} }Z^{H^\dagger} = m^2_{H,diag}$. We choose the numerical values of the parameters in such a way that the lightest component becomes the SM-like Higgs boson. We calculate the radiatively corrected Higgs mass up to two-loop for the top and stop sector as given in the Ref~\cite{Babu:2014vba}. The theoretical error in Higgs boson mass calculation allows for a mass range of $122-128$ GeV~\cite{MassH}. In our study, the lightest mass eigenstate for the CP-even Higgs boson is mostly composed of the bidoublet scalar fields. This is quite natural as the bidoublet fields are responsible for the EW symmetry breaking once they acquire non-zero VEVs at that scale.

Similarly, the imaginary component of the neutral Higgs fields produce the pseudo--scalar (CP-odd) states. Their mass--squared matrix in the gauge basis $(Im[{ \delta^{c^{0}} }], Im[ { \overline{\delta}^{c^{0}} }], Im[{\phi^{0}_{2}}], Im[{\phi^{0}_{1}}],\\ Im[{ S }])$ is given as 
%%%%%%%%%%%%%%%%%%%%%%%%%%%%%%%%%%%%%%%%%%%%%%%%%%%%%%%%%%%%%%%%%%%%%%
%%%%%%%%%%%%%%%%%%%%%%%%%%%%%%%%%%%%%%%%%%%%%%%%%%%%%%%%%%%%%%%%%%%%%%
\begin{equation} 
{\bf  m^2_{A}   } = \left( 
\begin{array}{ccccc}
M_{{ \delta^{c^{0}} }{ \delta^{c^{0}} }} & M_{{ \overline{\delta}^{c^{0}} }{ \delta^{c^{0}} }} &-\lambda^c~ \lambda' v_R v_u  &-  \lambda^c  \lambda' v_d v_R  & M_{{ S }{ \delta^{c^{0}} }}\\ 
 M_{{ \delta^{c^{0}} }{ \overline{\delta}^{c^{0}} }} & M_{{ \overline{\delta}^{c^{0}} }{ \overline{\delta}^{c^{0}} }} &-  \lambda^c  \lambda' \bar{v}_R v_u  &-  \lambda^c  \lambda' v_d \bar{v}_R  & M_{{ S }{ \overline{\delta}^{c^{0}} }}\\ 
-  \lambda^c  \lambda' v_R v_u  &-  \lambda^c  \lambda' \bar{v}_R v_u  & M_{{ \phi^{0}_{2} }{ \phi^{0}_{2} }} & M_{{ \phi^{0}_{1} }{ \phi^{0}_{2} }} & M_{{ S }{ \phi^{0}_{2} }}\\ 
-  \lambda^c  \lambda' v_d v_R  &- \lambda^c  \lambda' v_d \bar{v}_R  & M_{{ \phi^{0}_{2} }{ \phi^{0}_{1} }} & M_{{ \phi^{0}_{1} }{ \phi^{0}_{1} }} & M_{{ S }{ \phi^{0}_{1} }}\\ 
 M_{{ \delta^{c^{0}} }{ S }} & M_{{ \overline{\delta}^{c^{0}} }{ S }} &M_{{ \phi^{0}_{2} }{ S }} & M_{{ \phi^{0}_{1} }{ S }} & M_{{ S }{ S }}\end{array} 
\right),
 \end{equation}
%%%%%%%%%%%%%%%%%%%%%%%%%%%%%%%%%%%%%%%%%%%%%%%%%%%%%%%%%%%%%%%%%%%%%%
%%%%%%%%%%%%%%%%%%%%%%%%%%%%%%%%%%%%%%%%%%%%%%%%%%%%%%%%%%%%%%%%%%%%%%
where the elements of the above matrix are
%%%%%%%%%%%%%%%%%%%%%%%%%%%%%%%%%%%%%%%%%%%%%%%%%%%%%%%%%%%%%%%%%%%%%%
%%%%%%%%%%%%%%%%%%%%%%%%%%%%%%%%%%%%%%%%%%%%%%%%%%%%%%%%%%%%%%%%%%%%%%
{\allowdisplaybreaks
\begin{align} 
 M_{{ \delta^{c^{0}} }{ \delta^{c^{0}} }} &=    g_V ^{2} (- v_{R}^{2}  + \bar{v}_{R}^{2}) + \frac{g_{R}^{2}}{2} (2 \bar{v}_{R}^{2}  -2 v_{R}^{2}  - v_{d}^{2}  + v_{u}^{2}) +  \lambda^{c2}  v_{R}^{2}  +  m_1^2 +( \lambda^c  v_s  +  \mu_2 )^{2},\nn\\ 
%%%%%%%%%%%%%%%%%%%%%%%%
%%%%%%%%%%%%%%%%%%%%%%%
 M_{{ \delta^{c^{0}} }{ \overline{\delta}^{c^{0}} }} &= - {\mu_1^*B_2}  +  \lambda^c (\lambda' v_d v_u  -\mu_s v_s  -{k} v_{s}^{2}  +M^2 ) - v_s { \lambda A_{\lambda} },\nn \\ 
%%%%%%%%%%%%%%%%%%%%%%%%
%%%%%%%%%%%%%%%%%%%%%%%
 M_{{ \overline{\delta}^{c^{0}} }{ \overline{\delta}^{c^{0}} }} &=  g_V ^{2} ( v_{R}^{2}- \bar{v}_{R}^{2} ) + \frac{g_{R}^{2}}{2} ( 2 v_{R}^{2}  -2 \bar{v}_{R}^{2}  - v_{u}^{2}  + v_{d}^{2}) +  \lambda^{c2}  \bar{v}_{R}^{2}  +  m_2^2 + ( \lambda^c  v_s  +  \mu_2 )^{2},\nn\\ 
%%%%%%%%%%%%%%%%%%%%%%%%
%%%%%%%%%%%%%%%%%%%%%%%
 M_{{ \phi^{0}_{2} }{ \phi^{0}_{2} }} &= \lambda'^2 v_{u}^{2}  + ( \lambda' v_s  + \mu)^{2}\Big) + \frac{g_{2}^{2}}{4}(v_{d}^{2}- v_{u}^{2}) + \frac{g_{R}^{2}}{4}( 2 v_{R}^{2}  -2 \bar{v}_{R}^{2} - v_{u}^{2}  + v_{d}^{2}) + m_5^2,\nn\\
%%%%%%%%%%%%%%%%%%%%%%%%
%%%%%%%%%%%%%%%%%%%%%%% 
 M_{{ \phi^{0}_{2} }{ \phi^{0}_{1} }} &= \lambda' (\mu_s v_s  + {k} v_{s}^{2}  +  \lambda^c  v_R \bar{v}_R  - M^2) +2 v_s {\lambda'A_{\lambda'}}  +2 { \mu B },\nn\\ 
%%%%%%%%%%%%%%%%%%%%%%%%
%%%%%%%%%%%%%%%%%%%%%%%
 M_{{ \phi^{0}_{1} }{ \phi^{0}_{1} }} &=  \lambda'^2 v_{d}^{2}  + (\lambda' v_s  + \mu)^{2} + \frac{g_{2}^{2}}{4} ( v_{u}^{2}- v_{d}^{2})+ \frac{g_{R}^{2}}{4} (2 \bar{v}_{R}^{2}  -2 v_{R}^{2}  - v_{d}^{2}  + v_{u}^{2}) + m_5^2,\nn\\ 
%%%%%%%%%%%%%%%%%%%%%%%%
%%%%%%%%%%%%%%%%%%%%%%%
 M_{{ \delta^{c^{0}} }{ S }} &=  \lambda^c v_R (2{k} v_s  + \mu_s) - { \lambda A_{\lambda} }v_R,~~~M_{{ \overline{\delta}^{c^{0}} }{ S }} = \lambda^c \bar{v}_{R} (2{k} v_s  + \mu_s) - { \lambda A_{\lambda} }\bar{v}_{R},\nn\\
%%%%%%%%%%%%%%%%%%%%%%%%
%%%%%%%%%%%%%%%%%%%%%%%
 M_{{ \phi^{0}_{2} }{ S }} &=  -\lambda' v_u (2{k} v_s  +  \mu_s) + 2 v_u {\lambda'A_{\lambda'}},~~~ M_{{ \phi^{0}_{1} }{ S }} = -\lambda' v_d (2{k} v_s  +  \mu_s) + 2 v_d {\lambda'A_{\lambda'}},\nn\\
%%%%%%%%%%%%%%%%%%%%%%%%
%%%%%%%%%%%%%%%%%%%%%%%
 M_{{ S }{ S }} &= m_S^2-2 \lambda^c k v_R \bar{v}_R  +  \lambda^{c2} (v_{R}^{2} + \bar{v}_{R}^{2})+18 k_{s}^{2} v_{s}^{2} +2 \lambda'{k} v_d v_u  + \lambda'^2 (v_{d}^{2} + v_{u}^{2}) +  \mu_s^{2}\nonumber\\
 &~~~+2k (\mu_s v_s  +  M^2)-2 {\mu_s B_s} -6 v_s { k A_k }. \nn
\end{align}
} 
%%%%%%%%%%%%%%%%%%%%%%%%%%%%%%%%%%%%%%%%%%%%%%%%%%%%%%%%%%%%%%%%%%%%%%
%%%%%%%%%%%%%%%%%%%%%%%%%%%%%%%%%%%%%%%%%%%%%%%%%%%%%%%%%%%%%%%%%%%%%%
This pseudo--scalar mass--squared matrix can be diagonalized by the rotation matrix $Z^A$ as $Z^{A} {\bf  m^2_{A}   } Z^{A^\dagger} = m^2_{A,diag}$. After rotating the gauge fields into mass basis, we get three physical mass eigenstates $A_1, A_2$ and $ A_3$. The remaining two neutral states ($G_1^0, G_2^0$) become the massless Goldstone bosons which are absorbed as the longitudinal components of the corresponding gauge bosons $Z$ and $Z_R$ respectively. 

\subsubsection{Sfermionic sectors} 
The sfermions in our model consist of the scalar superpartners of the up- and down-type quarks and charged and neutral leptons. The existence of a right-handed neutrino and hence its superpartner is guaranteed by the extended gauge symmetry in this model, which leads to all the sfermion mass-squared matrices (including sneutrinos) being $6\times6$ matrices in general. We calculate the mass-squared matrices for the scalar down-type squarks, up-type squark, charged slepton and the sneutrino in the $(\widetilde{d}_{L}, \widetilde{d}_{R})$, $(\widetilde{u}_{L}, \widetilde{u}_{R})$, $(\widetilde{e}_L, \widetilde{e}_R)$ and $(\widetilde{\nu}_L, \widetilde{\nu}_R)$ gauge basis respectively. Thus, one can write the mass-squared matrices for the squarks as

%%%%%%%%%%%%%%%%%%%%%%%%%%%%%%%%%%%%%%%%%%%%%%%%%%%%%%%%%%%%%%%%%%%%%%
%%%%%%%%%%%%%%%%%%%%%%%%%%%%%%%%%%%%%%%%%%%%%%%%%%%%%%%%%%%%%%%%%%%%%%
\begin{minipage}{6.5cm}
\begin{equation} 
{\bf m^2_{\widetilde{d}} } = \left( 
\begin{array}{cc}
m_{\widetilde{d}_L\widetilde{d}_L^*} & m_{\widetilde{d}_L\widetilde{d}_R^*}\\ 
m_{\widetilde{d}_L\widetilde{d}_R^*}^\dagger  &m_{\widetilde{d}_R\widetilde{d}_R^*}\end{array} 
\right)
\label{Mmatrix-dquarks}
 \end{equation} 
\end{minipage}
\begin{minipage}{7.5cm}
\begin{equation} 
{\bf m^2_{\widetilde{u}} }= \left( 
\begin{array}{cc}
m_{\widetilde{u}_L\widetilde{u}_L^*} &m_{\widetilde{u}_L\widetilde{u}_R^*}  \\ 
m_{\widetilde{u}_L\widetilde{u}_R^*}^\dagger  &m_{\widetilde{u}_R\widetilde{u}_R^*}\end{array} 
\right) 
 \label{Mmatrix-uquarks}
 \end{equation}
\end{minipage}
\vspace{0.2cm}
%%%%%%%%%%%%%%%%%%%%%%%%%%%%%%%%%%%%%%%%%%%%%%%%%%%%%%%%%%%%%%%%%%%%%%
%%%%%%%%%%%%%%%%%%%%%%%%%%%%%%%%%%%%%%%%%%%%%%%%%%%%%%%%%%%%%%%%%%%%%%

where each matrix element is itself a $3\times3$ matrix given as 
{\allowdisplaybreaks
\begin{align} 
m_{\widetilde{d}_L\widetilde{d}_L^*} &= \delta_{{\alpha_1} {\beta_1}} (v_{d}^{2} {y_q  y_{q}^{T}}  + M_{QR}^2) +  \frac{  g_V ^{2}}{6} ( v_{R}^{2}- \bar{v}_{R}^{2}) + \frac{g_{2}^{2}}{4} ( v_{u}^{2}- v_{d}^{2}) \delta_{{\alpha_1} {\beta_1}}, \nn\\ 
m_{\widetilde{d}_R\widetilde{d}_R^*} &= \delta_{{\alpha_2} {\beta_2}} (v_{d}^{2} {y_{q}^{T}  y_q}  + M_{QR}^2) +  \frac{ g_V ^{2}}{6}(\bar{v}_{R}^{2}- v_{R}^{2}) + \frac{ g_{R}^{2}}{4} (2 \bar{v}_{R}^{2}  -2 v_{R}^{2}  - v_{d}^{2}  + v_{u}^{2})  \delta_{{\alpha_2} {\beta_2}}, \nn\\
m_{\widetilde{d}_L\widetilde{d}_R^*} &=  \delta_{{\alpha_1} {\beta_2}} (\lambda' v_s  + \mu)v_u y_q  - v_d \lambda T^{yq}),\nn\\
%%%%%%%%%%%%%%%%%%%%%%%%%%%%%%%%%%%%%%
%%%%%%%%%%%%%%%%%%%%%%%%%%%%%%%%%%%%%
m_{\widetilde{u}_L\widetilde{u}_L^*} &= \delta_{{\alpha_1} {\beta_1}} (v_{u}^{2} {y_q  y_{q}^{T}}  + M_{QR}^2) + \frac{  g_V ^{2}}{6} ( v_{R}^{2}- \bar{v}_{R}^{2}) + \frac{g_{2}^{2}}{4} ( v_{d}^{2}- v_{u}^{2})  \delta_{{\alpha_1} {\beta_1}}, \nn\\ 
m_{\widetilde{u}_R\widetilde{u}_R^*} &= \delta_{{\alpha_2} {\beta_2}} (v_{u}^{2} {y_{q}^{T}  y_q}  + M_{QR}^2) + \frac{ g_V ^{2}}{6} ( \bar{v}_{R}^{2}- v_{R}^{2}) + 3 \frac{g_{R}^{2}}{4} ( 2 v_{R}^{2}  -2 \bar{v}_{R}^{2} - v_{u}^{2}  + v_{d}^{2})  \delta_{{\alpha_2} {\beta_2}},\nn\\
m_{\widetilde{u}_L\widetilde{u}_R^*} &= \delta_{{\alpha_1} {\beta_2}} (-v_d(\lambda' v_s  + \mu)y_q  + v_u T^{yq}).
\end{align}}
Here $\alpha_1$, $\alpha_2$, $\beta_1$ and $\beta_2$ represents the color indices. These $3\times3$ matrices in general can be non-diagonal with the off-diagonal elements allowing for mixing between the various flavors. We do not consider flavor violating process in our study and hence, for simplicity, we will just consider the case where the matrices are diagonal and their elements are real. 

Similarly, the slepton mass-squared matrices are given as

%%%%%%%%%%%%%%%%%%%%%%%%%%%%%%%%%%%%%%%%%%%%%%%%%%%%%%%%%%%%%%%%%%%%%%
%%%%%%%%%%%%%%%%%%%%%%%%%%%%%%%%%%%%%%%%%%%%%%%%%%%%%%%%%%%%%%%%%%%%%%
\begin{minipage}{6.5cm}
\begin{equation} 
{\bf  m^2_{\widetilde{e}} }= \left( 
\begin{array}{cc}
m_{\widetilde{e}_L\widetilde{e}_L^*} & m_{\widetilde{e}_L\widetilde{e}_R}  \\ 
m_{\widetilde{e}_L\widetilde{e}_R^*}^\dagger  &m_{\widetilde{e}_R\widetilde{e}_R^*}\end{array} 
\right) 
 \label{Mmatrix-csleptons}
 \end{equation} 
\end{minipage}
\begin{minipage}{7.5cm}
\begin{equation} 
{\bf  m^2_{\widetilde{\nu}}} = \left( 
\begin{array}{cc}
m_{\widetilde{\nu}_L\widetilde{\nu}_L^*} & m_{\widetilde{\nu}_L\widetilde{\nu}_R} \\ 
m_{\widetilde{\nu}_L\widetilde{\nu}_R^*}^\dagger   &m_{\widetilde{\nu}_R\widetilde{\nu}_R^*}\end{array} 
\right) 
 \label{Mmatrix-nusleptons}
 \end{equation}
\end{minipage}
\vspace{0.2cm}
%%%%%%%%%%%%%%%%%%%%%%%%%%%%%%%%%%%%%%%%%%%%%%%%%%%%%%%%%%%%%%%%%%%%%%
%%%%%%%%%%%%%%%%%%%%%%%%%%%%%%%%%%%%%%%%%%%%%%%%%%%%%%%%%%%%%%%%%%%%%%

with the matrix elements in the slepton sector being
{\allowdisplaybreaks
\begin{align} 
m_{\widetilde{e}_L\widetilde{e}_L^*} &= \frac{  g_V ^{2}}{2} ( \bar{v}_{R}^{2}- v_{R}^{2}) + \frac{g_{2}^{2}}{4} ( v_{u}^{2}- v_{d}^{2})   + v_{d}^{2} {y_l  y_{l}^{T}}  + M_{LL}^2,\nn\\ 
m_{\widetilde{e}_R\widetilde{e}_R^*} &= \frac{  g_V ^{2}}{2} ( v_{R}^{2}- \bar{v}_{R}^{2}) + \frac{ g_{R}^{2}}{4} (2 \bar{v}_{R}^{2} -2 v_{R}^{2}  - v_{d}^{2}  + v_{u}^{2})   + v_{d}^{2} {y_{l}^{T}  y_l}  + M_{LR}^2,\nn\\
m_{\widetilde{e}_L\widetilde{e}_R} &=y_l (\lambda' v_s  +  \mu)v_u  - v_d T^{yl},\nn\\
%%%%%%%%%%%%%%%%%%%%%%%%%%%%%%%
%%%%%%%%%%%%%%%%%%%%%%%%%%%%%%%%%%%
m_{\widetilde{\nu}_L\widetilde{\nu}_L^*} &=  \frac{  g_V ^{2}}{2} ( \bar{v}_{R}^{2}- v_{R}^{2}) + \frac{ g_{2}^{2}}{4}(v_{d}^{2}- v_{u}^{2})   + v_{u}^{2} {y_l  y_{l}^{T}}  + M_{LL}^2\nn,\\ 
m_{\widetilde{\nu}_R\widetilde{\nu}_R^*} &=  \bar{v}_{R}^{2} {f^{cT} f^c}  + \frac{  g_V ^{2}}{2} ( v_{R}^{2}- \bar{v}_{R}^{2}) + \frac{ g_{R}^{2}}{4}( 2 v_{R}^{2}  -2 \bar{v}_{R}^{2}  - v_{u}^{2}  + v_{d}^{2})   + v_{u}^{2} {y_{l}^{T}  y_l}  + M_{LR}^2,\nn\\
m_{\widetilde{\nu}_L\widetilde{\nu}_R} &=- y_l v_d (\lambda' v_s  + \mu) + v_u T^{yl}.\nn
\end{align}
} 
%%%%%%%%%%%%%%%%%%%%%%%%%%%%%%%%%%%%%%%%%%%%%%%%%%%%%%%%%%%%%%%%%%%%%%
%%%%%%%%%%%%%%%%%%%%%%%%%%%%%%%%%%%%%%%%%%%%%%%%%%%%%%%%%%%%%%%%%%%%%%

The sfermionic mass matrices namely the down-squarks, up-squarks, charged-sleptons and sneutrinos given in Eqns.~\ref{Mmatrix-dquarks}-\ref{Mmatrix-nusleptons} can be diagonalized by the rotation matrices $U^{DL}$, $U^{UL}$, $U^{EL}$ and $U^{VL}$ respectively. 

\subsubsection{Electroweakino sectors}

The particle spectrum of our model allows for a large number of physical chargino and neutralino states (together referred to as electroweakinos from here on) which arise from the mixing of the charged and neutral gauginos and higgsinos respectively. Since R-parity is naturally conserved in this model, the lightest neutralino is stable and can be a good dark matter candidate. The electroweakinos are also very important for our study as the primary SUSY decay channels for the heavy gauge bosons will be into these particles, as will be seen in the next section. 

The chargino mass matrix in the basis $ (\widetilde{W}_R^-, \widetilde{W}_L^-, \widetilde{\delta}^{c^-}, \widetilde{\phi}_2^-)$ and $ (\widetilde{W}_R^+, \widetilde{W}_L^+, \widetilde{\overline{\delta}}^{c^+}, \widetilde{\phi}_2^+)$ can be written as 
%%%%%%%%%%%%%%%%%%%%%%%%%%%%%%%%%%%%%%%%%%%%%%%%%%%%%%%%
%%%%%%%%%%%%%%%%%%%%%%%%%%%%%%%%%%%%%%%%%%%%%%%%%%%%%%%%
\begin{equation} 
{\bf  m_{{\chi}_c}} = \left(  \begin{array}{c c c c} \widetilde{W}_R^-& \widetilde{W}_L^- & \widetilde{\delta}^{c^-}& \widetilde{\phi}_2^-\end{array} \right)
%%%%%%%%%%%%%%%%%%%%%%
\left( 
\begin{array}{cccc}
\frac{M^{W_R}_{{1 1}} + M^{W_R}_{{2 2}}}{2} &0 &\sqrt{2} g_R v_R  &g_R v_d \\ 
0 &\frac{1M^{W_L}_{{1 1}} + M^{W_L}_{{2 2}}}{2} &0 &g_2 v_u \\ 
- \sqrt{2} g_R \bar{v}_R  &0 & \lambda^c  v_s  +  \mu_2   &0\\ 
g_R v_u  &g_2 v_d  &0 & ( \lambda' v_s  + \mu)\end{array} 
\right)
%%%%%%%%%%%%%%%%%%%%%%%%%%%
\left( \begin{array}{c}
 \widetilde{W}_R^+ \\
 \widetilde{W}_L^+ \\
 \widetilde{\overline{\delta}}^{c^+}\\
 \widetilde{\phi}_2^+
\end{array}\right)
\label{Mmatrix-charginos}
 \end{equation}
%%%%%%%%%%%%%%%%%%%%%%%%%%%%%%%%%%%%%%%%%%%%%%%%%%%%%%%
%%%%%%%%%%%%%%%%%%%%%%%%%%%%%%%%%%%%%%%%%%%%%%%%%%%%%%
It is easy to see that this chargino mass matrix is asymmetric and can only be diagonalized by a bi-unitary transformation with $m_{{\chi}_c,diag} = U^{{Lm}} {\bf  m_{{\chi}_c} }U^{{Rp}^\dagger}$. Please note that the left-handed triplet higgsinos remain decoupled from these charginos and neutralinos due to left-handed triplet Higgs boson not acquiring any VEV. Hence we have a total of eight neutral electroweakinos which mix among each other in the gauge basis
\begin{equation}
(\widetilde B, \widetilde{W_R^0}, \widetilde{W_L^0}, \widetilde{\delta}^{c^{0}}, \widetilde{\bar{\delta}}^{c^{0}}, \widetilde{\phi}_2^0, \widetilde{\phi}_1^0, \widetilde{S}), 
\label{Eq:NetrBasis}
\end{equation}
 with their corresponding mass matrix as
%%%%%%%%%%%%%%%%%%%%%%%%%%%%%%%%%%%%%%%%%%%%%%%%%%%%%%%
%%%%%%%%%%%%%%%%%%%%%%%%%%%%%%%%%%%%%%%%%%%%%%%%%%%%%%
\begin{equation} 
\hspace*{-0.2cm}
{\bf  m_{\chi}} = \left( 
\begin{array}{cccccccc}
M_1 &0 &0 &m_{\widetilde{\delta}^{c^{0}}\widetilde{H}_d^0} &\sqrt{2}  g_V  v_R  &0 &0 &0\\ 
0 &M^{W_R}_{{3 3}} &0 &\sqrt{2} g_R \bar{v}_R  &m_{\widetilde{\bar{\delta}}^{c^{0}}\widetilde{W_R^0}} &m_{\widetilde{\phi}_2^0\widetilde{W_R^0}} &\frac{1}{\sqrt{2}} g_R v_u  &0\\ 
0 &0 &M^{W_L}_{{3 3}} &0 &0 &\frac{1}{\sqrt{2}} g_2 v_d  &m_{\widetilde{\phi}_1^0\widetilde{W_L^0}} &0\\ 
m_{\widetilde{H}_d^0\widetilde{\delta}^{c^{0}}} &\sqrt{2} g_R \bar{v}_R  &0 &0 &m_{\widetilde{\bar{\delta}}^{c^{0}}\widetilde{\delta}^{c^{0}}} &0 &0 & \lambda^c  v_R \\ 
\sqrt{2}  g_V  v_R  &m_{\widetilde{W_R^0}\widetilde{\bar{\delta}}^{c^{0}}} &0 &m_{\widetilde{\delta}^{c^{0}}\widetilde{\bar{\delta}}^{c^{0}}} &0 &0 &0 & \lambda^c  \bar{v}_R \\ 
0 &m_{\widetilde{W_R^0}\widetilde{\phi}_2^0} &\frac{1}{\sqrt{2}} g_2 v_d  &0 &0 &0 &m_{\widetilde{\phi}_1^0\widetilde{\phi}_2^0} &-\lambda' v_u \\ 
0 &\frac{1}{\sqrt{2}} g_R v_u  &m_{\widetilde{W_L^0}\widetilde{\phi}_1^0} &0 &0 &m_{\widetilde{\phi}_2^0\widetilde{\phi}_1^0} &0 &-\lambda' v_d \\ 
0 &0 &0 & \lambda^c  v_R  & \lambda^c  \bar{v}_R  &-\lambda' v_u  &-\lambda' v_d  &m_{\widetilde{S}\widetilde{S}}\end{array} 
\right).
\label{Mmatrix-Neutralino}
\end{equation}  
%%%%%%%%%%%%%%%%%%%%%%%%%%%%%%%%%%%%%%%%%%%%%%%%%%%%%%%
%%%%%%%%%%%%%%%%%%%%%%%%%%%%%%%%%%%%%%%%%%%%%%%%%%%%%%
Here
%%%%%%%%%%%%%%%%%%%%%%%%%%%%%%%%%%%%%%%%%%%%%%%%%%%%%%%
%%%%%%%%%%%%%%%%%%%%%%%%%%%%%%%%%%%%%%%%%%%%%%%%%%%%%%
{\allowdisplaybreaks
\begin{align} 
m_{\widetilde{H}_d^0\widetilde{\delta}^{c^{0}}} &= - \sqrt{2}  g_V  \bar{v}_R,~~m_{\widetilde{W_R^0}\widetilde{\bar{\delta}}^{c^{0}}} = - \sqrt{2} g_R v_R,~~m_{\widetilde{\delta}^{c^{0}}\widetilde{\bar{\delta}}^{c^{0}}} =  \lambda^c  v_s  +  \mu_2,~~m_{\widetilde{W_R^0}\widetilde{\phi}_2^0} = - \frac{1}{\sqrt{2}} g_R v_d,~~\nn  \\ 
m_{\widetilde{W_L^0}\widetilde{\phi}_1^0} &= - \frac{1}{\sqrt{2}} g_2 v_u,~~m_{\widetilde{\phi}_2^0\widetilde{\phi}_1^0} = -(\lambda' v_s  +\mu ),~~
m_{\widetilde{S}\widetilde{S}} = (2 {k} v_s  +  \mu_s).
\end{align}
}
%%%%%%%%%%%%%%%%%%%%%%%%%%%%%%%%%%%%%%%%%%%%%%%%%%%%%%%
%%%%%%%%%%%%%%%%%%%%%%%%%%%%%%%%%%%%%%%%%%%%%%%%%%%%%%
This matrix is diagonalized by $Z^{fN}$ rotation matrix as $ m_{\chi,diag} = Z^{{fN}} {\bf  m_{\chi}} Z^{{fN}^\dagger}$.
Also in this model, it is possible to get two types of doubly--charged chargino particles, one from the $SU(2)_L$ triplet and another from $SU(2)_R$ triplet sectors. These do not mix among each other, resulting in the left-handed triplets being quite massive while the right-handed doubly-charged higgsinos can remain light with a mass of
%%%%%%%%%%%%%%%%%%%
%%%%%%%%%%%%%%%%%%%%
\begin{equation}
M_{\chi^{\pm\pm}}=  \lambda^c  v_s  +  \mu_2.  \,\,
\label{MassXD}
\end{equation}
%%%%%%%%%%%%%%%%%%%%
%%%%%%%%%%%%%%%%%%%%

\section{Heavy gauge boson decays for different LSP compositions} \label{BP}

In this study, we mainly concentrate on the heavy $W_R$ and $Z_R$ gauge bosons. Depending on the numerical values of the new parameters such as gauge couplings $g_R,g_V$ and the vacuum expectation values, i.e., the minimum of the potential in the particular scalar field directions (except the singlet scalar field), the masses of these gauge bosons could change.
In the presence of light LRSUSY particles, and if kinematically allowed, these heavy gauge bosons can decay to these particles with a significantly high branching fraction. In order to estimate these BSM decays, we choose the parameters in such a way that all the sparticle sectors are sufficiently heavy except the electroweakinos (see Eqn.~\ref{Eq:NetrBasis}) so that the heavy gauge bosons can only decay into these light electroweakinos. We specifically focus on this sector as the SUSY decays of the heavy gauge bosons is maximum here. First we consider different benchmark points where the light electroweakinos are primarily composed of only one type of fermionic fields (a particular type of gaugino or higgsino). Then we allow for mixing between the various fermionic states such that the decay branching fractions of the gauge bosons can change significantly. We also vary the masses of the squarks and sleptons so as to open up the gauge bosons decay channels into sfermionic final states. Thus in this section we study a host of scenarios with various possible final states for the heavy gauge boson decays and study the corresponding decay BRs. 

A systematic study of the heavy gauge bosons decay channels requires one to deal with a large number of free parameters in the minimal LRSUSY model that has been considered in the paper. The experimentally measured particle masses and other low energy observables can be used to constrain the Yukawa sector of the model to a certain extend. The scalar couplings, on the other hand, have some bounds coming from the measured Higgs boson properties. We further require the lightest neutralino to be the LSP as it can then be a good dark matter candidate. Yet there are a large number of free parameters in the model, most of which do not have any significant effect on our results. We thus keep the numerical values of these parameter to be constant for the rest of our analysis as can be seen in Tab.~\ref{table-Fix}. Varying the rest of the parameters, we consider different field combinations for the electroweakino sector while also varying the sfermion masses to study the corresponding gauge boson decays.

%%%%%%%%%%%%%%%%%%%%%%%%%%%%%%%%%%%%%%%%%%%%%
\begin{table}[h!]
\begin{center}\scalebox{0.8}{
\begin{tabular}{||C{20cm}||}
\hline
\hline
Parameters\\
\hline
\hline
\\
%%%%%%%%%%%
$M^2_{QL,11}=6.75\times10^7$, $M^2_{QL,22}=6.74\times10^7$, $M^2_{QL,33}=4.90\times10^7$,
%%%%%%%%%%%%%%
$M^2_{QR,11}=9.76\times10^7$, $M^2_{QR,22}=9.75\times10^7$, $M^2_{QR,33}=2.90\times10^7$,\\
%%%%%%%%%%%%%%
\\
$M^2_{LL,11}=9.60\times10^7$, $M^2_{LL,22}=9.20\times10^7$, $M^2_{LL,33}=9.0\times10^7$,
%%%%%%%%%%%%%%
$M^2_{LR,11}=9.20\times10^7$, $M^2_{LR,22}=9.10\times10^7$, $M^2_{LR,33}=9.0\times10^7$,\\
%%%%%%%%%%%%%%
\\
$ y_{l, 11}= 3.34\times10^{-5}$, $ y_{l, 22}=0.007$,  $ y_{l, 33}=0.118$,
%%%%%%%%%%%%%%%%%
$T^{yl}_{11}= -1000$, $T^{yl}_{22}=-1000$,  $T^{yl}_{33}=  -2000$,\\
$ y_{l, ij}=0=T^{yl}_{ij}$, for $i\neq j$,\\
%%%%%%%%%%%%%%
\\
$ y_{q, 11}=1.45\times10^{-5}$, $ y_{q, 22}=0.0073$,  $ y_{q, 33}=1.0006$,
%%%%%%%%%%%%%%
$T^{yq}_{11}= -3000$, $T^{yq}_{22}= -1000 $,\\
$ y_{q, ij}=0=T^{yq}_{ij}$, for $i\neq j$,\\
%%%%%%%%%%%%%%
\\
$ f^c_{11}= 1.4$, $ f^c_{22}=1.6$,  $ f^c_{33}= 1.8$,
%%%%%%%%%%%%%
$T^{f^c}_{11}= -12100$, $T^{f^c}_{22}=-3200$,  $T^{f^c}_{33}= -4200$,\\
$ f^c_{ij}=0=T^{f^c}_{ij}$, for $i\neq j$.\\
%%%%%%%%%%%%%%
\\
$ { \lambda' } =0.10$, $ { k } =1.8$,\\
\\
${\mu_1^* B_2 }=-2000$, ${\mu_s B_s }=10^{10}$, ${ M^2}=-10^8$, ${ \lambda C_{\lambda} }= 10^3$,
%%%%%%%%%%%%%%
${ \lambda A_{\lambda} }=-100$, ${ k A_{k} }=10000$, ${ \lambda' A_{\lambda'} }=-3500$, $M_G=6000$,\\
\\
$v_u=173.457$, $v_d=15$, $v_R=6365$, $\bar{v}_R=3000$, $v_s=3500$,
$g_V=0.36$, $g_2=g_R=0.64$\\
\hline
\hline
\end{tabular}}
\end{center}
\caption{ These parameters remain fixed throughout this section. The unit of the Mass parameter is in GeV and Mass-squared is in ${\rm GeV^2}$.}
\label{table-Fix}
\end{table}
%%%%%%%%%%%%%%%%%%%%%%%%%%%%%%%%%%%%%%%%%%%%%%%%%%%%%%%%

\subsection{Case-1: Single component LSP}
%%%%%%%%%%%%%%%%%%%%%

We first identify the parameter spaces where the LSP is mostly composed of only one type of component among the neutral fermion fields in the basis given in Eqn.~\ref{Eq:NetrBasis}. We make sure all other SUSY particles are much heavier so that the heavy gauge bosons do not decay into these states. It can be seen from the Eqn.~\ref{MassWZ} that the mass of the neutral $Z_R$ boson always remains $\sqrt{2 (1+ g_R^2 g_V^{-2})}$ times heavier than the mass of $W_R$ boson. 
We keep the numerical values of the VEVs fixed at $v_u=173.457$ GeV, $v_d=15$ GeV, $v_R=6365$ GeV, $\bar{v}_R=3000$ GeV, and $v_s=3500$ GeV and the gauge couplings $g_V=0.36$, $g_2=g_R=0.64$. Thus the masses of the heavy gauge bosons remain unchanged at $M_{W_R}=4.5$ TeV and $M_{Z_R}=7.3$ TeV throughout this section. One could also choose different values of the VEVs and the gauge couplings to get different gauge bosons masses. We however choose relatively heavy masses for the gauge bosons to evade the experimental bounds~\cite{Aaboud:2017buh,Sirunyan:2017ukk,Khachatryan:2016jww}. As discussed earlier, we have fixed the numerical values of a number of parameters in the model which are shown in Tab.~\ref{table-Fix}. Note that the large values of the right-handed Yukawa couplings $ f^c_{ii}$ in the table results in the right-handed neutrinos being heavier than the $W_R$ boson mass. Thus it is impossible for the heavy gauge bosons to decay into right-handed neutrino final states. The dominant contribution to the LSP from each fermion field is shown in the Tab.~\ref{table-100LSP} and will be discussed in details below.

%%%%%%%%%%%%%%%%%%%%%%%%%%%%%%%%%%%%%%%%%%%%%
\begin{table}[h!]
\begin{center}\scalebox{0.7}{
\begin{tabular}{|c||c|c|c|}
\hline
\hline
~~~~ &  &\multicolumn{2}{c|}{ }\\
~~LSP-Type~~ & Benchmark points &\multicolumn{2}{c|}{ Branching Ratio of $W_R,~Z_R$ into different BSM fields}\\
~~~~ &  &\multicolumn{2}{c|}{ }\\
\cline{3-4} 
 &~~~~~~~~~~~~~ &   $W_R^\pm \rightarrow BSM$       &   $Z_R   \rightarrow BSM  $\\
\hline
1a.Bino   (95 $\%$) &   $ M_1 = 0$, $ M^{W_L}_{  33,22,11} = 60$ TeV, $ M^{W_R}_{  33,22,11} = 60$ TeV,   & &  \\
&$ { \lambda^c }  = -0.2$, $ { \mu_2 }  = 21$ TeV,  $ {\mu}  = -17$ TeV,&$0\%$         &   $0.01\%$ ($Z_R \rightarrow   { \chi^0_1 }      { \chi^0_1 }   $)\\
&$ { \mu_s }  = -20$ TeV,  $ T^{yq}_{  33}=82$ TeV&&\\
&&&\\
&$M_{\chi_1^0}=475$ GeV, $M_{\chi_2^0}=7.4$ TeV, $M_{\chi_1^\pm}=16.6$ TeV&&\\
\hline
1b.Bino   (95 $\%$) &   $ M_1 = -300$ GeV, $ M^{W_L}_{  33,22,11} = 60$ TeV, $ M^{W_R}_{  33,22,11} = 60$ TeV,   &&\\
&$ { \lambda^c }  = -0.2$, $ { \mu_2 }  = 21$ TeV,  $ {\mu}  = -17$ TeV,& $0\%$         &   $0.01\%$ ($Z_R \rightarrow   { \chi^0_1 }      { \chi^0_1 }   $)\\
&$ { \mu_s }  = -20$ TeV,  $ T^{yq}_{  33}=82$ TeV&&\\
&&&\\
&$M_{\chi_1^0}=184$ GeV, $M_{\chi_2^0}=7.4$ TeV, $M_{\chi_1^\pm}=16.6$ TeV&&\\
\hline
2. Wino-L  (99.9 $\%$) &  $ M_1 = 60$ TeV, $ M^{W_L}_{  33,22,11} = 500$ GeV, $ M^{W_R}_{  33,22,11} = 60$ TeV,   &&\\
& $ { \lambda^c }  = -0.2$, $ { \mu_2 }  = 21$ TeV,  $ {\mu}  = -17$ TeV,&  $0\%$          &   $0\%$\\
&$ { \mu_s }  = -20$ TeV,  $ T^{yq}_{  33}=82$ TeV&&\\
&&&\\
&$M_{\chi_1^0}=500$ GeV, $M_{\chi_2^0}=7.4$ TeV, $M_{\chi_1^\pm}=500.1$ GeV&&\\
\hline
3a. Wino-R   (95 $\%$)    &$ M_1 = 60$ TeV, $ M^{W_L}_{  33,22,11} = 60$ TeV, $ M^{W_R}_{  33} = -1350$, &&\\
 &   $ M^{W_R}_{  22,11} = 60$ TeV, $ { \lambda^c }  = -0.2$, $ { \mu_2 }  = 21$ TeV,  & $0\%$        &   $0.1\%$ ($Z_R \rightarrow   { \chi^0_1 }      { \chi^0_1 }   $) \\
&  $ {\mu}  = -17$ TeV, $ { \mu_s }  = -20$ TeV,  $ T^{yq}_{  33}=82$ TeV&&\\
          &    &         &   \\
&$M_{\chi_1^0}=172$ GeV, $M_{\chi_2^0}=7.4$ TeV, $M_{\chi_1^\pm}=16.6$ TeV&&\\
\hline
3b. Wino-R   (95 $\%$)     & $ M_1 = 60$ TeV, $ M^{W_L}_{  33,22,11} = 60$ TeV, $ M^{W_R}_{  33,22,11} = -1350$, &&\\
&   $ { \lambda^c }  = -0.2$, $ { \mu_2 }  = 21$ TeV,  $ {\mu}  = -17$ TeV,   &   $28.9\%$ ($W_R^\pm \rightarrow   { \chi^0_1 }      \chi^\pm_1  $) &         $24.9\%$ ($Z_R \rightarrow   \chi^\mp_1     \chi^\pm_1  $)\\
&$ { \mu_s }  = -20$ TeV,  $ T^{yq}_{  33}=82$ TeV&&\\
       &        &   &\\
&$M_{\chi_1^0}=172$ GeV, $M_{\chi_2^0}=7.4$ TeV, $M_{\chi_1^\pm}=551$ GeV&&\\
\hline
3c. Wino-R   (95 $\%$)  &$ M_1 = 60$ TeV, $ M^{W_L}_{  33,22,11} = 60$ TeV, $ M^{W_R}_{  33} = -1350$,&&\\
  &   $ M^{W_R}_{  22,11} = -955$, $ { \lambda^c }  = -0.2$, $ { \mu_2 }  = 21$ TeV,    &   $29.3\%$ ($W_R^\pm \rightarrow   { \chi^0_1 }      \chi^\pm_1  $)      &   $25\%$ ($Z_R \rightarrow   \chi^\mp_1     \chi^\pm_1  $)\\
&$ {\mu}  = -17$ TeV,  $ { \mu_s }  = -20$ TeV,  
$ T^{yq}_{  33}=82$ TeV&&\\
       &        &&\\
&$M_{\chi_1^0}=172$ GeV, $M_{\chi_2^0}=7.4$ TeV, $M_{\chi_1^\pm}=175$ GeV&&\\
\hline
4.Higgsino-($\phi_{1,2}$)     &  $ M_1 = 60$ TeV, $ M^{W_L}_{  33,22,11} = 60$ TeV, $ M^{W_R}_{  33,22,11} = 60$ TeV,    &   $9\%$ ($W_R^\pm \rightarrow   { \chi^0_1 }      \chi^\pm_1  $)         &  $8\%$ ($Z_R \rightarrow   { \chi^0_1 }      { \chi^0_2 }   $)\\
(49.9 $\%$ each)&$ { \lambda^c }  = -0.2$, $ { \mu_2 }  = 21$ TeV,  $ {\mu}  = -1$ TeV, &$9\%$ ($W_R^\pm \rightarrow   { \chi^0_2 }      \chi^\pm_1  $)&$8\%$ ($Z_R \rightarrow   \chi^\mp_1     \chi^\pm_1  $)\\
&$ { \mu_s }  = -20$ TeV,  $ T^{yq}_{  33}=79$ TeV&&\\
&&&\\
&$M_{\chi_1^0}=649.9$ GeV, $M_{\chi_2^0}=650$ GeV, $M_{\chi_1^\pm}=650$ GeV&&\\
\hline
5.Higgsino-($\Delta_R^{1,2}$)  & $ M_1 = 60$ TeV, $ M^{W_L}_{  33,22,11} = 60$ TeV, $ M^{W_R}_{  33,22,11} = 60$ TeV, & $12\%$ ($W_R^\pm \rightarrow   { \chi^0_1 }      \chi^\pm_1  $)& $33.7\%$ ($Z_R \rightarrow   { \chi^0_1 }      { \chi^0_2 }   $)\\
(57.7 and 41.6 $\%$)        &  $ { \lambda^c }  = -0.05$, $ { \mu_2 }  = -500$ GeV,  $ {\mu}  = -17$ TeV,   &   $12\%$ ($W_R^\pm \rightarrow   { \chi^0_2 }      \chi^\pm_1  $)         & $2.02\%$ ($Z_R \rightarrow   \chi^\mp_1     \chi^\pm_1  $) \\
&  $ { \mu_s }  = -20$ TeV,  $ T^{yq}_{  33}=78.2$ TeV &$21.12\%$ ($W_R^pm \rightarrow   \chi^\mp_1   \chi^{\pm\pm}$)& $9.45\%$ ($Z_R \rightarrow \chi^{\mp\mp} \chi^{\pm\mp}$)\\
&&& \\
&&&\\
&$M_{\chi_1^0}=434$ GeV, $M_{\chi_2^0}=698.5$ GeV, $M_{\chi_1^\pm}=648.9$ GeV&&\\
\hline
6.Singlino    (99.9 $\%$)     & $ M_1 = 60$ TeV, $ M^{W_L}_{  33,22,11} = 60$ TeV, $ M^{W_R}_{  33,22,11} = 60$ TeV, &&\\
&$ { \lambda^c }  = -0.2$, $ { \mu_2 }  = 21$ TeV,  $ {\mu}  = -17$ TeV, & $0\%$   &        $0.0001\%$ ($Z_R \rightarrow   { \chi^0_1 }      { \chi^0_1 }   $)\\
&$ { \mu_s }  = -12$ TeV,  $ T^{yq}_{  33}=86.4$ TeV,&&\\
&&&\\
&$M_{\chi_1^0}=521$ GeV, $M_{\chi_2^0}=16.6$ TeV, $M_{\chi_1^\pm}=16.6$ TeV&&\\
\hline
\end{tabular}}
\end{center}
\caption{ The LSP is mostly composed of only one type of component among the neutral fermion fields in the basis given in Eqn.~\ref{Eq:NetrBasis}. $ M_{\chi_i} ( \chi_i = \chi_{1,2}^0, \chi_1^\pm)$ stands for the masses of the electroweakinos for these benchmark points. It is to be noted that the other parameters are fixed as in the Tab.~\ref{table-Fix}.} 
\label{table-100LSP}
\end{table}
%%%%%%%%%%%%%%%%%%%%%%%%%%%%%%%%%%%%%%%%%%%%%%%%%%%%%%%%

\subsubsection{Bino-like LSP}
Let us first consider the case where the neutralino LSP is mostly composed of $\widetilde B$. As is quite evident from the neutralino mass matrix given in Eqn.~\ref{Mmatrix-Neutralino}, a bino-type LSP would require one to choose a small value of the parameter $M_1$. Choosing a positive value of $M_1$ though can only result in a lower bound of the bino-type LSP mass with $M_{  { \chi^0_1 }} \geq 475$ GeV. So even if we choose $M_1$ = 0 GeV, the lightest neutralino is around 475 GeV. This is because the off-diagonal $(1,5)$ and $(1,6)$ elements of the ${\bf   {m_\chi}   }$ matrix which are $\sqrt{2} g_V  v_R$ and $\sqrt{2} g_V  \bar{v}_R$ respectively are large producing significant corrections to the bino mass. However as $M_1$~\footnote{ One can also choose different numerical values of the VEVs and gauge couplings to get the lighter LSP.} is a free parameter in this model we are free to choose a negative value for it and thus can bring down the LSP mass to any value we want. In Tab.~\ref{table-100LSP} we show two BPs 1.a and 1.b having $M_1=0$ and $=-300$ GeV. This results in a neutralino LSP mass $M_{{ \chi^0_1 }}= 475$ and $184$ GeV respectively with almost 95\% of the LSP composed of $\widetilde B$. This small variation in $M_1$ does not alter the other particles masses significantly.

The other neutralinos and charginos become heavier than $7$ TeV for these choices of BPs. Therefore a heavy vector boson $W_R$ of mass $M_{W_R}=4.5$ TeV cannot decay to any combination of neutralinos plus charginos in the final state. Similarly, it cannot decay to any combination of down-squarks plus up-squarks and/or charge-sleptons plus sneutrinos in the final state since these particles are heavier than $4.6$ TeV. The right-handed neutrinos remain heavy due to the choices of large VEVs $v_R$, $\bar{v}_R$ and coupling $f$, hence $W_R$ decay into right-handed neutrinos is also forbidden. $W_R$ thus mostly decays into the final state SM quarks. The corresponding vertices are given in Eqn.~\ref{coupWRFud} of Appendix~\ref{App:WR} and we find the decay width is approximately $36$ GeV in each up-type plus down-type quarks final state. The decay of $W_R$ into Higgs boson final states would occur only for the states which are primarily consisting of the bidoublet fields. The bidoublet kinetic term leads to vertices with both left-handed and right-handed gauge bosons and hence the heavy $W_R$ can decay into a light gauge boson from these vertices. So the relevant decay channels would be $W_R^\pm \rightarrow h_{1,3} W^\pm$, $W_R^\pm \rightarrow A_{1} W^\pm$ and $W_R^\pm \rightarrow  H_1^\pm  Z$. Here all of these particles primarily consist of bidoublet fields with $H_1 \sim Re[\phi^0_1]$, $H_3 \sim Re[\phi^0_2]$, $ A_1  \sim Im[\phi^0_2]$ and $H_1^\pm \sim \phi^\pm_2$ with their masses being $M_{h_1}=125$ GeV, $M_{h_3}=3.61$ TeV, $M_{ A_1 }=3.61$ TeV and $M_{ H^\pm _1 }=3.64$ TeV respectively. 
We have provided these vertices in Eqns.~\ref{coupWRHWL},~\ref{coupWRAWL},~\ref{coupWRZHm} of the Appendix~\ref{App:WR}.  
The branching ratios of these channels are still negligibly small due to the coupling suppression or the phase space suppression or a combination of both. All other combinations of the Higgs boson final states come out to be heavier than the $W_R$ mass and hence are forbidden in this case. 

Similarly, $Z_R$ boson of mass $M_{Z_R}=7.3$ TeV mostly decays into the final state SM quarks and charged leptons but cannot decay to any combination neutrinos. This is because the decay into two right-handed neutrinos is kinematically forbidden for our choice of parameters while the decay into two left-handed light neutrinos is suppressed by the extremely small mixing between the light and heavy neutrinos. The $Z_R$ boson decay to the sfermions are also kinematically forbidden since they are heavy.
The $Z_R$ boson can decay into two LSPs, however the decay branching is negligibly small due the suppression in the $Z_R  { \chi^0_1 } { \chi^0_1 }   $ vertex given in Eqn.~\ref{coupZRXX}. This coupling strength is $\sim\mathcal{O}(10^{-3})$ for these choices of the BPs 1.a and 1.b. The $Z_R$ boson decays to any combination of other neutralinos, charginos are kinematically disallowed as well due to the combined heavier mass in the final state. The decay mode $Z_R \rightarrow h_{1} Z$ is allowed but has a small branching $\sim 1\%$ due to the suppressed coupling in the vertex as can be seen in Eqn.~\ref{coupZRHZ}. Although the coupling strength in the $Z_R W^\pm W^\mp$ vertex (see Eqn.~\ref{coupZRWmWp}) is small, it is a momentum dependent term and can result in the branching of $W^+ W^-$ final state reaching upto $2\%$ due to the large momentum of the final state particles. The branching ratios to the other Higgs bosons and/or vector bosons final state decay channels are negligibly small due to the small coupling strengths (see Eqns.~\ref{coupZRHgamma}, \ref{coupZRHZ}, \ref{coupZRHmWp}, \ref{coupZRHmWRp}) or the phase space suppression due to the large final state particle masses.

\subsubsection{$SU(2)_L$ Wino-like LSP}
$SU(2)_L$ wino (or wino-L as we will refer to them) dominated LSP primarily consists of $\widetilde W_{L}^0$ gauge boson field. The corresponding BP is given as wino-L case in the Tab.~\ref{table-100LSP}. Since the $SU(2)_L$ symmetry is broken much below the SUSY breaking scale, the soft masses corresponding to all the gaugino partners of $SU(2)_L$ gauge bosons must be equal and is denoted as $M_{WL}$ in our model. In this particular choice of BP, we use a value of the parameter $M_{WL}=500$ GeV. The lightest neutralino and the lightest chargino masses become almost degenerate as both of them are dominantly coming from the $SU(2)_L$ wino multiplet. The lightest neutralino which is the LSP in this case gets a mass $M_{\chi^0_1}=500$ GeV while the lightest chargino, the NLSP here, has a mass of $M_{  \chi^\pm_1  }=500.1$ GeV. 

The $W_R$ boson decay into a ${ \chi^0_1 } $ and a $\chi^\pm_1$ final state, though kinematically allowed, yields an extremely small decay branching ratio owing to the coupling suppression in the $W^\pm_R  { \chi^0_1 } \chi^\mp_1$ vetrex which comes out to  $\mathcal{O}(10^{-4})$ as obtained from Eqn.~\ref{coupWRFXXm}. This coupling would naturally vanish for a pure $SU(2)_L$ wino-like LSP but the presence of small mixings in the electroweakino sectors result in a non-zero but highly suppressed coupling. Similarly, the coupling strengths of the $Z_R$ boson to the lightest neutralino or chargino vertices (see Eqn.~\ref{coupZRXX} and \ref{coupZRXmXp}) are also $\mathcal{O}(10^{-4})$. Hence the branching ratios of the decay channels $Z_R \rightarrow   { \chi^0_1 } { \chi^0_1 }$ and $Z_R \rightarrow \chi^\pm_1 \chi^\pm_1$ are very small. The other neutralinos, charginos, squarks, sleptons and the right-handed neutrinos remain heavy in this case. As a result the heavy $W_R$ and $Z_R$ vector bosons cannot decay into any combination of electroweakinos in the final state.

Similar to the previous case, the $W_R$ and $Z_R$ bosons mostly decay into SM quarks and charged leptons through $W_R^\pm \rightarrow q \overline q'$ and $Z_R \rightarrow q \overline{q}, l^+ l^-$ channels. The decays of $Z_R \rightarrow h_{1} Z$ (BR $ \sim 1\%$), $Z_R \rightarrow W^\pm W^\mp$ (BR $\sim 2\%$) and in other channels remain almost identical as in the previous case (bino-like LSP) due to similar couplings in the respective vertices.

\subsubsection{$SU(2)_R$ Wino-like LSP}
As a consequence of the right-handed symmetry being broken above the SUSY breaking scale, one can choose different values of the soft SUSY breaking terms for the charged and the neutral components of the $SU(2)_R$ wino fields. The parameter $ M^{W_R}_{33}$ can thus be adjusted to get a neutralino LSP primarily consisting of neutral $SU(2)_R$ wino (or wino-R) field. Again, in this case, due to the off-diagonal (2,4) and (2,5) entries in the neutralino mass matrix ($\sqrt{2}g_{R} \bar{v}_R$ and $\sqrt{2}g_{R} v_R$ respectively as can be seen from Eqn.~\ref{Mmatrix-Neutralino}), the lightest neutralino mass $M_{\chi^0_1}\geq 1.4$ TeV for any positive value of $ M^{W_R}_{33}$ for our chosen parameters. Similarly for positive values of the parameters $ M^{W_R}_{ii}(i=1,2)$, the wino-R-like chargino mass remains $M_{\chi^\pm_1} \geq 735$ GeV due to the off-diagonal (1,3) and (3,1) terms in the chargino mass matrix given in Eqn.~\ref{Mmatrix-charginos}. In fact if all the three soft terms are equal and positive, the chargino mass always remain lighter than the neutralino mass for a $SU(2)_R$ wino-like LSP. Thus we are compelled to choose either negative or unequal numerical values for $ M^{W_R}_{ii}(i=1-3)$. To analyze the distinct regions in the parameter space, we choose three corresponding BPs 3.a, 3.b and 3.c as shown in Tab.~\ref{table-100LSP}. The Higgs bosons, right-handed neutrinos, squarks and sleptons masses are unaltered since all other parameters remain same as in the previous cases. 

In BP 3.a we choose $ M^{W_R}_{33}=-1350$ GeV while $ M^{W_R}_{11}= M^{W_R}_{22}=60$ TeV. Thus the neutralino mass matrix, after diagonalization, gives the lightest neutralino mass of $M_{\chi^0_1}=172$ GeV. The lightest chargino mass, on the other hand, is obtained as $M_{\chi^\pm_1}=16.65$ TeV owing to the large values chosen for the chargino mass parameters. As a result $W_R$ and $Z_R$ bosons maximally ($\sim100\%$) decay into final state SM particles for this choice of BP.

For BP 3.b and 3.c we keep the numerical value of $ M^{W_R}_{  33}$ same as in BP 3.a while choosing the charged wino-R soft masses to be $-1350$ GeV and $-955$ GeV respectively. This gives us a lightest chargino mass of 551 GeV for BP 3.b and 175 GeV for BP 3.c. The lightest neutralino mass is kept at 172 GeV for all the three benchmarks we have taken here. The $W_R^\pm \rightarrow   { \chi^0_1 }      \chi^\pm_1  $ and $Z_R \rightarrow   \chi^\pm_1     \chi^\mp_1  $ channels open up resulting in significant branching ratios into these decay channels. Both these cases have similar couplings for $W_R^\pm   { \chi^0_1 }      \chi^\mp_1$ which comes out $\sim (0.587 P_L + 0.605 P_R)$ while $Z_R   \chi^\pm_1     \chi^\mp_1$ coupling is $\sim ( 0.515 P_L + 0.547 P_R)$, where $P_L$ and $P_R$ are the left and right chiral projection operators of the fermions. As a result the branching ratio of $W_R^\pm \rightarrow   { \chi^0_1 }      \chi^\pm_1  $ is around 29\% while $Z_R \rightarrow   \chi^\pm_1     \chi^\mp_1  $ remains around 25\% for both BP 3.b and 3.c as can be seen from Tab.~\ref{table-100LSP}.
The branching ratios to the other Higgs bosons and/or vector bosons final state decay channels remain negligibly small as in the previous cases.

\subsubsection{Bidoublet Higgsino-like LSP}
We now consider the case where the LSP is primarily composed of the bidoublet higgsino fields. This can be achieved by choosing a relatively small value for the off-diagonal bidoublet higgsino cross-term $| { \frac{\lambda'}{2} }  v_s  +  \frac{\mu}{2} |$ while keeping the other mass terms in the neutralino mass matrix to be quite large. As the diagonal mass terms for $\widetilde{\phi}_1^0$ and $\widetilde{\phi}_2^0$ in the ${\bf   m_\chi   }$ matrix is zero, their entire mass thus comes from the off-diagonal term mentioned above. This results in two light neutralino mass eigenstates which are almost degenerate in mass and composed of maximally mixed states of $\widetilde \phi^0_1$ and $\widetilde \phi^0_2$ while their mixing with the other states are negligibly small. Hence, both the LSP and the NLSP arise from an equal mixing ($\sim 49.99\%$) of $\widetilde{\phi}_1^0$ and $ \widetilde{\phi}_2^0$ with their masses being 649.9 GeV and 650 GeV respectively. The lightest chargino also primarily consists of bi-doublet charged fields and has a mass of 650 GeV. Our choice of parameters (the BP 4 in the Tab.~\ref{table-100LSP}) directly affect the tree-level and loop-level Higgs mass, we thus slightly modify the numerical value of $ T^{yq}_{33}$ to get the lightest Higgs mass of $M_{h_1} = 125$ GeV.

The branching ratio of $W_R^\pm\rightarrow   { \chi^0_1 }      \chi^\pm_1  $ and $W_R^\pm\rightarrow  \chi^0_2    \chi^\pm_1  $ become identical ($9\%$ in each channel) since their coupling strengths ($\sim 0.31 P_L + 0.31 P_R $) are also equal here. This is because the composition of $  { \chi^0_1 }   $ and $  { \chi^0_2 }   $ are almost the same.
The $\widetilde{\phi}_1^0 \widetilde{\phi}_1^0 Z_R$ and $\widetilde{\phi}_2^0 \widetilde{\phi}_2^0 Z_R$ terms in the gauge basis appear with equal and opposite sign couplings and as a result, in the mass basis, the $Z_R   { \chi^0_1 }      { \chi^0_1 }    $ and $Z_R   { \chi^0_2 }      { \chi^0_2 }   $ couplings vanish in this case. This fact can also be seen from Eqn.~\ref{coupZRXX}. Hence, these are suppressed (as $\chi_{1,2}$ has tiny contributions from the other neutral higgsino and gaugino fields) and the corresponding decay BRs remain negligibly small. However, the vertices $Z_R   { \chi^0_1 }      { \chi^0_2 }$ and $Z_R   \chi^\pm_1     \chi^\pm_1$ have quite large couplings resulting in a sizable branching ratio of around 8\% in each of these channels.

\subsubsection{Right-handed gauge triplet Higgsino-like LSP}
The right-handed triplet higgsino-like neutralino LSP can be obtained by adjusting the value of the parameter  $ \mu_2$ ,  $ \lambda^c$  and $v_s$ appearing in the (4,5) and the (5,4) matrix elements of the neutralino mass matrix given in Eqn.~\ref{Mmatrix-Neutralino}.  We change  $ \mu_2$  and  $ \lambda^c$ while keeping $v_s$ constant. As these parameters directly affect the tree-level Higgs mass, we change loop-corrected Higgs mass by adjusting the value of the $ T^{yq}_{  33}$ parameter to achieve a lightest Higgs boson mass of 125 GeV. In this case, we can get large splitting between the neutralinos $  { \chi^0_1 }   $ and $  { \chi^0_2 }$ due to the large off-diagonal (2,4), (2,5), (4,2) and (5,2) elements of the neutralino mass matrix (see Eqn.~\ref{Mmatrix-Neutralino}). This also allows us to have quite different contributions from $\widetilde{\delta}^{c^{0}}$ and $\widetilde{\bar{\delta}}^{c^{0}}$ in ${ \chi^0_1 }$ eigenstate unlike the previous case of bidoublet higgsinos. We show this as BP 5 in the Tab.~\ref{table-100LSP}. The light neutralino masses are $M_{  { \chi^0_1 }   }=434$ GeV and $M_{  { \chi^0_2 }   }=698.5$ GeV while the lightest chargino mass is $M_{\chi^\pm_1}=648.9$ GeV for this benchmark point. An important consequence of our chosen benchmark point is that the doubly-charged higgsino also remains light here as it is a part of the same fermion multiplet which gives the LSP in this scenario.  We get the mass of the doubly-charged higgsino $M_{\chi^{\pm\pm}}=675$ GeV.

As ${ \chi^0_1 }   $, $  { \chi^0_2 }   $ and $  \chi^\pm_1  $ are all composed of right-handed triplet states, their couplings with $W_R$ are relatively large compared to other cases. The coupling strengths of $ W_R^\pm   { \chi^0_1 }     \chi^\pm_1$ vertex is $\sim 0.47 P_L + 0.40 P_R$ while that of $ W_R^\pm   { \chi^0_2 }     \chi^\pm_1$ vertex is   $\sim 0.40 P_L + 0.47 P_R$ (see the second term of both the lines of Eqn.~\ref{coupWRFXXm}) for the BP 5. Thus the branching ratios of $W_R^\pm \rightarrow   { \chi^0_1 }      \chi^\pm_1  $ and $W_R^\pm \rightarrow   { \chi^0_2 }      \chi^\pm_1  $ are almost same here and equal to around 12\% each. The doubly-charged fermion being light allows an additional decay channel for $W_R^\pm \rightarrow   \chi^\mp_1   \chi^{\pm\pm}$ and gives an even larger branching ratio of $21.12\%$ for this channel owing to the larger coupling of $ W_R^\pm   \chi^\mp_1   \chi^{\pm\pm}$ vertex as given in Eqn.~\ref{coupWRFXppXm}. 

The neutral $Z_R$ boson has several possible decay modes into electroweakino final states in this case. The $\widetilde{\delta}^{c^{0}} \widetilde{\delta}^{c^{0}} Z_R$ and $ \widetilde{\bar{\delta}}^{c^{0}} \widetilde{\bar{\delta}}^{c^{0}} Z_R$  terms in the gauge basis appear with equal and opposite sign as in the previous case. Hence the $Z_R   { \chi^0_1 }      { \chi^0_1 }   $ and $Z_R   { \chi^0_2 }      { \chi^0_2 }   $ couplings are again small due to cancellations in the mass basis. On the other hand, the $Z_R \rightarrow   { \chi^0_1 }    \chi_2^0$ channel has a large branching ratio of $33.7\%$ due to a large coupling of $\sim 0.63 P_L + 0.63 P_R$ (see Eqn.~\ref{coupZRXX}) in this vertex. The $Z_R \rightarrow   \chi^\pm_1     \chi^\mp_1  $ remains quite small at around 2\%. A sizable $9.45\%$ branching in the channel $Z_R \rightarrow \chi^{\mp\mp} \chi^{\pm\pm}$ is also obtained in this scenario. 

\subsubsection{Singlino-like LSP}
The singlino-like neutralino LSP primarily consists of $\widetilde{S}$ field. In this case, we have adjusted $\mu_s$ parameter to get the singlino-like LSP. The choice of BP 6 in the Tab.~\ref{table-100LSP}, gives a singlino-like LSP with a mass $M_{  { \chi^0_1 }   }=521$ GeV. The other neutralinos and charginos remain very massive here and cannot be produced from the decay of the heavy gauge bosons. Since the singlino has no gauge interactions it is quite natural that the $W_R$ and $Z_R$ bosons do not decay into these states. Hence the entire decay of the heavy gauge bosons will be into SM particles in this scenario.

%%%%%%%%%%%%%%%%%%%%%%%%%%%%%%%%%%%%%%%%%%%%%%%%%%
\subsection{Case-II : Mixed LSP}
%%%%%%%%%%%%%%%%%%%%%%%%%%%%%%%%%%%%%%%%%%%%%%%%%%

%%%%%%%%%%%%%%%%%%%%%%%%%%%%%%%%%%%%%%%%%%%%%
\begin{table}[h!]
\begin{center}\scalebox{0.7}{
\begin{tabular}{|c||c|c|c|}
\hline
\hline
~~~~ &  &\multicolumn{2}{c|}{ }\\
~~LSP-Type~~ & Benchmark points &\multicolumn{2}{c|}{ Branching Ratio of $W_R^\pm,~Z_R$ into different BSM fields}\\
~~~~ &  &\multicolumn{2}{c|}{ }\\
\cline{3-4} 
 &~~~~~~~~~~~~~ &   $W_R^\pm \rightarrow BSM$       &   $Z_R   \rightarrow BSM  $\\
\hline
1. Bino  (49.9 $\%$) &  $ M_1 = -272.534$, $ M^{W_L}_{  33,22,11} = 200$ GeV, $ M^{W_R}_{  33,22,11} = 60$ TeV,   &        &  \\
and & $ { \lambda^c }  = -0.2$, $ { \mu_2 }  = 21$ TeV,  $ {\mu}  = -17$ TeV,&  $0\%$          &   $0\%$\\
Wino-L  (49.9 $\%$)&$ { \mu_s }  = -20$ TeV,  $ T^{yq}_{  33}=82$ TeV&&\\
&&&\\
&$M_{\chi_1^0}=200$ GeV, $M_{\chi_2^0}=200$ GeV, $M_{\chi_1^\pm}=200$ GeV&&\\
\hline
2.  Bino    (49.9 $\%$)   & $ M_1 = 50$, $ M^{W_L}_{  33,22,11} = 60$ TeV, $ M^{W_R}_{  33,22,11} = 60$ TeV, &&\\
and &$ { \lambda^c }  = -0.2$, $ { \mu_2 }  = 21$ TeV,  $ {\mu}  = -17$ TeV, & $0\%$   &        $0\%$ \\
Singlino    (49.9 $\%$)&$ { \mu_s }  = -12$ TeV,  $ T^{yq}_{  33}=86.4$ TeV,&&\\
&&&\\
&$M_{\chi_1^0}=366$ GeV, $M_{\chi_2^0}=676$ GeV, $M_{\chi_1^\pm}=16.6$ TeV&&\\
\hline
3.  Wino-L    (49.9 $\%$)   & $ M_1 = 60$ TeV, $ M^{W_L}_{  33,22,11} = 521.59$, $ M^{W_R}_{  33,22,11} = 60$ TeV, &&\\
and &$ { \lambda^c }  = -0.2$, $ { \mu_2 }  = 21$ TeV,  $ {\mu}  = -17$ TeV, & $0\%$   &        $0\%$ \\
Singlino    (49.9 $\%$)&$ { \mu_s }  = -12$ TeV,  $ T^{yq}_{  33}=86.4$ TeV,&&\\
&&&\\
&$M_{\chi_1^0}=521.5$ GeV, $M_{\chi_2^0}=521.7$ GeV, $M_{\chi_1^\pm}=521.6$ GeV&&\\
\hline
4. Bino   (49.9 $\%$) & $ M_1 = -250$, $ M^{W_L}_{  33,22,11} = 60$ TeV, $ M^{W_R}_{  33,22,11} = -1350$, & &\\
 and   &   $ { \lambda^c }  = -0.2$, $ { \mu_2 }  = 21$ TeV,  $ {\mu}  = -17$ TeV,   &   $15.57\%$ ($W_R^\pm \rightarrow   { \chi^0_1 }      \chi^\pm_1  $) &         $24.9\%$ ($Z_R \rightarrow   \chi^\mp_1     \chi^\pm_1  $)\\
Wino-R   (49.9 $\%$) &$ { \mu_s }  = -20$ TeV,  $ T^{yq}_{  33}=82$ TeV&$12.31\%$ ($W_R^\pm \rightarrow  \chi^0_2    \chi^\pm_1  $)&\\
       &        &   &\\
&$M_{\chi_1^0}=645$ GeV, $M_{\chi_2^0}=980.7$ GeV, $M_{\chi_1^\pm}=551.9$ GeV&&\\
\hline
5. Wino-L   (49.9 $\%$) & $ M_1 = 60$ TeV, $ M^{W_L}_{  33,22,11} = 172.265$, $ M^{W_R}_{  33,22,11} = -1350$, &&\\
 and   &   $ { \lambda^c }  = -0.2$, $ { \mu_2 }  = 21$ TeV,  $ {\mu}  = -17$ TeV,   & $12.31\%$ ($W_R^\pm \rightarrow  \chi^0_2    \chi^\pm_1  $) &         $24.9\%$ ($Z_R \rightarrow   \chi^\mp_2     \chi^\pm_2  $)\\
Wino-R   (49.9 $\%$) &$ { \mu_s }  = -20$ TeV,  $ T^{yq}_{  33}=82$ TeV& $12.67\%$ ($W_R^\pm \rightarrow  \chi^0_2    \chi^\pm_2  $) &\\
       &        &   $12.31\%$ ($W_R^\pm \rightarrow  \chi^0_2    \chi^\pm_1  $) &\\
&$M_{\chi_1^0}=172.2$ GeV, $M_{\chi_2^0}=172.3$ GeV, $M_{\chi_1^\pm}=172.3$ GeV&&\\
\hline
6.  Wino-R    (49.9 $\%$)   & $ M_1 = 60$ TeV, $ M^{W_L}_{  33,22,11} = 60$ TeV, $ M^{W_R}_{  33,22,11} = -1010$, &&\\
and &$ { \lambda^c }  = -0.2$, $ { \mu_2 }  = 21$ TeV,  $ {\mu}  = -17$ TeV, &  $15.44\%$ ($W_R^\pm \rightarrow   { \chi^0_1 }      \chi^\pm_1  $)&        $24.7\%$ ($Z_R \rightarrow   \chi^\mp_1     \chi^\pm_1  $) \\
Singlino    (49.9 $\%$)&$ { \mu_s }  = -12$ TeV,  $ T^{yq}_{  33}=86.4$ TeV,&$12.16\%$ ($W_R^\pm \rightarrow  \chi^0_2    \chi^\pm_1  $)  &\\
&&&\\
&$M_{\chi_1^0}=230$ GeV, $M_{\chi_2^0}=765$ GeV, $M_{\chi_1^\pm}=230.1$ GeV&&\\
\hline
\hline
\end{tabular}}
\end{center}
\caption{ 
The LSP is mostly composed of only two type of the neutral fermion fields in the basis given in Eqn.~\ref{Eq:NetrBasis}. The other parameters are fixed as in the Tab.~\ref{table-Fix}.}
\label{table-50LSP}
\end{table}
%%%%%%%%%%%%%%%%%%%%%%%%%%%%%%%%%%%%%%%%%%%%%%%%%%%%%%%%

In the previous section, we analyzed the regions of the parameter space where the major contribution to the LSP was primarily coming from only one type of component of the neutral gaugino or higgsino fields. In this section, we discuss the parameter space where the LSP consists of two type of components (see Eqn.~\ref{Eq:NetrBasis}) and study its effect on the heavy gauge boson decay channels. As before the rest of the BSM particle spectrum (squarks, sleptons, right-handed neutrinos, etc.) have been kept much heavier to make sure that the heavy gauge bosons do not decay into them. Cases with dominant contribution to the LSP from two gaugino fields are shown in the Tab.~\ref{table-50LSP} while the cases with mixed gaugino and higgsino fields relevant for our study are presented in Tab.~\ref{table-33LSP}. The higgsinos, as well, can mix among each other resulting in the LSP being composed of equal parts coming from the bidoublet fields and the right-handed triplet fields. The choice of parameters leading to this mixed higgsino-like LSP is given in Tab.~\ref{table-25LSP}. We now present a detailed discussion of each of these cases.

\subsubsection{ Combination of two of the Bino, Wino-L and Singlino}

The first three benchmark points in the Tab.~\ref{table-50LSP}  (BPs-1,2 and 3) represents the cases with the LSP comprising of a maximal mixing of the bino and wino-L field, the bino and singlino field and the wino-L and singlino fields respectively. In each case, the next lightest neutralino is almost same as the LSP in both mass and composition. As has been discussed earlier, the heavy $W_R$ and $Z_R$ gauge bosons do not have significant couplings to either a pair of binos, wino-Ls or singlinos. Thus, even here, the decay of these heavy gauge bosons into the light neutralinos are non-existent. Thus they decay primarily into SM particles with almost no branching into BSM particles.

\subsubsection{ Combination of Wino-R and one of the Bino, Wino-L and Singlino}

The last three benchmark points in the Tab.~\ref{table-50LSP} are much more interesting as the LSP here are all composed of around 50\% from $SU(2)_R$ Wino which does interact quite significantly with the heavy gauge bosons. BPs 4, 5 and 6 in the Tab.~\ref{table-50LSP} give a lightest neutralino LSP $  { \chi^0_1 }   $ consisting of $\sim 50\%$ wino-R and $\sim 50\%$ from either a Bino or wino-L or singlino field respectively. The lightest chargino $  \chi^\pm_1  $ in BP 4 and 6 primarily consists of charged wino-R fields. Thus for these two points the decay of the $W_R^\pm \rightarrow \chi^0_1 \chi^\pm_1$ and $W_R^\pm \rightarrow \chi^0_2 \chi^\pm_1$ are quite significant with a BR around 15\% and 12\% respectively. The coupling strengths of the vertices $W_R^\pm \chi^0_1 \chi^\pm_1$ and $W_R^\pm \chi^0_2 \chi^\pm_1$ become almost equal and half of the coupling strength $W_R^\pm \chi^0_1 \chi^\pm_1$ of the pure wino-R case as in the Tab.~\ref{table-100LSP}. These strengths become half due the changes of the mixing angles $Z^{fN}$ in the vertices (see Eqn.~\ref{coupWRFXXm}) for the extra combination of Bino or wino-L or singlino fields in the LSP.  As the combinations of the lightest chargino $  \chi^\pm_1  $ remains similar as in the wino-R case as in the Tab.~\ref{table-100LSP}, hence the coupling strength, the $Z_R$ boson decays significantly into $\chi^\pm_1   \chi^\mp_1  $ with a BR of around 25\% here.
For BP 5 its actually the second lightest chargino $\chi^\pm_2$ which is coming from the charged component of the $SU(2)_R$ Wino. Also in this case, $W_R$ thus decays into $  \chi^\pm_2   \chi^0_1$ and $  \chi^\pm_2   \chi^0_2$ with BR of around 15\% and 12\% respectively while $Z_R$ boson decays into $\chi^\pm_2   \chi^\mp_2  $ with a BR of around 25\%.

\subsubsection{ Combination of Wino-R and Higgsinos of bidoublet/triplet}

The only interesting cases here will be the ones where the bidoublet or $SU(2)_R$ triplet higgsinos mix with the wino-R state as the other cases will be very similar to pure higgsino LSP, only with reduced branchings into SUSY final states. We show these points in the Tab.~\ref{table-33LSP}. 

%%%%%%%%%%%%%%%%%%%%%%%%%%%%%%%%%%%%%%%%%%%%%
\begin{table}[h!]
\begin{center}\scalebox{0.7}{
\begin{tabular}{|c||c|c|c|}
\hline
\hline
~~~~ &  &\multicolumn{2}{c|}{ }\\
~~LSP-Type~~ & Benchmark points &\multicolumn{2}{c|}{ Branching Ratio of $W_R^\pm,~Z_R$ into different BSM fields}\\
~~~~ &  &\multicolumn{2}{c|}{ }\\
\cline{3-4} 
 &~~~~~~~~~~~~~ &   $W_R^\pm \rightarrow BSM$       &   $Z_R   \rightarrow BSM  $\\
\hline
&$ M_1 = 60$ TeV, $ M^{W_L}_{  33,22,11} = 60$ TeV, $ M^{W_R}_{  33,22,11} = -795$, &$9.56\%$ ($W_R^\pm \rightarrow   { \chi^0_1 }      \chi^\pm_1  $)         & $4.08\%$ ($Z_R \rightarrow   { \chi^0_1 }     \chi^0_2 $)\\
1.Higgsino-($\phi_{1,2}$)     &  $ { \lambda^c }  = -0.2$, $ { \mu_2 }  = 21$ TeV,  $ {\mu}  = -1$ TeV,     &   $15.23\%$ ($W_R^\pm \rightarrow   { \chi^0_3 }      \chi^\pm_1  $)          &  $2.11\%$ ($Z_R \rightarrow   { \chi^0_2 }     \chi^0_3 $)\\
(33.3 $\%$ each)& $ { \mu_s }  = -20$ TeV,  $ T^{yq}_{  33}=79$ TeV &$3.52\%$ ($W_R^\pm \rightarrow   { \chi^0_1 }      \chi^\pm_2  $)&$21.64\%$ ($Z_R \rightarrow   \chi^\mp_1     \chi^\pm_1  $)\\
Wino-R&& $6.76\%$ ($W_R^\pm \rightarrow   { \chi^0_2 }      \chi^\pm_2  $)& $6.38\%$ ($Z_R \rightarrow   \chi^\mp_2     \chi^\pm_2  $)\\
&$M_{\chi_1^0}=614.2$ GeV, $M_{\chi_2^0}=652.4$ GeV, $M_{\chi_3^0}=717$ GeV&$3.37\%$ ($W_R^\pm \rightarrow   { \chi^0_3 }      \chi^\pm_2  $)&\\
&$M_{\chi_1^\pm}=214$ GeV, $M_{\chi_2^\pm}=659.3$ GeV&&\\
\hline
2.Higgsino-($\Delta_R^{1,2}$)         & $ M_1 = 60$ TeV, $ M^{W_L}_{  33,22,11} = 60$ TeV, $ M^{W_R}_{  33,22,11} = -9150$, & $1.16\%$ ($W_R^\pm \rightarrow   { \chi^0_1 }      \chi^\pm_1  $)& $7.05\%$ ($Z_R \rightarrow   { \chi^0_1 }      { \chi^0_1 }   $) \\
(70.7\%, 10.0 $\%$) & $ { \lambda^c }  = -0.1$, $ { \mu_2 }  = 3$ TeV,  $ {\mu}  = -17$ TeV, &   $5.75\%$ ($W_R^\pm \rightarrow \chi_2^0   \chi^\pm_1  $) & $10.73\%$ ($Z_R \rightarrow   { \chi^0_1 }    \chi_3^0$)\\
and~Wino-R (18.7$\%$)&$ { \mu_s }  = -20$ TeV,  $ T^{yq}_{  33}=78.6$ TeV&$6.0\%$ ($W_R^\pm \rightarrow \chi_3^0   \chi^\pm_1  $)& $5.64\%$ ($Z_R \rightarrow \chi_2^0 \chi_3^0$)\\
&&$3.0\%$ ($W_R^\pm \rightarrow \chi_3^0   \chi^\pm_2  $)& $4.79\%$ ($Z_R \rightarrow   \chi^\pm_1     \chi^\mp_1  $)\\
&$M_{\chi_1^0}=366.5$ GeV, $M_{\chi_2^0}=650.1$ GeV, $M_{\chi_3^0}=651.9$ GeV&$11.17\%$ ($W_R^\pm \rightarrow \chi^{\pm \pm}   \chi^\mp_1  $)& $9.29\%$ ($Z_R \rightarrow \chi^{\pm \pm} \chi^{\mp \mp}$)\\
&$M_{\chi_1^\pm}=641.3$ GeV, $M_{\chi_2^\pm}=762.4$ GeV&&\\
\hline
\hline
\end{tabular}}
\end{center}
\caption{ The bidoublet or triplet higgsinos mix with the wino-R to form the lightest neutalino.}
\label{table-33LSP}
\end{table}
%%%%%%%%%%%%%%%%%%%%%%%%%%%%%%%%%%%%%%%%%%%%%%%%%%%%%%%%

BP 1 in the Tab.~\ref{table-33LSP} represents the case where the bidoublet higgsinos mix with the wino-R to form the lightest neutalino $  { \chi^0_1 }   $ consisting almost 33\% of each $\widetilde \phi_1^0$, $\widetilde \phi_2^0$ and $\widetilde W_R$. In order to get the LSP with equal contributions from the two bidoublet higgsinos and the wino-R we had to make sure that all the mass terms related to these states remain small. This in turn results in two other neutralino states remaining light. The LSP $  { \chi^0_1 }   $ has a mass of 614 GeV for our chosen parameters.
The next lightest neutralino $  { \chi^0_2 }   $ is mostly coming from the bi-doublet fields and has a mass of 652 GeV. The next heavier member in the neutralino spectrum  $ \chi^0_3 $ gets a large contribution from the wino-R field and has a mass of  714 GeV. The chargino $ \chi^\pm_1 $ primarily consist of charged component of wino-R field whereas  $ \chi^\mp_2  $ consist of bi-doublet fields. They have masses in the same order as the neutralino masses as expected.
Thus a number of different channels open up for the $W_R$ and $Z_R$ decay albeit with smaller branching ratios compared to the pure higgsino or pure wino-R case. However the total branching into the final state BSM particles become large since a host of new decay channels has now opened up. The largest non-SM branching ratio in the $W_R$ decay is in the $W_R^\pm \rightarrow  \chi^0_3    \chi^\pm_1  $ channel with $15.23\%$ (much smaller than pure wino-R case with BR of $29.3\%$) due to the mixing suppression in the neutralino and the chargino sectors. Similarly the $W_R^\pm \rightarrow \chi_{1,2}^0   \chi^\pm_2  $ channels are smaller here than the previous case as can be seen in the Tab.~\ref{table-33LSP} and~\ref{table-100LSP} respectively. A few other new combinations of neutralino plus chargino final states open up due to the smaller mass of the final state particles and mixing between bi-doublet and wino-R fields. We find the total branching of $W_R$ into BSM final state particles to be roughly $38.44\%$. Notably the $W_R^\pm \rightarrow \chi^0_2   \chi^\pm_1  $ channel is very small since there is no direct $W_R^\pm \widetilde W_R^\mp \widetilde \Phi$ coupling in the gauge basis and only arises from mixings which are quite small. Similarly $Z_R$ has a number of possible decay channels including $Z_R \rightarrow   { \chi^0_1 }    \chi_2^0$, $Z_R \rightarrow \chi_2^0 \chi_2^0$, $Z_R \rightarrow   \chi^\pm_1     \chi^\mp_1  $ and $Z_R \rightarrow   \chi^\pm_2     \chi^\mp_2  $ with 4.1\%, 2.1\%, 21.6\% and 6.4\% BR in each channel respectively. The coupling strength of the vertex $Z_R   \chi^\pm_1     \chi^\pm_1 $ is slightly smaller than the pure wino-R case due to due to mixing with higgsino states in the lightest chargino for this case.

One can adjust the parameters $ M^{W_R}_{33}$,  $ \lambda^c$  and  $ \mu_2$  to get neutralino LSP which consists of neutral fermionic components wino-R and two higgsino from triplet fields. In this case, it not possible to get $33\%$ contribution to the LSP from each of the fields due to the large off-diagonal ($g_{V} v_R$, $g_{R} v_R$, $g_V \bar{v}_R$ and $g_R \bar{v}_R$) terms in the neutralino mass matrix (see Eqn.~\ref{Mmatrix-Neutralino}).
We represent this as BP 2 and the corresponding contributions of each of these fields are shown in Tab.~\ref{table-33LSP}. Similar to the previous case with light triplet, the doubly charged chargino remains light here as well. We get the light neutralino and chargino particles masses as $M_{  { \chi^0_1 }   }=367$ GeV, $M_{\chi_2^0}=650$ GeV, $M_{\chi_3^0}=651$ GeV, $M_{  \chi^\pm_1  }=641$ GeV, $M_{  \chi^\pm_2}=762$ and $M_{\chi^{\pm\pm}}=2.65$ TeV for this choice of BP. 
The $  { \chi^0_1 }   $ and $\chi_2^0$ are both coming from a combination of the triplet and the wino-R fields. The lightest chargino $  \chi^\pm_1  $ has around 90\% higgsino components from the triplet and 10\% from $SU(2)_R$ wino component. The decay channels for the $W_R$ and the $Z_R$ fields are given in BP2 of Tab.~\ref{table-33LSP}. As the $\chi_3^0,  \chi^\pm_2  $ and $\chi^{\pm\pm}$ are mostly coming from the triplet sector, the corresponding decay of the heavy gauge bosons into these particles remain large. The partial decay widths (and hence the branching ratios), through are slightly smaller compared to the previous case with pure triplet higgsino LSP due to larger mixing with the $SU(2)_R$ wino.

%%%%%%%%%%%%%%%%%%%%%%%%%%%%%%%%%%%%%%%%%%%%%
\begin{table}[h!]
\begin{center}\scalebox{0.7}{
\begin{tabular}{|c||c|c|c|}
\hline
\hline
~~~~ &  &\multicolumn{2}{c|}{ }\\
~~LSP-Type~~ & Benchmark point &\multicolumn{2}{c|}{ Branching Ratio of $W_R^\pm,~Z_R$ into different BSM fields}\\
~~~~ &  &\multicolumn{2}{c|}{ }\\
\cline{3-4} 
 &~~~~~~~~~~~~~ &   $W_R^\pm \rightarrow BSM$       &   $Z_R   \rightarrow BSM  $\\
\hline
\hline
&$ M_1 = 60$ TeV, $ M^{W_L}_{  33,22,11} = 60$ TeV, $ M^{W_R}_{  33,22,11} = 60$ TeV,&$2.96\%$ ($W_R^\pm \rightarrow   { \chi^0_1 }      \chi^\pm_1  $)         & $2.44\%$ ($Z_R \rightarrow   { \chi^0_1 }     \chi^0_2 $)\\
1.Higgsino-($\phi_{1,2}$     &   $ { \lambda^c }  = -0.2$, $ { \mu_2 }  = -718.9$,  $ {\mu}  = -1$ TeV,   &   $5.5\%$ ($W_R^\pm \rightarrow   { \chi^0_2 }      \chi^\pm_1  $)          &  $14.41\%$ ($Z_R \rightarrow   { \chi^0_1 }     \chi^0_4 $)\\
and $\Delta_R^{1,2}$)& $ { \mu_s }  = -20$ TeV,  $ T^{yq}_{  33}=78.8$ TeV &$2.53\%$ ($W_R^\pm \rightarrow   { \chi^0_3 }      \chi^\pm_1  $)&$2.13\%$ ($Z_R \rightarrow  \chi^0_2   \chi^0_3 $)\\
(25 $\%$ each)& & $5.03\%$ ($W_R^\pm \rightarrow   { \chi^0_1 }      \chi^\pm_2  $)& $16.60\%$ ($Z_R \rightarrow  \chi^0_3   \chi^0_4 $)\\
&&$5.85\%$ ($W_R^\pm \rightarrow   { \chi^0_3 }      \chi^\pm_2  $)& $4.58\%$ ($Z_R \rightarrow   \chi^\mp_1     \chi^\pm_1  $)\\
&&$10.87\%$ ($W_R^\pm \rightarrow   { \chi^0_4 }      \chi^\pm_2  $)&$1.83\%$ ($Z_R \rightarrow   \chi^\mp_2     \chi^\pm_2  $)\\
&$M_{\chi_1^0}=649.7$ GeV, $M_{\chi_2^0}=650$ GeV, $M_{\chi_3^0}=650.3$ GeV&$17.19\%$ ($W_R^\pm \rightarrow \chi^{\pm\pm}   \chi^\mp_2  $)&$8.55\%$ ($Z_R \rightarrow \chi^{\mp\mp} \chi^{\pm\pm}$)\\
&$M_{\chi_1^\pm}=650$ GeV, $M_{\chi_2^\pm}=867.7$ GeV&&\\
\hline
\hline
\end{tabular}}
\end{center}
\caption{Combination of bidoublet and triplet higgsinos: The LSP are composed of 25\% from each of the higgsino fields.}
\label{table-25LSP}
\end{table}
%%%%%%%%%%%%%%%%%%%%%%%%%%%%%%%%%%%%%%%%%%%%%%%%%%%%%%%%

\subsubsection{ Combination of bidoublet and triplet Higgsinos}

We now discuss a parameter choice where the contribution to the LSP are mainly coming from the mixing between the bi-doublet and right-handed triplet higgsino fields. In this case, we get four light neutralinos having masses $M_{  { \chi^0_1 }   }=649.9$ GeV, $M_{\chi_2^0}=650$ GeV, $M_{\chi_3^0}=650.2$ GeV and $M_{\chi_4^0}=914.5$ GeV. Here $  { \chi^0_1 }   $ and $\chi_3^0$ are composed of 25\% from each of the higgsino fields while $\chi_2^0$ is almost entirely coming from the bidoublet field while $\chi_4^0$ from the triplet fields. The lightest chargino mainly has charged bidoublet higgsino contribution with $M_{  \chi^\pm_1  }=650.2$ GeV and the second lightest chargino state is mainly from the triplet higgsino with mass $M_{  \chi^\pm_2  }=863$. The light doubly-charged chargino arising from the $SU(2)_R$ triplet higgsino has $M_{\chi^{\pm\pm}}=893$ GeV.

A host of new final state decay channels for the heavy gauge bosons into these light electroweakinos are obtained here. The $W_R^\pm   { \chi^0_i }     { \chi^\mp_j }   $ and $W_R^\pm \chi^{\mp\mp}   { \chi^\pm_j }   $ couplings are given in Eqns.~\ref{coupWRFXXm},\\ \ref{coupWRFXppXm} while the $Z_R   { \chi^0_i }     { \chi^0_j}   $ and $Z_R   { \chi^\pm_i }      { \chi^\mp_j }   $ and $Z_R \chi^{\pm\pm} \chi^{\mp\mp}$ are presented in Eqns.~\ref{coupZRXX},\ref{coupZRXmXp},\ref{coupZRXmmXpp}. These coupling strengths become smaller compared to cases with pure bidoublet or triplet higgsinos due to the mixing between them in this case. Since the gauge bosons couple more strongly with the triplets compared to the bidoublets, the branching ratio of $W_R$ into $\chi_4^0   \chi^\pm_2  $ and $\chi^{\pm\pm}   \chi^\pm_2  $ are quite large of the order of 11\% and 17\% respectively. As stated earlier, $\chi_4^0$, $ \chi^\pm_2  $ and $\chi^{\pm\pm}$ are mostly coming from the triplet fields. Similarly $Z_R$ decays involving $\chi_4^0$ in the final state are much larger than the others. All the important BSM decay channels of the heavy gauge bosons for this case are given in the Tab.~\ref{table-25LSP}. Combining all these channels, the total branching ratios of the $W_R$ and $Z_R$ bosons into BSM final states become 49.93\% and 50.54\% respectively, which are larger than the scenarios with almost pure bidoublet or triplet higgsino-like LSP.

%%%%%%%%%%%%%%%%%%%%%%%%%%%%%%%%%%%%%%%%%%%%%%%%%%
\subsection{Case-III : Pure LSP and light pair of squarks}
%%%%%%%%%%%%%%%%%%%%%%%%%%%%%%%%%%%%%%%%%%%%%%%%%%
In the previous sections, the squarks and sleptons masses were larger than the $W_R$ and $Z_R$ masses so that their decays into final state squarks and sleptons were kinematically disallowed. We now study the effect of low mass sfermions on the possible decays of the heavy gauge bosons. In order to achieve this, we significantly decrease the values of the soft masses for the squarks and sleptons so that they are light enough for the heavy gauge bosons to decay into these particles. The small stop squark mass will significantly affect the Higgs boson loop-corrected mass and we have to adjust the mixing parameter $ T^{yq}_{  33}$ to account for this change. One has to choose a reasonably large $T^{yq}$ for this, resulting in the left-handed squarks becoming quite massive (more than $M_{W_R}/2$). The right-handed squarks though can remain light since the corrections to the right-handed squarks coming from the triplet VEV through the D-term is negative for our choice of parameters. Thus the heavy gauge bosons can decay into these right-handed squark final states.

Here we change the values of the soft mass-squared terms $M_{lL}^2$, $M_{lR}^2$, $ M_{QL}^2$, $ M_{QR}^2$ and the mixing $ T^{yq}_{ii}$ ($i=1,2,3$) between the left and right handed squarks sectors to get lighter ($\mathcal{O}(1)$ TeV) squarks and sleptons. Therefore the heavy vector boson decays into final state squarks and sleptons are kinematically allowed. It must be noted here that the heavy sneutrino masses have not been altered and hence sneutrino final states are still absent. The relevant heavy gauge boson couplings with the squarks and sleptons are given in Eqns.~\ref{coupWRsdsu}, \ref{coupZRsdsd}, \ref{coupZRsusu} and \ref{coupZRsese}. It is easy to see that these coupling strengths are proportional to the momentum of the final state particles. Hence the decay widths significantly depend on the momentum of the final state particles which are  
\begin{equation}
|\vec{p}_{cm}|=\frac{1}{2 M_V} \sqrt{M_V^4 + M_{\tilde{q}_i}^4 + M_{\tilde{q}_i}^4 - 2 M_V^2 M_{\tilde{q}_i}^2 -2 M_V^2 M_{\tilde{q}_j}^2 - 2 M_{\tilde{q}_i}^2 M_{\tilde{q}_j}^2}.
\end{equation}
The masses of the light up-type squarks (primarily consisting of right-handed squark fields) remain always larger than the light down-type squarks (also right-handed) for our choice of parameters due to the opposite sign contribution from the $SU(2)_R$ D-term. This term is additive for the $m_{\widetilde u_R \widetilde u_R^*}$ term while it is subtractive for the $m_{\widetilde d_R \widetilde d_R^*}$.

Let us consider the benchmark point 1.b in the Tab.~\ref{table-100LSP} with a Bino-like LSP. We keep the numerical values of all the parameters same except we now choose $ M_{QL,ii}^2 =  M_{QR,ii}^2=  M_{lR,ii}^2=9.61\times 10^{6}$ ${\rm GeV}^2$ ($i=1,2$), $M_{QL,33}^2 =  M_{QR,33}^2=M_{lR,33}^2=9.60\times 10^{6}$ ${\rm GeV}^2$ and $ T^{yq}_{  ii}=48$ TeV GeV ($i=1,2,3$) to get lightest down-squark $m_{\tilde{d}_1}=1$ TeV while the other two have $m_{\tilde{d}_{2,3}}=1.5$ TeV.
The lightest up-type squark mass $m_{\tilde{u}_1}=1.98$ TeV while $m_{\tilde{u}_{2,3}}=2.03$ TeV. The charged slepton masses are $m_{\tilde{e}_1}=2.14$ TeV, $m_{\tilde{e}_{2}}=2.19$ TeV and $m_{\tilde{e}_{3}}=2.25$ TeV. These relatively light squarks and sleptons are mostly composed of the right-handed fields. All other squarks and sleptons masses remain heavier than $3.3$ TeV.
Hence the $W_R$ boson decay into $\tilde{u}_{i}$ plus $\tilde{d}_{i}$ ($i=1,2,3$) final state particles become accessible while slepton final states are still forbidden owing to the large sneutrino masses. We find that the decay channels of the $W_R$ boson into squarks ($W_R^\pm \rightarrow \widetilde {q_i} \widetilde {q_j}^*$) each have a BR of around 2\%. The $Z_R$ boson, on the other hand, can decay into the final states with $\tilde{u}_{i}\tilde{u}^*_{i}$, $\tilde{d}_{i}\tilde{d}^*_{i}$ or $\tilde{e}_{i}\tilde{e}^*_{i}$ ($i=1,2,3$) particles. We get the $\sim 5\%$ branching in the final state $\tilde{d}_{i}\tilde{d}^*_{i}$ particles channels. The final state $\tilde{u}_{i}\tilde{u}^*_{i}$, $\tilde{e}_{i}\tilde{e}^*_{i}$ decay channels are suppressed due to the small couplings and larger masses of the final state squarks and sleptons in these cases. We show the total branching ratio into the final states squarks in the Tab.~\ref{table-100LSPSquarks}.
Also in this case,
the masses of the neutralinos and chargino remain unaltered and the $W_R$ boson decay into $  { \chi^0_1 }   $ plus $  \chi^\pm_1  $ final state remains negligibly small due the suppression
in the vertex $W^\pm_R  { \chi^0_1 }     \chi^\mp_1  \sim\mathcal{O}(10^{-4})$ (see Eqn.~\ref{coupWRFXXm}).
Similar to the previous case as in the Tab.~\ref{table-100LSP}, the $W_R$ and $Z_R$ bosons mostly decay into SM quarks and charged leptons through $W_R^\pm \rightarrow q \overline q'$ and $Z_R \rightarrow q \overline{q}, l^+ l^-$ channels. The other neutralinos, charginos, squarks, sleptons and the right-handed neutrinos remain heavy. As a result the heavy $W_R$ and $Z_R$ vector bosons cannot decay into any combination of these particles in the final state.

%%%%%%%%%%%%%%%%%%%%%%%%%%%%%%%%%%%%%%%%%%%%%
\begin{table}[h!]
\begin{center}\scalebox{0.7}{
\begin{tabular}{|c||c|c|}
\hline
\hline
~~~~ &  \multicolumn{2}{c|}{ }\\
~~LSP-Type~~  &\multicolumn{2}{c|}{ Branching Ratio of $W_R^\pm,~Z_R$ into different BSM fields for the common choice of}\\
~~~~   &\multicolumn{2}{c|}{ $M_{QL,ii}^2 =  M_{QR,ii}^2=9.61~{\rm TeV^2}$($i=1,2$) and $M_{QL,33}^2 =  M_{QR,33}^2=9.60~{\rm TeV^2}$ }\\
\cline{2-3} 
 &~~~~~~~~~~~~~~~~~~~~~~    $W_R^\pm \rightarrow BSM$   ~~~~~~~~~~~~~~~~~~~~~~~    &   $Z_R   \rightarrow BSM  $\\
\hline
Bino, Wino-L, Singlino  &&$\sim16\%$ ($Z_R \rightarrow \sum \tilde{d}_i \tilde{d}_i$)\\
$ T^{yq}_{  ii}=48$ TeV ($i=1,2,3$)    & $\sim4.5\%$ ($W_R^\pm \rightarrow \sum \tilde{u}_i \tilde{d}_i$)           &   $\sim 1\%$ ($Z_R \rightarrow \sum \tilde{u}_i \tilde{u}_i$) \\
&&\\
\hline
Wino-R   (3b)       &   $27.91\%$ ($W_R^\pm \rightarrow   { \chi^0_1 }      \chi^\pm_1  $) &         $21.61\%$ ($Z_R \rightarrow   \chi^\mp_1     \chi^\pm_1  $)\\
$ T^{yq}_{  ii}=48$ TeV ($i=1,2,3$) & $3\%$ ($W_R^\pm \rightarrow \sum \tilde{u}_i \tilde{d}_i$)&$12\%$ ($Z_R \rightarrow \sum \tilde{d}_i \tilde{d}_i$)\\
&&$\sim 1\%$ ($Z_R \rightarrow \sum \tilde{u}_i \tilde{u}_i$)\\
\hline
Higgsino-($\phi_{1,2}$)       &   $8.43\%$ ($W_R^\pm \rightarrow   { \chi^0_1 }      \chi^\pm_1  $)         &  $6.58\%$ ($Z_R \rightarrow   { \chi^0_1 }     \chi^0_2 $)\\
$ T^{yq}_{  ii}=45$ TeV ($i=1,2,3$)&$8.43\%$ ($W_R^\pm \rightarrow   { \chi^0_2 }      \chi^\pm_1  $)&$6.57\%$ ($Z_R \rightarrow   \chi^\mp_1     \chi^\pm_1  $)\\
&$6.38\%$ ($W_R^\pm \rightarrow \sum \tilde{u}_i \tilde{d}_i$)&$15.67\%$ ($Z_R \rightarrow \sum \tilde{d}_i \tilde{d}_i$)\\
&&$1.5\%$ ($Z_R \rightarrow \sum \tilde{u}_i \tilde{u}_i$)\\
\hline
Higgsino-($\Delta_R^{1,2}$)  & $11.75\%$ ($W_R^\pm \rightarrow   { \chi^0_1 }      \chi^\pm_1  $)& $30.37\%$ ($Z_R \rightarrow   { \chi^0_1 }     \chi^0_2 $)\\
$ T^{yq}_{  ii}=45$ TeV ($i=1,2,3$)          &   $11.77\%$ ($W_R^\pm \rightarrow   { \chi^0_2 }      \chi^\pm_1  $)         & $1.82\%$ ($Z_R \rightarrow   \chi^\mp_1     \chi^\pm_1  $) \\
&$20.70\%$ ($W_R^pm \rightarrow   \chi^\mp_1   \chi^{\pm\pm}$)& $8.49\%$ ($Z_R \rightarrow \chi^{\mp\mp} \chi^{\pm\mp}$)\\
&$1.97\%$ ($W_R^\pm \rightarrow \sum \tilde{u}_i \tilde{d}_i$)&$8.9\%$ ($Z_R \rightarrow \sum \tilde{d}_i \tilde{d}_i$) \\
&&$\sim 1\%$ ($Z_R \rightarrow \sum \tilde{u}_i \tilde{u}_i$)\\
\hline
\end{tabular}}
\end{center}
\caption{ The remaining other parameters are fixed as in the Tab.~\ref{table-Fix}. The electroweakino masses remain same as in the pure LSP (see the Tab.~\ref{table-100LSP}).}
\label{table-100LSPSquarks}
\end{table}
%%%%%%%%%%%%%%%%%%%%%%%%%%%%%%%%%%%%%%%%%%%%%%%%%%%%%%%%

The left-handed wino-like LSP primarily consists of $\widetilde W_{L}^0$ gauge boson fields.
We use similar numerical values of the parameters as in the previous case to get the lightest down-type squark with $m_{\tilde{d}_1}=1$ TeV. The decay width and branching of the heavy vector bosons $W_R$ and $Z_R$ into the final state squarks remain almost similar as in the previous Bino-like LSP case. 

The singlino-like LSP case, we use similar values of the mass-squared parameters $M_{QL,ii}^2$ and $ M_{QR,ii}^2$. As $ms$ is different for this choice of BP 6 in the Tab.~\ref{table-100LSP}, we take different $ T^{yq}_{  ii}=52.5$ TeV ($i=1,2,3$) to get the Higgs mass $125$ GeV and lightest down-squark mass $m_{\tilde{d}_1}=1$ TeV. Also in this case, the decay width and branching of the heavy vector bosons $W_R$ and $Z_R$ into the final state squarks remain almost similar. 

In above cases, the branching of the $W_R$ and $Z_R$ bosons into the SM particles final states decreases due to newly accessible squarks final states. Similarly for the choice of BP 3b and 3c in the Tab.~\ref{table-100LSP} including alike $M_{QL,ii}^2$, $ M_{QR,ii}^2$ and $ T^{yq}_{  ii}$ give almost same decay width of the heavy vector bosons $W_R$ and $Z_R$ into the final state squarks channels. Here the branching of the channels $W_R^\pm \rightarrow   { \chi^0_1 }      \chi^\pm_1  $ and $Z_R\rightarrow   \chi^\pm_1     \chi^\mp_1  $ are decreased by $\sim1\%$ and $\sim5\%$ respectively. Whereas the SM final states are changed by $\sim4\%$ and $\sim10\%$ respectively. As  $ \lambda^c$ ,  $ \mu_2$  and  $ \mu $  have different numerical values for the choice of BP 4 and 5, we use $ T^{yq}_{  ii}= 45$ TeV to get lightest squarks $m_{\tilde{d}_1}=1$ TeV and Higgs-like scalar mass at 125 GeV. Also in these cases, the decay width into new final state squark channels remain almost same and hence the branching ratios of the both SM and BSM, i.e., combination of neutralinos and/or charginos final state have changed.

%%%%%%%%%%%%%%%%%%%%%%%%%%%%%%%%%%%%%%%%%%%%%
\begin{table}[h!]
\begin{center}\scalebox{0.7}{
\begin{tabular}{|c||c|c|c|}
\hline
\hline
~~~~ &  &\multicolumn{2}{c|}{ }\\
~~LSP-Type~~ & Benchmark points &\multicolumn{2}{c|}{ Branching Ratio of $W_R^\pm,~Z_R$ into different BSM fields}\\
~~~~ & for collider analysis &\multicolumn{2}{c|}{ }\\
\cline{3-4} 
 &~~~~~~~~~~~~~ &   $W_R^\pm \rightarrow BSM$       &   $Z_R   \rightarrow BSM  $\\
\hline
      &       & $29.01\%$ ($W_R^\pm \rightarrow   { \chi^0_1 }      \chi^\pm_1  $) & $25.17\%$ ($Z_R \rightarrow   \chi^\mp_1     \chi^\pm_1  $)\\
& $ M_1 = 60$ TeV, $ M^{W_L}_{  33,22,11} = 60$ TeV, $ M^{W_R}_{ 22,11} = 550$ GeV, &$23.46\%$ ($W_R^+ \rightarrow   u  \ov d  $) & $8.62\%$ ($Z_R \rightarrow   u \ov u  $)\\
1. Wino-R   (95 $\%$)    & $ M^{W_R}_{ 33} = -550$ GeV, $ { \mu_2 }  = 21$ TeV,  $ {\mu}  = -17$ TeV,  $ { \lambda^c }  = -0.2$,  &  $23.46\%$ ($W_R^+ \rightarrow   c \ov s  $)  &         $13.1\%$ ($Z_R \rightarrow d \ov d$)\\
{\bf BP1}&$ { \mu_s }  = -20$ TeV,  $ T^{yq}_{  33}=82$ TeV&  $23.46\%$ ($W_R^+ \rightarrow   t \ov b  $) &$8.62\%$ ($Z_R \rightarrow c \ov c$)\\
&&$0\%$ ($W_R^\pm \rightarrow   l \nu  $)&$13.1\%$ ($Z_R \rightarrow s \ov s$)\\
&&&$8.62\%$ ($Z_R \rightarrow t \ov t$)\\
&&&$13.1\%$ ($Z_R \rightarrow b \ov b$)\\
&&&$4.04\%$ ($Z_R \rightarrow l l$, $l=e,\mu$)\\
\cline{3-4}
       &        &  \multicolumn{2}{c|}{}\\  
       &   $M_{  { \chi^0_1 }   }=  903$ GeV      &   \multicolumn{2}{c|}{}\\      
       &   $M_{  \chi^\pm_1  }=  1261$ GeV      &  \multicolumn{2}{c|}{$100\%$ ($  \chi^\pm_1   \rightarrow   { \chi^0_1 }    W^\pm$)}\\
       &   $M_{\chi_2}=  7468$ GeV       &  \multicolumn{2}{c|}{}\\
       &   $M_{\chi^{\pm\pm}}=  20300.0$ GeV   &  \multicolumn{2}{c|}{}\\
       &        &   \multicolumn{2}{c|}{}\\
&&\multicolumn{2}{c|}{}\\
\hline
&& $12.59\%$ ($W_R^\pm \rightarrow   { \chi^0_1 }      \chi^\pm_1  $)& $34.34\%$ ($Z_R \rightarrow   { \chi^0_1 }    \chi^0_2$)\\
2.Higgsino-($\Delta_R^{1,2}$)         & $ M_1 = 60$ TeV, $ M^{W_L}_{  33,22,11} = 60$ TeV, $ M^{W_R}_{  33,22,11} = 60$ TeV, &   $12.48\%$ ($W_R^\pm \rightarrow   { \chi^0_2 }      \chi^\pm_1  $)         & $2.02\%$ ($Z_R \rightarrow   \chi^\mp_1     \chi^\pm_1  $) \\
(57.7 and 41.6 $\%$)& $ { \mu_2 }  = -1$ TeV,  $ {\mu}  = -17$ TeV, $ { \lambda^c }  = -0.05$,&$16.25\%$ ($W_R^\pm \rightarrow   \chi^\mp_1   \chi^{\pm\pm}$)& $9.40\%$ ($Z_R \rightarrow \chi^{\mp\mp} \chi^{\pm\mp}$)\\
{\bf BP2}& $ { \mu_s }  = -20$ TeV,  $ T^{yq}_{  33}=78.2$ TeV & $19.39\%$ ($W_R^+ \rightarrow   u \ov d  $)   &    $6.15\%$ ($Z_R \rightarrow u \ov u$)\\
 &    &  $19.39\%$ ($W_R^+ \rightarrow   c \ov s  $)  &     $9.34\%$ ($Z_R \rightarrow d \ov d$)\\
 &    & $19.39\%$ ($W_R^+ \rightarrow   t \ov b $)   &     $6.15\%$ ($Z_R \rightarrow c \ov c$)\\
 &    &   $0\%$ ($W_R^+ \rightarrow   l \nu  $) &     $9.34\%$ ($Z_R \rightarrow s \ov s$)\\
 &    &   &     $6.15\%$ ($Z_R \rightarrow t \ov t$)\\
 &    &   &     $9.34\%$ ($Z_R \rightarrow b \ov b$)\\
 &    &    &     $2.86\%$ ($Z_R \rightarrow l l$, $l=e,\mu$)\\
\cline{3-4}
       &        &   \multicolumn{2}{c|}{}\\
       &   $M_{  { \chi^0_1 }   }=  928$ GeV      &  \multicolumn{2}{c|}{ $2.40\%$ ($\chi^0_2 \rightarrow   { \chi^0_1 }    h$)} \\       
       &   $M_{  \chi^\pm_1  }=  1148$ GeV      &   \multicolumn{2}{c|}{ $97.57\%$ ($\chi^0_2 \rightarrow   { \chi^0_1 }    Z$)}\\
       &   $M_{\chi_2}=  1193$ GeV       &   \multicolumn{2}{c|}{$100\%$ ($  \chi^\pm_1   \rightarrow   { \chi^0_1 }    W^\pm$)}\\
       &   $M_{\chi^{\pm\pm}}=  1175$ GeV     &  \multicolumn{2}{c|}{}\\
       &        &  \multicolumn{2}{c|}{$18.75\%$ ($\chi^{\pm\pm} \rightarrow   { \chi^0_1 }    l l$)}\\
       &        &  \multicolumn{2}{c|}{$18.68\%$ ($\chi^{\pm\pm} \rightarrow   \chi^\pm_1   l \nu_l$)}\\
       &        &  \multicolumn{2}{c|}{$62.57\%$ ($\chi^{\pm\pm} \rightarrow   \chi^\pm_1   u d$)}\\
\hline
\end{tabular}}
\end{center}
\caption{The branching fraction of relevant processes for these benchmark points to analyze the mono-$X$ ($X=W,Z$) plus $\slashed{E}_T$ and dilepton signatures through the cascade and direct decay of the heavy gauge bosons. The other input parameters and the heavy gauge bosons masses are fixed as in the previous section.
}
\label{table-NLSP}
\end{table}
%%%%%%%%%%%%%%%%%%%%%%%%%%%%%%%%%%%%%%%%%%%%

\section{Collider signature of the heavy gauge bosons} \label{coll}

The decay modes of the heavy gauge bosons into pairs of leptons or jets can be considered a direct probe of these heavy gauge bosons in the context of collider searches. It has already been studied in the literature for various kinds of models with extra gauge bosons~\cite{Aaboud:2017yvp,Sirunyan:2017ukk,Aaboud:2017buh,Khachatryan:2016jww}. The heavy gauge bosons in our model can also have significant branching ratios into final states involving quarks and charged leptons. Processes involving right--handed neutrino final states are absent here due to these particles being heavier than the heavy gauge bosons. Hence, similar to the previous studies, one can probe the gauge bosons in the context of this LRSUSY model in the dilepton and dijet final states. However, in the presence of light supersymmetric particles, various other decay channels with significant branching are also present. It would be interesting to study these new channels involving the light electroweakinos as they may reveal interesting features of this model in the context of collider searches. These may even be helpful in suggesting new direction to search for these extra gauge bosons. To address these questions, we analyze several possible decay channels for the heavy gauge bosons in the context of the High Luminosity LHC (HL-LHC) with $\sqrt{s}=14$ TeV energy and a luminosity of 3000\,${\rm fb}^{-1}$ and the High Energy LHC (HE-LHC) with 27 TeV com energy and luminosity of 3000\,${\rm fb}^{-1}$. Before we discuss the new SUSY decay channels, we would first like to analyze the familiar dilepton and dijet final states arising from the decay of the heavy gauge bosons in our model. 

We consider two benchmark points (BP1 \& BP2) in Tab.~\ref{table-NLSP} corresponding to two distinctly different compositions for the LSP. In both of these two BPs, the electroweakino masses are taken around a TeV so that one could avoid the present constraints. The BPs are chosen such that the heavy gauge bosons have significant decay BRs into SUSY final states as they will also be used later in our analysis of SUSY decays of $W_R$ and $Z_R$ bosons.  

%%%%%%%%%%%%%%%%%%%%%%%%%%%%%%%%%%%%%%%%%%%%
\begin{figure}
\begin{center}
\includegraphics[width=8.2cm]{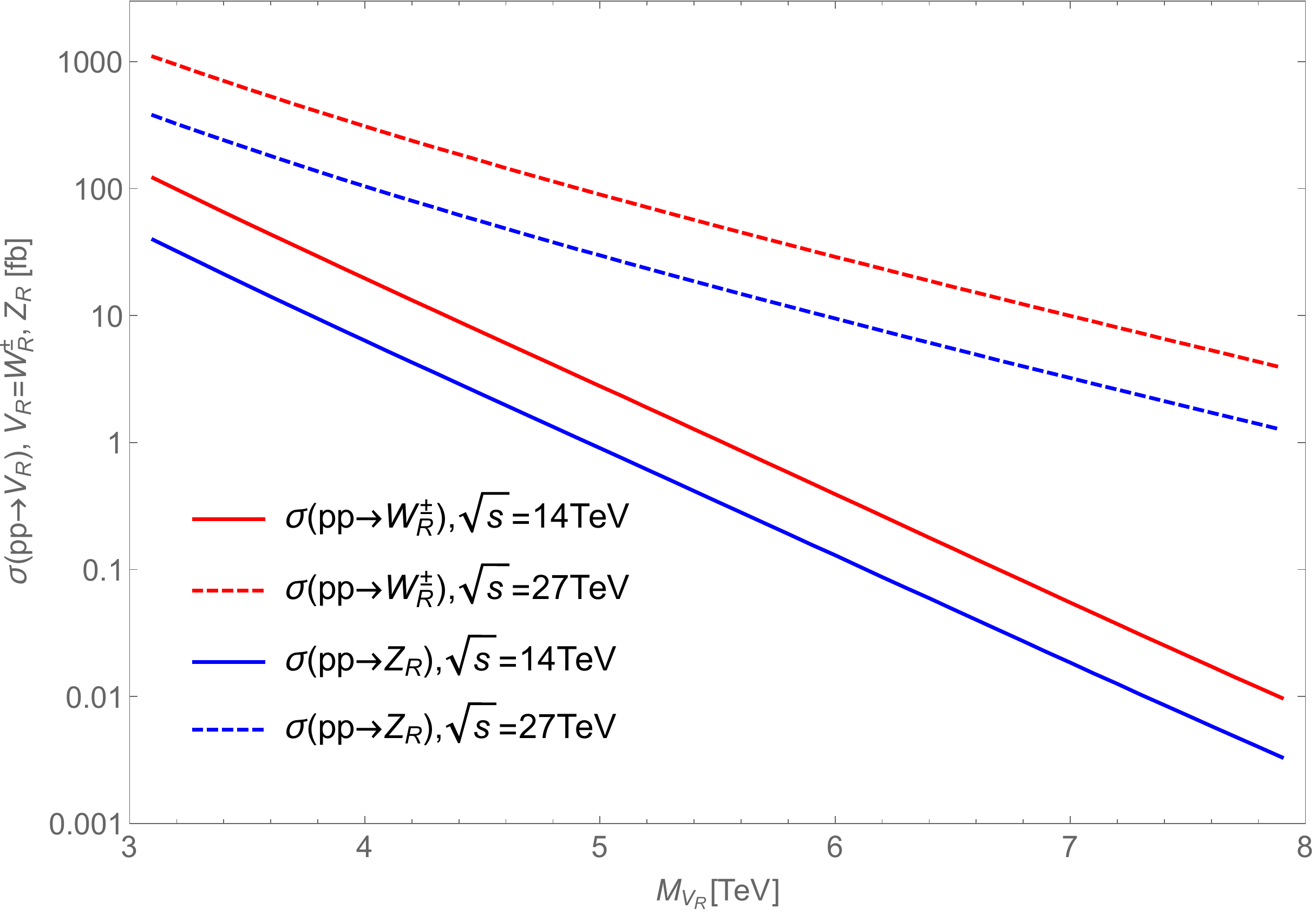}
\end{center}
\caption{Production cross-section for the heavy $W_R$ and $Z_R$ bosons at the 14 and 27 TeV LHC experiment.}
\label{fig:cs}
\end{figure}
%%%%%%%%%%%%%%%%%%%%%%%%%%%%%%%%%%%%%%%%%%%%%%%%%%%

The gauge bosons are produced via the s-channel quark-quark scattering processes. The production cross--sections of the heavy gauge boson as a function of their masses are shown in Fig.~\ref{fig:cs} for the 14 and 27 TeV center-of-mass energies at the LHC. We also list the respective production cross-sections for gauge bosons masses of $M_{W_R}=4.5$ TeV and $M_{Z_R}=7.3$ TeV in the Tab.~\ref{table-S1}, as those are the ones we consider extensively for our analysis. We use {\tt SARAH-4.8.6}~\cite{Staub:2012pb,Staub:2015kfa} to get the input codes for {\tt SPheno-4.0.3}~\cite{Porod:2003um,Porod:2011nf} and {\tt MadGraph-2.3.3}~\cite{Alwall:2014hca}. The SPheno software helps to produce the particle spectrum. Using this spectrum into the {\tt MadGraph-2.3.3}, we calculate the production cross-section of the heavy gauge bosons. We use {\tt MadGraph-2.3.3} to generate the
signal as well as background events and {\tt  PYTHIA-6.4.28}~\cite{Sjostrand:2006za} for showering and hadronization. All generated signal and background events are processed through a fast simulation package {\tt Delphes-3.4.1}~\cite{deFavereau:2013fsa} and we choose ALTAS configuration card for the analysis.

%%%%%%%%%%%%%%%%%%%%%%%%%%%%%%%%%%%%%%%%%%%%
\begin{table}
\begin{center}\scalebox{0.7}{
\begin{tabular}{|c||c|c|c|}
\hline
\hline
Sl. no. &  { Processes}~~      &   \multicolumn{2}{c|}{cross-section in [fb]}\\
\cline{3-4}
 &   &   $14$ TeV &  $27$ TeV\\
\hline
&&&\\
1~~&~   $  pp \rightarrow W_R^\pm$   ~&~        7.03     ~&~      163         ~\\
&&&\\
2~~&~   $pp\rightarrow Z_R $              ~&~      0.01           ~&~   2.3    ~\\
&&&\\
\hline
\end{tabular}}
\end{center}
\caption{ Production cross-sections of the heavy gauge bosons with mass $M_{W_R}=4.5$ TeV and $M_{Z_R}=7.3$ TeV at $\sqrt{s} =  14$ and $27$ TeV respectively.
}
\label{table-S1}
\end{table}
%%%%%%%%%%%%%%%%%%%%%%%%%%%%%%%%%%%%%%%%%%%%%%%%%%%%%%%

\subsection{Dilepton searches}
The major contributions to the dilepton final states from the heavy gauge bosons in this model are coming from the $Z_R$ bosons decaying into the same-flavor opposite-sign (SFOS) leptonic final states with BRs of $4.04\%$ and $2.86\%$ for BP1 and BP2 respectively. The charged gauge boson $W_R$ decaying into leptonic final states $W_R^\pm\rightarrow l\nu$, $l=e,\mu$ are negligibly small due to the heavier right-handed neutrinos as was discussed earlier. We thus analyze the dilepton (electron or muon) final state signals at the 14 and 27 TeV LHC to investigate the possibility of identifying possible dilepton signal from the $Z_R$ decays in this model. To cleanly identify the final state leptons we use several selection criteria for the isolated leptons. The charged lepton isolation demands that there is no other charged particle with $p_T > 0.5$ GeV within a cone of
$\Delta R = \sqrt{\Delta \eta^2 +  \Delta\Phi^2} < 0.5$ centered on the cell associated to the charged lepton. Here $p_T$, $\eta$ and $\phi$ are the transverse momentum, pseudo-rapidity and the polar angle of charged leptons respectively. In addition, the ratio of the scalar sum of the transverse momenta of all tracks to the $p_T$ of the lepton (chosen for isolation) is less than $0.12$ ($0.25$) for electron (muon). The event is selected with each isolated lepton (electron or muon) having transverse momentum $p_T$ larger than 30 GeV. Also the candidate electron(muon) is required to satisfy the rapidity cut $|\eta| < 2.5$. Another important variable is the dilepton (SFOS pair) invariant mass distribution $M_{ll}$ which will be a useful probe to search for the $Z_R$ gauge boson in this case.

Several SM processes can contribute as background for the dilepton signal arising from the decay of the $Z_R$ boson. Among them, the $pp\rightarrow Z/\gamma^* \rightarrow ll$ channels become dominant due to the presence of virtual photon mediated processes. The other processes like $pp\rightarrow ZZ ~(Z \rightarrow ll, Z\rightarrow \nu \overline{\nu})$, $pp\rightarrow ZW~(Z \rightarrow ll, W\rightarrow l \nu )$, $pp \rightarrow WW~ (W \rightarrow l \nu)$ and $t\overline{t}, t \rightarrow W b$ also add to the SM background. Invariant mass distribution for the signal and background events are shown in Fig.~\ref{fig:3ev} for LHC energy of $\sqrt{s}=14$ TeV (left) and  $\sqrt{s}=27$ TeV (right). 
%%%%%%%%%%%%%%%%%%%%%%%%%%%%%%%%%%%%%%%%%%%%%%%%%%%%%
%%%%%%%%%%%%%%%%%%%%%%%%%%%%%%%%%%%%%%%%%%%%%%%%%%%%%%
%%%%%%%%%%%%%%%%%%%%%%%%%%%%%%%%%%%%%%%%%%%%%%%%%%%%%
\begin{figure}[h!]
\centering
\subfloat[]{\includegraphics[width=3.0in,height=2.8in, angle=0]{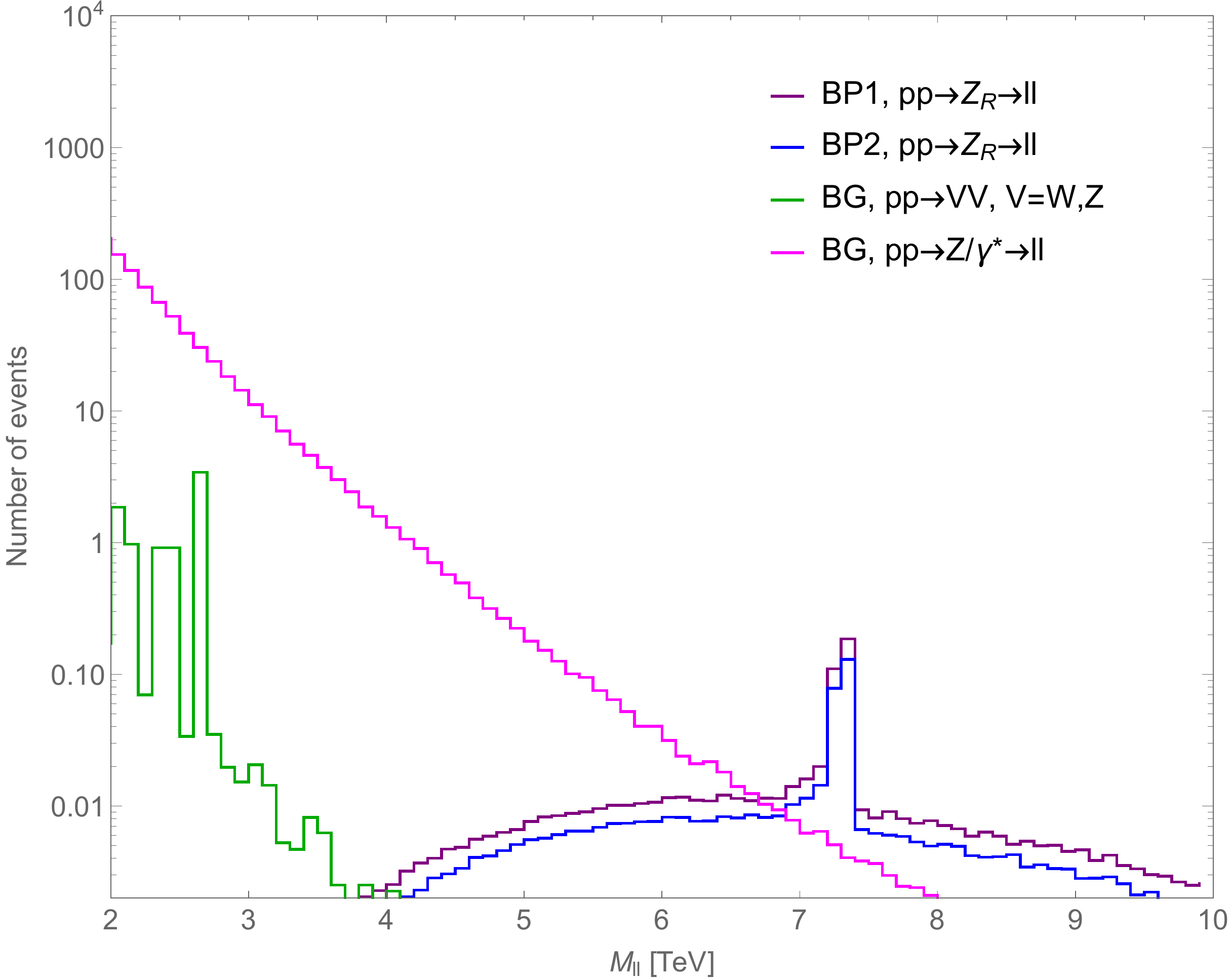}} 
\subfloat[]{\includegraphics[width=3.0in,height=2.8in, angle=0]{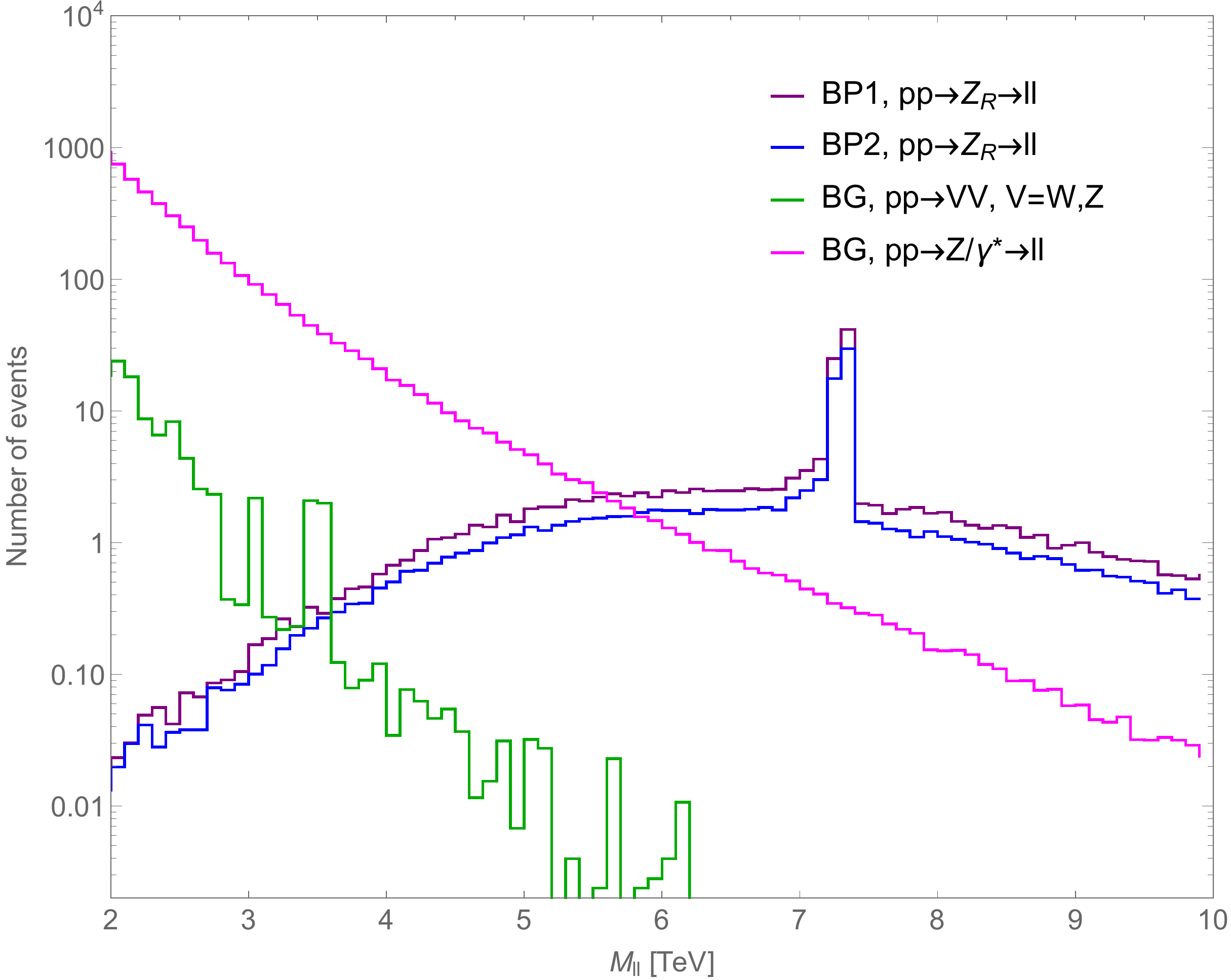}}
\caption{ The invariant mass distribution of the SFOS pair of leptons. The signal as well as the backgrounds are selected with $p_T^l>30$ GeV ($l=e,\mu$) and $|\eta|<2.5$. 
The purple and blue lines stand for the signal $pp\rightarrow ll$ events corresponding to BP1 and BP2 respectively, whereas the backgrounds are indicated by the green and magenta lines. The left plot is in the context of the LHC at $14$ TeV and right plot is for $27$ TeV. The integrated luminosity  $3000~{\rm fb^{-1}}$ remains same for all these processes.} 
\label{fig:3ev} 
\end{figure} 
%%%%%%%%%%%%%%%%%%%%%%%%%%%%%%%%%%%%%%%%%%%%%%%%%%%%% 
%%%%%%%%%%%%%%%%%%%%%%%%%%%%%%%%%%%%%%%%%%%%%%%%%%%%%
%%%%%%%%%%%%%%%%%%%%%%%%%%%%%%%%%%%%%%%%%%%%%%%%%%%%%
%%%%%%%%%%%%%%%%%%%%%%%%%%%%%%%%%%%%%%

The shape of the invariant mass distribution in the dilepton plots shown in Fig.~\ref{fig:3ev} can be understood as an effect of the lepton smearing and the final state radiation (FSR).
%%%%%%%%%%%%%%%%%%%%%%%%%%%%%%%%%%%%%%%%%%%%
\begin{figure}
\begin{center}
\includegraphics[width=8.2cm]{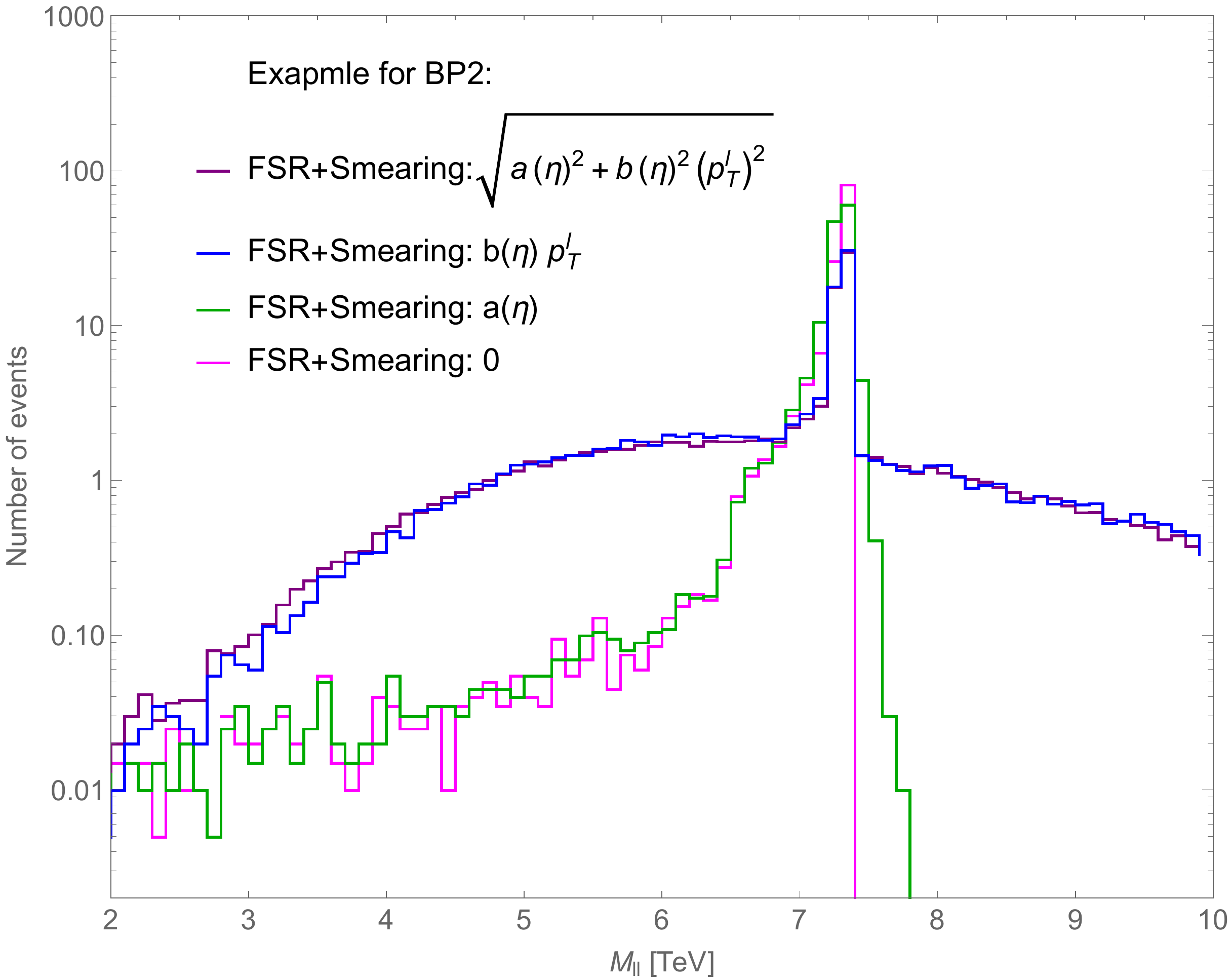}
\end{center}
\caption{The effect of the smearing and the final state radiation (FSR) in the invariant mass distribution for the dilepton signal events at 27 TeV with $L=3000 ~{\rm fb^{-1}}$ for BP2.}
\label{fig:semar}
\end{figure}
%%%%%%%%%%%%%%%%%%%%%%%%%%%%%%%%%%%%%%%%%%%%%%%%%%%
The final state lepton momentum is obtained by a Gaussian smearing of the initial 4-momentum vector. The resolution, which depends on the interaction with detector(s), is parametrized as a function of $p_T^l$ and $\eta$ given as 
\begin{equation}
F(\eta, p_T^l)=\sqrt{a^2(\eta) + b^2(\eta) {(p^l_T)}^2}.
\label{eqn:res}
\end{equation}
The {\tt Delphes} ATLAS card provides the typical values of $a(\eta)$ ranging from $0.01$ to $0.15$ while $b(\eta)$ varies from $1.0  \times 10^{-4}$ to $3.1 \times 10^{-3} $. Thus, for events with large $p_T^l$, the second term in Eqn.~\ref{eqn:res} will dominate compared to the first term. This is especially true in our scenario where the leptons are produced from the decay of a heavy $Z_R$ boson. The separate invariant mass distributions has been demonstrated in Fig.~\ref{fig:semar} to visualize the FSR and smearing effects. The purple line represents the original signal events at 27 TeV LHC with $L=3000 ~{\rm fb^{-1}}$ for BP2 in which both the $a(\eta)$ and momentum dependent $b(\eta) p_T^l$ terms are present in the smearing resolution function $F(\eta, p_T^l)$. In the blue line, the $a(\eta)$ term has been taken to be zero resulting in the resolution function being $F(\eta, p_T^l)=b(\eta) p_T^l$. As the original resolution function was anyway dominated by the momentum dependent term, it is not surprising that the blue plot is fairly similar to the original purple plot. The green plot is obtained by choosing the momentum dependent term in the resolution function to be zero. Finally, if the smearing effect is completely neglected, i.e., $a(\eta)=b(\eta)=0$, then one can obtain the magenta line in the invariant mass-distribution. Although there are no events here in the $M_{ll}>M_{Z_R}$ range, some smearing-like effect is still there in the region $M_{ll}<M_{Z_R}$. The cause of this smearing is the final state radiation (FSR) of photons from the charged leptons. Since these FSR photons can only be radiated with energies less than the energy of the leptons, the smearing effect is only observed in the $M_{ll}<M_{Z_R}$ region. The nature of the dilepton invariant mass distribution plots arising from the decay of the heavy $Z_R$ boson can thus be easily understood.

A significant contribution to the dilepton background events can also arise from the $pp \rightarrow jj$ where the jets $j$ can fake as leptons ($0.5\%$ into electron, whereas $0.1\%$ for muon). In fact, the jets faking leptons background for $p_T^j>20$ GeV become larger than the $pp\rightarrow Z/\gamma^* \rightarrow ll$ events due to their large production cross-section. It could effectively be reduced by the large $p_T^j$ cut on the selected background events. The signal events $pp\rightarrow Z_R \rightarrow ll$ though will also be affected by the same $p_T^l$ cut. We have thus selected only the signal as well as the background events with $p_T^{j,l}>1$ TeV. The signal region $7.2<M_{ll} <7.4$ TeV is used to further reduce the backgrounds and optimize the signals. In this signal region with $p_T^l>1$ TeV, the $pp\rightarrow VV,t \ov t$ background become almost negligible. The number of signal and background events after implementing these cuts are shown in the Tab.~\ref{table-BGSD}.

The expected number of the signal events for both BP1 and BP2 at the 14 TeV run of the LHC with luminosity $L=3000 ~{\rm fb^{-1}}$ become less than unity because of the small production cross-section (see the Tab.~\ref{table-BGSD}). However, for the LHC run at 27 TeV with $L=3000 ~{\rm fb^{-1}}$, the dilepton final state channels produce a large number of signal events satisfying all the above mentioned cuts. The significance of the signal over background attains a value of 32.25(22.59) for BP1(BP2) for the HE-LHC. Hence, one can use these results to discover/exclude the heavy $Z_R$ boson through this channel.
 
%%%%%%%%%%%%%%%%%%%%%%%%%%%%%%%%%%%%%%%%%%%%%
\begin{table}[h!]
\begin{center}\scalebox{0.7}{
\begin{tabular}{|c|c|c|c|c|c|c|c|c|c|c|c|c|}
\hline
\hline
Energy & \multicolumn{6}{c|}{ SM background for the signal $pp \rightarrow Z_R \rightarrow l l$}&\multicolumn{2}{c|}{Total signal events}&\multicolumn{2}{c|}{Significance}\\
\cline{2-11}
[TeV]&$pp \rightarrow Z/\gamma^* \rightarrow ll$&~~ $pp \rightarrow jj$ ~& ~~$ZZ$~~& ~~$WZ$~~~& ~~~$WW$~~~ & ~~~~$t\overline{t}$~~~~&BP1&BP2&BP1&BP2\\
\cline{1-11}
&&&&&&&&&&\\
14&$\sim 0$&$\sim 0$&0&0&0&0&0.28&0.19&--&--\\
&&&&&&&&&&\\
27&2.77&1.53&0&0&0&0&65.73&46.72&32.25&22.59\\
&&&&&&&&&&\\
\hline
\end{tabular}}
\end{center}
\caption{
The signal as well as the backgrounds are selected with $p_T^l>1$ TeV ($l=e,\mu$) and $|\eta|<2.5$ to reduce the background contribution. 
The signal $pp \rightarrow Z_R \rightarrow ll$ and the background are obtained after optimization cuts  $7.2<M_{ll} <7.4$ TeV.
The dijet (jet misidentified as lepton) background are also reduced by these choice of selection $p_T^{j}>1$ TeV and optimization  $7.2<M_{jj} <7.4$ TeV.}
\label{table-BGSD}
\end{table}
%%%%%%%%%%%%%%%%%%%%%%%%%%%%%%%%%%%%%%%%%%%%%%%%%%%%%%%

In these searches, only the $Z_R\rightarrow l l$, ($l=e,\mu$) decay modes have been considered. However, the $Z_R\rightarrow \tau\tau$ can potentially enhance the signal as $\tau$ can give one lepton (electron or muon) in the final state through its decay via the virtual $W$ boson. This contribution remains small as the final state electron or muon can come via the decay of the virtual gauge bosons which is suppressed by the branching and the chosen selection and  optimization cuts $p_T^l>1$ TeV and $7.2<M_{ll} <7.4$ TeV. The number of events coming from the $\tau \tau$ channel at the 14 TeV run of the LHC is always less than one event and not considered further. On the other hand, the number of events for the $Z_R\rightarrow \tau\tau$ channel can be enhanced by 2.70(1.91) for BP1(BP2) at 27 TeV LHC and the corresponding significance  for BP1 and BP2 reaches values of 33 and 23.46 respectively.

\subsection{Dijet searches}

The heavy gauge bosons $Z_R$ and $W_R$ both can decay directly into two quarks giving rise to dijet final states which we study in the context of 14 and 27 TeV LHC experiments. Experimental search for the heavy gauge boson have already been performed~\cite{Aaboud:2017yvp} in the dijet channel and here we follow a similar strategy. The events are selected with at least two anti-$k_t$  jets with jet cone size 0.4 which having transverse momentum $p_T$ greater than 1 TeV. Also the candidate jet is required to satisfy a pseudo-rapidity cut of $|\eta| < 2.5$.
%%%%%%%%%%%%%%%%%%%%%%%%%%%%%%%%%%%%%%%%%%%%%%%%%%%%%
%%%%%%%%%%%%%%%%%%%%%%%%%%%%%%%%%%%%%%%%%%%%%%%%%%%%%%
%%%%%%%%%%%%%%%%%%%%%%%%%%%%%%%%%%%%%%%%%%%%%%%%%%%%%
\begin{figure}[h!]
\centering
\subfloat[]{\includegraphics[width=3.0in,height=2.8in, angle=0]{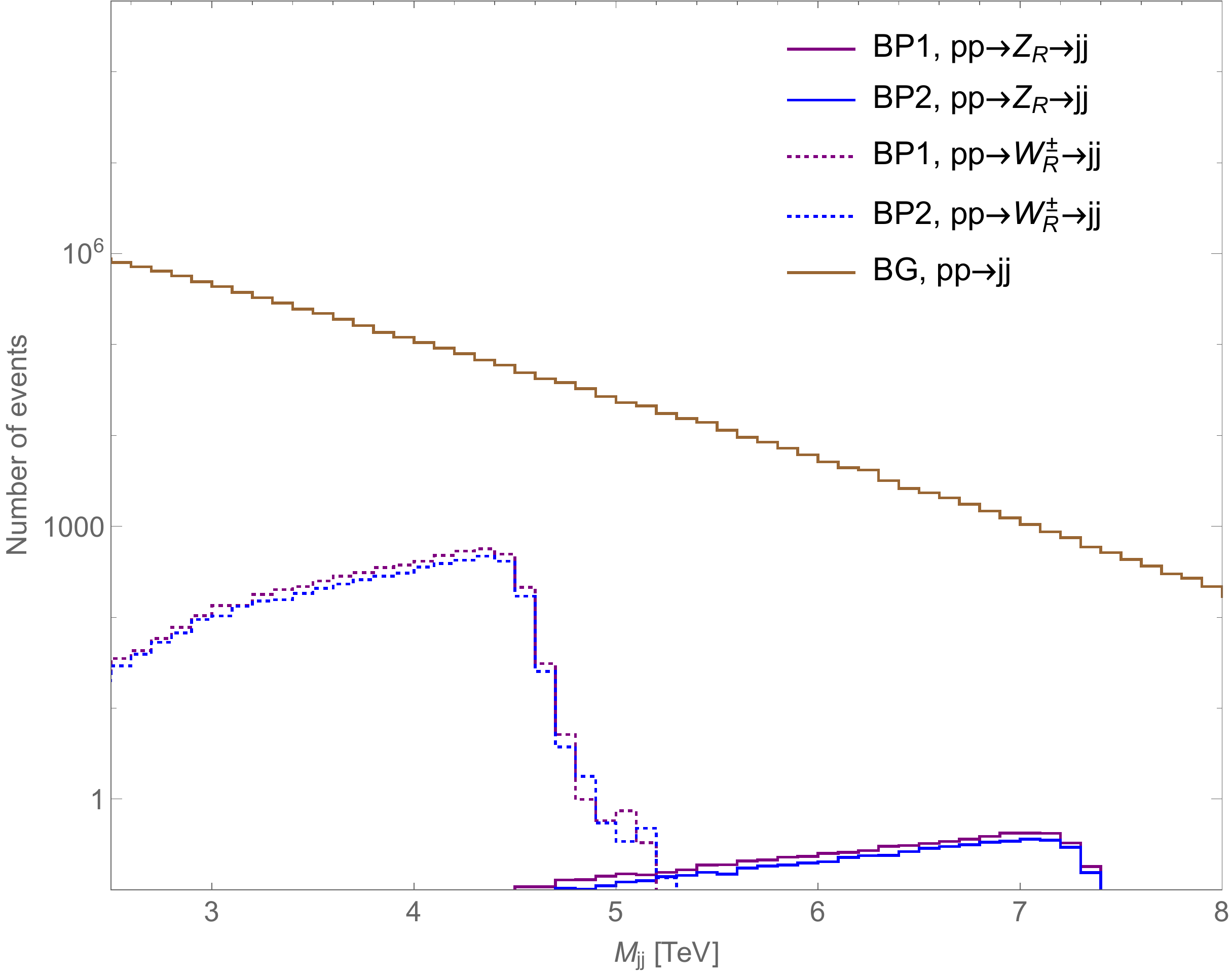}}
\hspace{0.1cm} 
\subfloat[]{\includegraphics[width=3.0in,height=2.8in, angle=0]{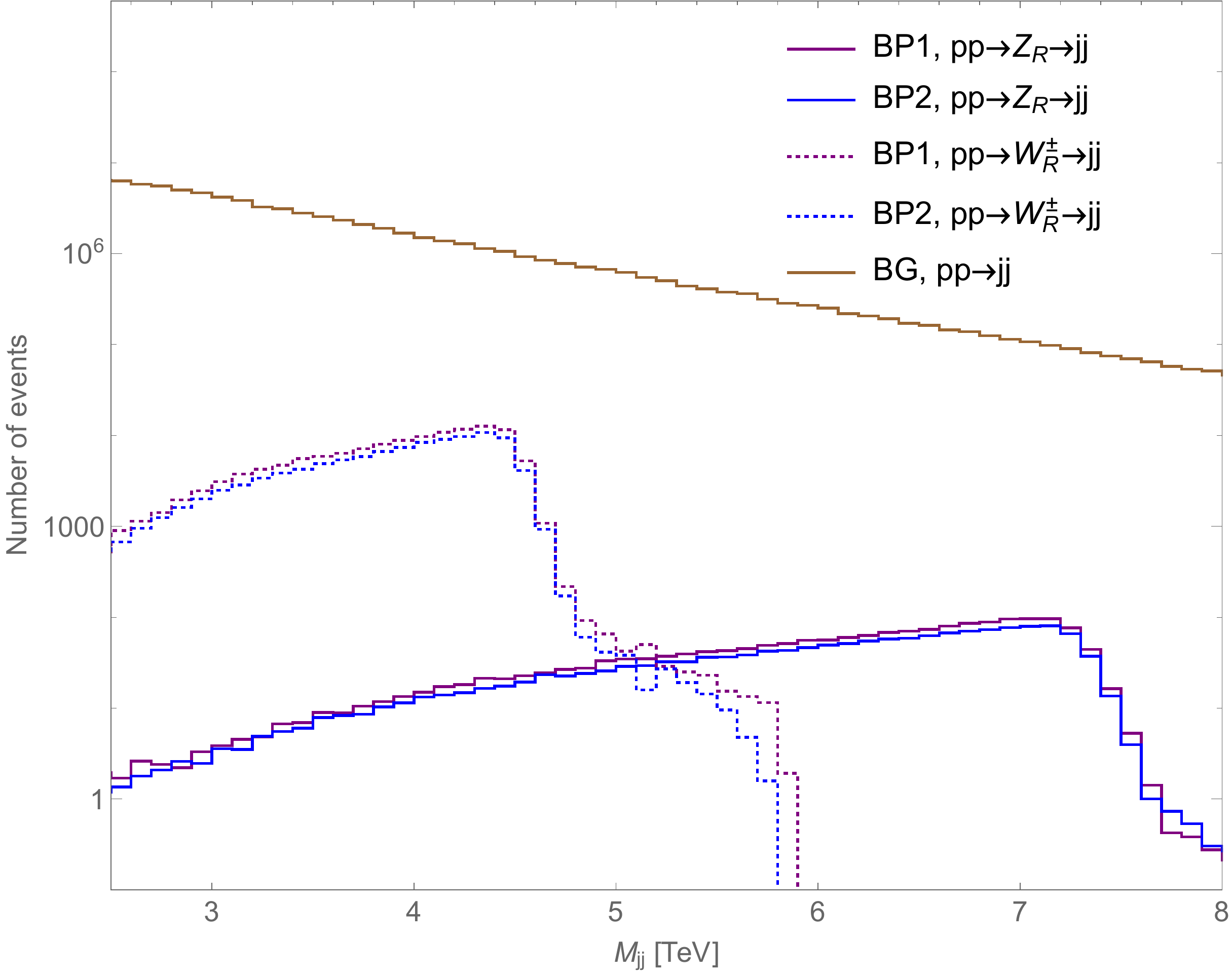}}
\caption{
The dijet invariant mass distribution the signal process $pp \rightarrow  W_R^\pm  \rightarrow jj$ are denoted by doted purple and blue lines. Whereas the solid line stand for the $pp \rightarrow  Z_R  \rightarrow jj $. The left plot is drown in the context of the LHC at $14$ TeV and right plot for  $27$ TeV. The integrated luminosity  $3000~{\rm fb^{-1}}$ remains same for all these processes. 
The dominant SM background $pp \rightarrow jj$ distribution is indicate by the brown line. Here in these plots, the signal as well as the background are selected with $p_T^j>1$ TeV and $|\eta|<2.5$ along with jet's con size 0.4.} 
\label{fig:4ev} 
\end{figure} 
%%%%%%%%%%%%%%%%%%%%%%%%%%%%%%%%%%%%%%%%%%%%%%%%%%%%% 
%%%%%%%%%%%%%%%%%%%%%%%%%%%%%%%%%%%%%%%%%%%%%%%%%%%%%
%%%%%%%%%%%%%%%%%%%%%%%%%%%%%%%%%%%%%%%%%%%%%%%%%%%%%
%%%%%%%%%%%%%%%%%%%%%%%%%%%%%%%%%%%%%%
Similar to the dilepton search, the dijet invariant mass distribution $M_{jj}$ can be a useful probe to search for $W_R$ and $Z_R$ gauge bosons. The dominant SM background for the dijet signal arises from the $pp \rightarrow jj$ process. Other processes including  $pp\rightarrow VV$, $V=W,Z$ and $t\overline{t}$ also add to the SM background but these contributions are extremely small and can be safely neglected. The invariant mass distribution of the signal as well as background events are shown in Fig.~\ref{fig:4ev} for LHC energy $\sqrt{s}=14$ TeV (left) and  $\sqrt{s}=27$ TeV (right). The purple and blue doted-lines indicate the hadronic decay of the $W_R$ whereas the purple and blue solid-lines are those for $Z_R$ gauge boson. A signal region $4.2<M_{jj} <4.55$ TeV has been used to optimize the significance for the $W_R$ gauge boson search. In this choice, the expected number of signal events attain values of 1728.37(1419.94) for BP1(BP2) at 14 TeV LHC with $L=3000 ~{\rm fb^{-1}}$. The significances become $3.60(2.95)$ due to the large irreducible SM background events with large number of events $N_{pp\rightarrow jj}=2.3\times10^{5}$. At the 27 TeV LHC with a luminosity of $3000 ~{\rm fb^{-1}}$, the signal events increase to  $3.92 \times10^4 $( $ 3.24\times10^ 4$) for BP1(BP2) while the background events become $3.91\times 10^6$. This results in an increase of the signal to background significance to $19.83(16.37)$ for BP1 (BP2) respectively in this case.  

A more stringent cut on the signal and background regions with $p_T^j>1.5$ TeV for $W_R$ search can result in a better significance. Though this cut will reduce the signal as well as background events, the background will be affected more since the signal jets are arising from the decay of a heavy resonance and can have larger $p_T$.%%%%%%%%%%%%%%%%%%%%%%%%%%%%%%%%%%%%%%%%%%%%%
\begin{table}[h!]
\begin{center}\scalebox{0.7}{
\begin{tabular}{|c|c|c|c|c|c|c|c|c|c|c|c|}
\hline
\hline
Energy & \multicolumn{5}{c|}{ SM background for the signal $W_R^\pm \rightarrow jj$}&\multicolumn{2}{c|}{Total signal events}&\multicolumn{2}{c|}{Significance}\\
\cline{2-10}
[TeV]&~~ $pp \rightarrow jj$~~& ~~$ZZ$~~& ~~$WZ$~~~& ~~~$WW$~~~ & ~~~$t\overline{t}$~~~&BP1&BP2&BP1&BP2\\
\cline{1-10}
&&&&&&&&&\\
14&$6.5\times10^4$&0&0&0&0&1399.53&1166.33&5.47&4.56\\
&&&&&&&&&\\
27&$1.6\times 10^6$&0&0&0&0& $ 3.19\times10^4 $ &$2.66\times10^4$&29.73&24.74\\
&&&&&&&&&\\
\hline
\end{tabular}}
\end{center}
\caption{Total number of events of the signal $pp \rightarrow W_R^\pm \rightarrow jj$ and SM background with $p_T^j>1.5$ TeV and $|\eta|<2.5$ along with con size 0.4 at the 14 and  27 TeV run of LHC with $L=3000 ~{\rm fb^{-1}}$ are obtained after the optimization cut $4.2<M_{jj} <4.55$ TeV. }
\label{table-BGWJJ}
\end{table}
%%%%%%%%%%%%%%%%%%%%%%%%%%%%%%%%%%%%%%%%%%%%%%%%%%%%%%%
If we choose the same signal region $4.2<M_{jj} <4.55$ TeV as before, the signal significances will attain values of 5.47 (4.56) for 14 TeV LHC with $L=3000 ~{\rm fb^{-1}}$ while the same becomes 29.73 (24.74) for 27 TeV LHC with $L=3000 ~{\rm fb^{-1}}$ for BP1 (BP2) parameter space. The corresponding number are quoted in Tab.~\ref{table-BGWJJ}.

The large mass of the $Z_R$ boson results in a small production cross-section at the 14 TeV LHC. Thus the expected numbers of the signal $pp\rightarrow Z_R \rightarrow jj$ events with $p_T^j>1$ TeV at the 14 TeV LHC with $L=3000 ~{\rm fb^{-1}}$ remain small as can be seen in the distribution plot in Fig.~\ref{fig:4ev}. For a favorably chosen signal region with a dijet invariant mass $6.7<M_{jj} <7.35$ TeV, the expected number of events are two signal events compared to around 7000 background events at the 14 TeV LHC with $3000 ~{\rm fb^{-1}}$ luminosity. The expected numbers for 27 TeV LHC are found to be around 500 signal events compared to $7.12\times10^5$ background events. Hence the significances are extremely poor and one will not be able to identify a $Z_R$ boson in this case. 
As before, we can try to increase the required $p_T$ of the jets to check if the signal significance improves. We select the signal and background events with $p_T^j>3$ TeV. This yields only one expected signal event compared to about 200 background events at the LHC 14 TeV (see Tab.~\ref{table-BGZJJ}). However, the expected number as well as the significance will increase at 27 TeV with $L=3000 ~{\rm fb^{-1}}$. The number of signal events is found to be 276.98 (229.83) and the corresponding significance attains a value of 1.85 (1.55) for BP1(BP2). Hence, we need large energy and luminosity to observe signatures of such heavy $7.3$ TeV neutral gauge boson through the dijet final state channel in this LRSUSY model.

%%%%%%%%%%%%%%%%%%%%%%%%%%%%%%%%%%%%%%%%%%%%%
\begin{table}[h!]
\begin{center}\scalebox{0.7}{
\begin{tabular}{|c|c|c|c|c|c|c|c|c|c|c|c|}
\hline
\hline
Energy & \multicolumn{5}{c|}{ SM background for the signal $Z_R \rightarrow jj$}&\multicolumn{2}{c|}{Total signal events}&\multicolumn{2}{c|}{Significance}\\
\cline{2-10}
[TeV]&~~ $pp \rightarrow jj$~~& ~~$ZZ$~~& ~~$WZ$~~~& ~~~$WW$~~~ & ~~~$t\overline{t}$~~~&BP1&BP2&BP1&BP2\\
\cline{1-10}
&&&&&&&&&\\
14&196.54&0&0&0&0&1.24&1.04&--&--\\
&&&&&&&&&\\
27&$2.17\times10^4$&0&0&0&0&276.98&229.83&1.85&1.55\\
&&&&&&&&&\\
\hline
\end{tabular}}
\end{center}
\caption{Total number of events of the signal $pp \rightarrow Z_R \rightarrow jj$ and background with $p_T^j>3$ TeV along with cone size 0.4 at the 14 TeV and  27 TeV  run of LHC with $L=3000 ~{\rm fb^{-1}}$ are obtained after the optimization cut $6.7<M_{jj} <7.35$ TeV. }
\label{table-BGZJJ}
\end{table}
%%%%%%%%%%%%%%%%%%%%%%%%%%%%%%%%%%%%%%%%%%%%%%%%%%%%%%%

In the above analysis, anti-$k_t$ jets with cone size 0.4 have been considered and the corresponding invariant mass distribution was shown in Fig.~\ref{fig:4ev} for the selected events with $p_T^j>1$ TeV and $|\eta|<2.5$. Now if we increase the jets' cone size to 1.0, the number of events corresponding to the signal as well as the background will be increased. To show the effect, we plot same distribution in Fig.~\ref{fig:4evC} for the BP1 only. The $pp \rightarrow  W_R^\pm \rightarrow jj$ process is demonstrated by the purple lines whereas $pp \rightarrow  Z_R  \rightarrow jj$ process is denoted by blue lines. The solid lines correspond to events with jets' cone size 0.4 (as in the Fig.~\ref{fig:4ev}) whereas the dotted lines indicate the events with the jets having cone size 1.0.
%%%%%%%%%%%%%%%%%%%%%%%%%%%%%%%%%%%%%%%%%%%%%%%%%%%%%
%%%%%%%%%%%%%%%%%%%%%%%%%%%%%%%%%%%%%%%%%%%%%%%%%%%%%%
%%%%%%%%%%%%%%%%%%%%%%%%%%%%%%%%%%%%%%%%%%%%%%%%%%%%%
\begin{figure}[h!]
\centering
\subfloat[]{\includegraphics[width=3.0in,height=2.8in, angle=0]{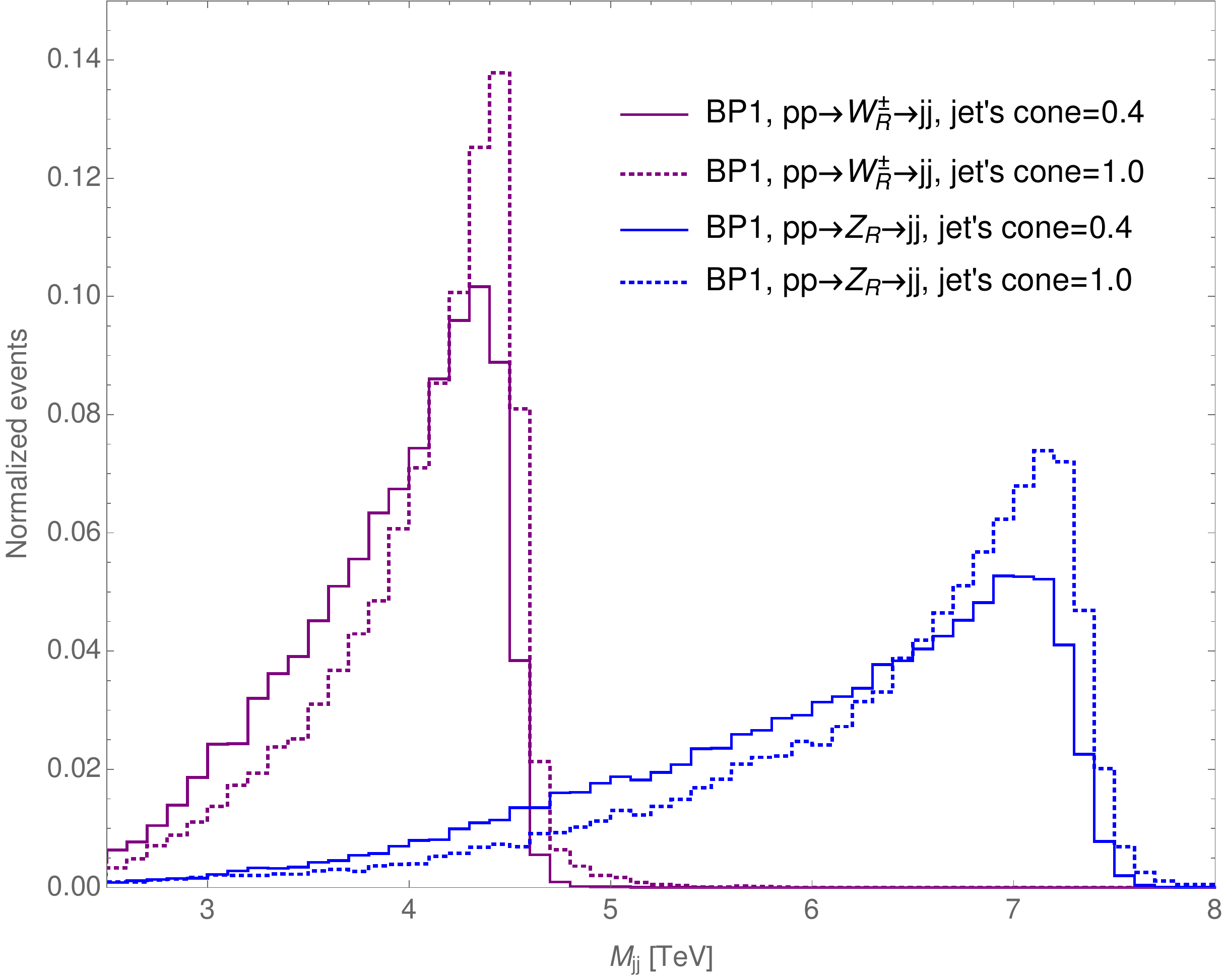}}
\hspace{0.1cm} 
\subfloat[]{\includegraphics[width=3.0in,height=2.8in, angle=0]{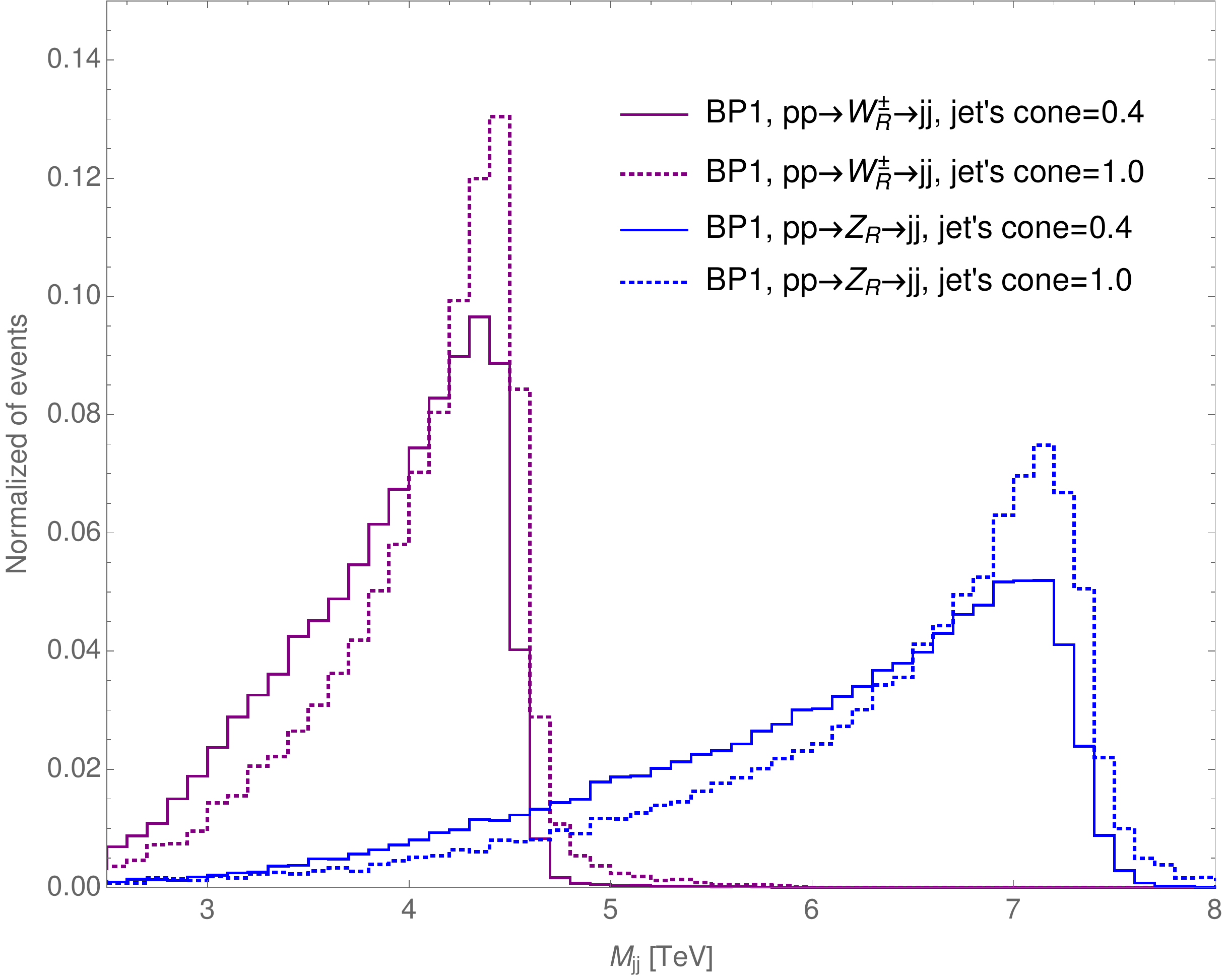}}
\caption{ 
The dijet invariant mass distribution of the signal processes $pp \rightarrow  W_R^\pm \rightarrow jj$ and $pp \rightarrow  Z_R  \rightarrow jj $ are denoted by the purple and blue lines respectively for the BP1. The brown lines stand for the SM background $pp\rightarrow jj$. The solid lines stand for the events with $p_T^j>1$ TeV and $|\eta|<2.5$ along with jet's con size 0.4 as in the Fig.~\ref{fig:4ev}.
Whereas the dotted lines indicate the events with $p_T^j>1$ TeV and $|\eta|<2.5$ along with jet's con size 1.0.
The left plot is drown in the context of the LHC at $14$ TeV and right plot for  $27$ TeV. The integrated luminosity  $3000~{\rm fb^{-1}}$ remains same for all these processes.}
\label{fig:4evC} 
\end{figure}

A similar cut on the signal and background regions with $p_T^j>1.5$ TeV and signal region $4.2<M_{jj} <4.55$ TeV for $W_R$ search yields better significance for a larger jet cone size of 1.0. The number of background events for $pp\rightarrow jj$ attains a values of $9.75 \times 10^4$ at the 14 TeV LHC with luminosity  $3000~{\rm fb^{-1}}$ whereas it become $1.74\times 10^6$ at 27 TeV. The other SM backgrounds $pp\rightarrow VV, t \ov t$ still remain zero. The signal $pp \rightarrow  W_R^\pm \rightarrow jj$ events with these selection and optimization cuts become 2009.74 (1660.52) for BP1 (BP2) at 14 TeV and $4.55 \times 10^4$ ($3.8\times 10^4$) at 27 TeV. The signal over background significance is enhanced and becomes 6.44 (5.32) for BP1 (BP2) at 14 TeV and 34.43 (28.79) at 27 TeV as can be seen from Tab.~\ref{table-BGWJJ2}.
%%%%%%%%%%%%%%%
\begin{table}[h!]
\begin{center}\scalebox{0.7}{
\begin{tabular}{|c|c|c|c|c|c|c|c|c|c|c|c|}
\hline
\hline
Energy & \multicolumn{5}{c|}{ SM background for the signal $W_R^\pm \rightarrow jj$}&\multicolumn{2}{c|}{Total signal events}&\multicolumn{2}{c|}{Significance}\\
\cline{2-10}
[TeV]&~~ $pp \rightarrow jj$~~& ~~$ZZ$~~& ~~$WZ$~~~& ~~~$WW$~~~ & ~~~$t\overline{t}$~~~&BP1&BP2&BP1&BP2\\
\cline{1-10}
&&&&&&&&&\\
14&$9.75 \times 10^4$&0&0&0&0&2009.74&1660.52&6.44&5.32\\
&&&&&&&&&\\
27&$1.74\times 10^6$&0&0&0&0& $4.55 \times 10^4$ &$3.8\times 10^4$&34.43&28.79\\
&&&&&&&&&\\
\hline
\end{tabular}}
\end{center}
\caption{Total number of events of the signal $pp \rightarrow W_R^\pm \rightarrow jj$ and SM background with $p_T^j>1.5$ TeV and $|\eta|<2.5$ along with con size 1.0 at the 14 and  27 TeV run of LHC with $L=3000 ~{\rm fb^{-1}}$ are obtained after the optimization cut $4.2<M_{jj} <4.55$ TeV. }
\label{table-BGWJJ2}
\end{table}
%%%%%%%%%%%%%%%%%%%%%%%%%%%%%%%%%%%%%%%%%%%%%%%%%%%%%%%

Similarly, the signal $pp \rightarrow  Z_R \rightarrow jj$ events with the selection and optimization cuts $p_T^j>3$ TeV and $6.7<M_{jj} <7.35$ TeV go to 1.62 (1.39) at 14 TeV and 363.36 (306.48) at 27 TeV for BP1 (BP2) with $3000~{\rm fb^{-1}}$ luminosity. The backgrounds number of events attain values of 281.34 in the context of the LHC at $14$ TeV and $3.1\times10^4$ at 27 TeV. The significance is slightly enhanced to 2.07 (1.75) at 27 TeV (see the Tab.~\ref{table-BGZJJ2}). Hence it is clear that if we increase the jet's cone size, it will give us larger signal significance for the dijet searches of the heavy gauge bosons in this model. However the observation of a heavy neutral gauge boson with $M_{Z_R} = 7.3$ TeV in the dijet channel is still highly challenging with the luminosity $3000 ~{\rm fb^{-1}}$ while it can easily be discovered in the dilepton channel. Further increasing the luminosity to something like $L=20000~{\rm fb^{-1}}$ with $\sqrt{s}=27$ TeV may be enough to observe this $Z_R$ boson in the dijet channel. On the other hand, a heavy charged gauge boson $W_R$ with a mass of 4.5 TeV may be observed in the dijet channel at the LHC with the luminosity $3000 ~{\rm fb^{-1}}$.

%%%%%%%%%%%%%%%%%%%%%%%%%%%%%%%%%%%%%%%%%%%%%
\begin{table}[h!]
\begin{center}\scalebox{0.7}{
\begin{tabular}{|c|c|c|c|c|c|c|c|c|c|c|c|}
\hline
\hline
Energy & \multicolumn{5}{c|}{ SM background for the signal $Z_R \rightarrow jj$ }&\multicolumn{2}{c|}{Total signal events}&\multicolumn{2}{c|}{Significance}\\
\cline{2-10}
[TeV]&~~ $pp \rightarrow jj$~~& ~~$ZZ$~~& ~~$WZ$~~~& ~~~$WW$~~~ & ~~~$t\overline{t}$~~~&BP1&BP2&BP1&BP2\\
\cline{1-10}
&&&&&&&&&\\
14&281.34&0&0&0&0&1.62&1.39&--&--\\
&&&&&&&&&\\
27&$3.1\times10^4$&0&0&0&0&363.36&306.48&2.07&1.75\\
&&&&&&&&&\\
\hline
\end{tabular}}
\end{center}
\caption{Total number of events of the signal $pp \rightarrow Z_R \rightarrow jj$ and background with $p_T^j>3$ TeV along with cone size 1.0 at the 14 TeV and  27 TeV  run of LHC with $L=3000 ~{\rm fb^{-1}}$ are obtained after the optimization cut $6.7<M_{jj} <7.35$ TeV. }
\label{table-BGZJJ2}
\end{table}
%%%%%%%%%%%%%%%%%%%%%%%%%%%%%%%%%%%%%%%%%%%%%%%%%%%%%%%

We are now ready to discuss the various SUSY decay channels for the heavy gauge bosons and their implications in context of the LHC experiment. The presence of light SUSY particles in the spectrum allows the heavy gauge bosons to decay into these light states which can lead to interesting new channels. The branching fraction of the heavy gauge bosons decaying into light electroweakinos can be quite large depending upon the choice of LRSUSY parameters. The charginos and neutralinos produced from the decay of heavy gauge bosons can themselves undergo cascade decays giving rise to final state leptons and jets the LSP remaining undetected. Thus these signals are quite different compared to conventional search channels due to the presence of large missing energy in the final state. A large ensemble of final states can arise from the SUSY decays of the heavy gauge bosons. Here we will mainly focus on the leptonic cascade decay modes arising from the mono-$X$ + $\slashed{E}_T$ ($X = W, Z$) channels~\cite{Barman:2016kgt} in the context of the HL-LHC and HE-LHC experiments. The main motivation for choosing these channels are:
\begin{itemize}

\item These final states are well understood as they have already been experimentally studied in the context of dark matter searches by the ATLAS and CMS collaborations~\cite{ATLAS:2014pna,Aad:2014vka}.
 
 \item We also restrict ourselves to leptonic decay channels for the SM gauge bosons as these produce relatively clean channels which are easy to identify in a hadron-rich environment like the LHC experiment. 

\end{itemize} 
Fig.~\ref{fig:cascadeWZ} depicts a couple of examples where the heavy gauge boson SUSY decays can lead to final states with a SM gauge boson plus large $\slashed{E}_T$.

%%%%%%%%%%%%%%%%%%%%%%%%%%%%%%%%%%%%%%%%%%%%%%%%%%%%%%
%%%%%%%%%%%%%%%%%%%%%%%%%%%%%%%%%%%%%%%%%%%%%%%%%%%%%
\begin{figure}[h!]
\centering
\subfloat[]{\includegraphics[width=2.8in,height=1.5in, angle=0]{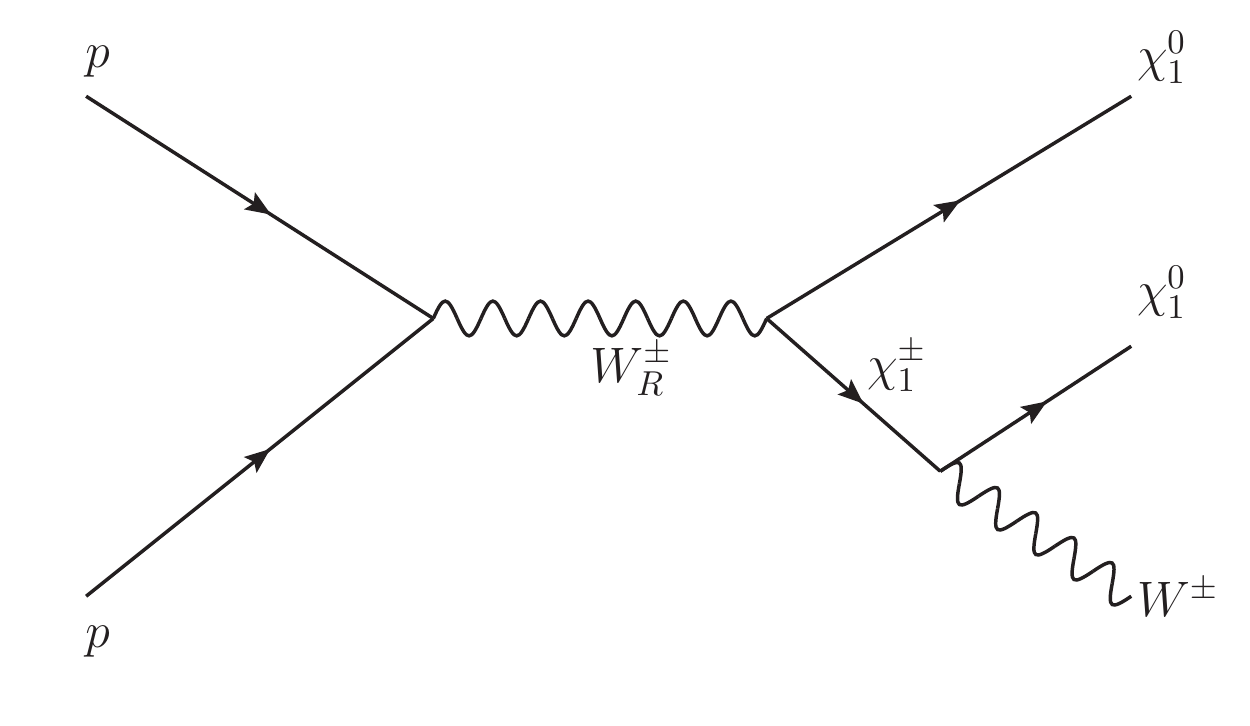}} 
\subfloat[]{\includegraphics[width=2.8in,height=1.5in, angle=0]{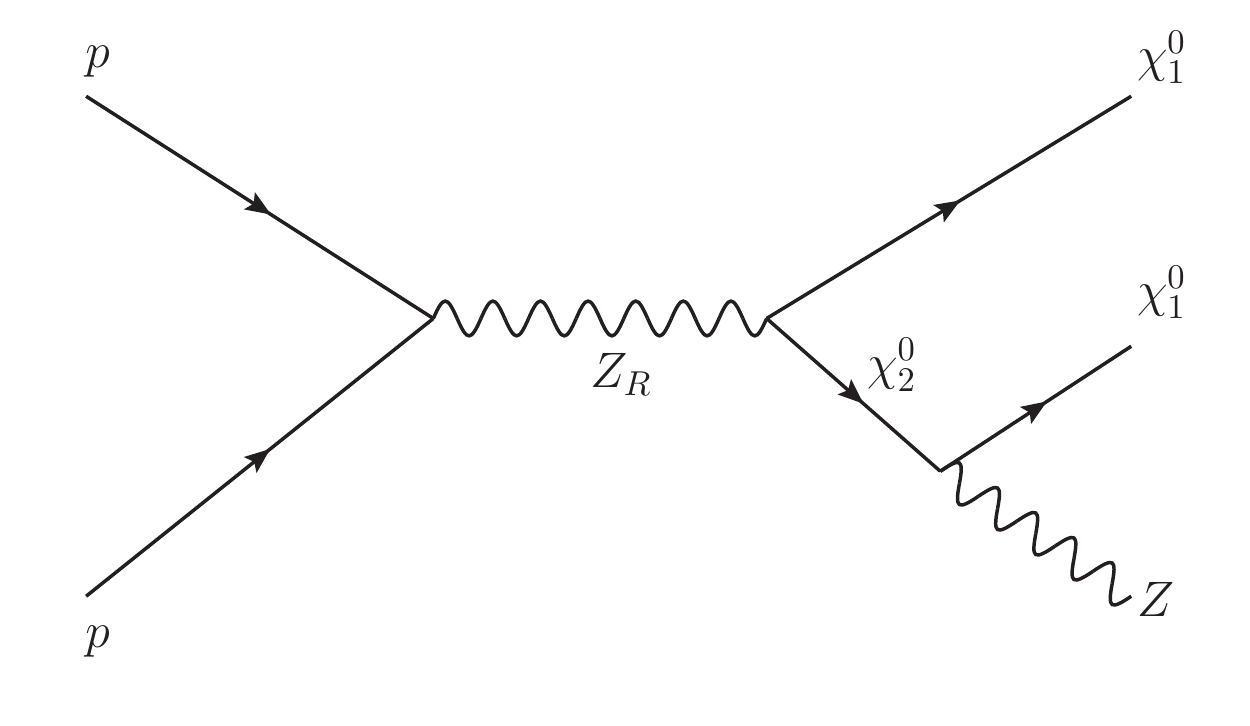}}
\caption{Mono-$X$ + $\slashed{E}_T$ ($X = W, Z$) channels through one-step cascade of the heavy gauge bosons.} 
\label{fig:cascadeWZ}
\end{figure} 
%%%%%%%%%%%%%%%%%%%%%%%%%%%%%%%%%%%%%%%%%%%%%%%%%%%%% 
%%%%%%%%%%%%%%%%%%%%%%%%%%%%%%%%%%%%%%%%%%%%%%%%%%%%%
%%%%%%%%%%%%%%%%%%%%%%%%%%%%%%%%%%%%%%%%%%%%%%%%%%%%%
%%%%%%%%%%%%%%%%%%%%%%%%%%%%%%%%%%%%%%

One can also encounter relatively complex cascade decays, with multiple decay chains in between, which may eventually lead to final states with multiple leptons and jets along with large missing transverse energy. As an example let us consider the following decay chains
$W_R^\pm \rightarrow   \chi^\mp_1   \chi^{\pm\pm} ~\text{or}~Z_R \rightarrow \chi^{\mp\mp} \chi^{\pm\pm},
 \text{with}~\chi^{\pm\pm} \rightarrow   { \chi^0_1 }    H^{\pm\pm}_1~\text{or}~\chi^{\pm\pm} \rightarrow   \chi^\pm_1   W^{\pm},
 \text{and finally}~H^{\pm\pm}_1 \rightarrow l^\pm l^\pm ~\text{while}~ \chi^\pm_1 \rightarrow W^\pm   { \chi^0_1 }   ~\text{with}~W^{\pm} \rightarrow l^\pm \nu_l / q q'$. 
The detailed collider analysis of these channels are beyond the scope of the current paper and will not be discussed here.

For our collider analysis we chose two benchmark points as was discussed before. The first benchmark points, BP1 is optimized for the mono-$W$ + $\slashed{E}_T$ searches, where the lightest neutralino and
the lightest chargino are $SU(2)_R$ wino dominated. In BP1,
BR($\chi^\pm_1 \rightarrow \chi^0_1 W^\pm $) = 100\% and the total
 contribution coming from $W^\pm_R$ to $\chi^\pm_1 \chi^0_1$ are about 29\%.
Our second benchmark point, BP2, is more suitable for the mono-$Z$ + $\slashed{E}_T$ searches 
where both $\chi^0_1$ and $\chi^0_2$ are mixture of substantial higgsinos components $\widetilde{\delta}^{c^{0}}$ and $\widetilde{\bar{\delta}}^{c^{0}}$ of the triplet Higgs bosons while the lightest chargino $\chi^\pm_1$ is higgsino (triplet) dominated.
It is because of the BR($W_R^\pm \rightarrow   { \chi^0_1 }      \chi^\pm_1  $) is only around 12\% while BR($Z_R \rightarrow   { \chi^0_1 }    \chi^0_2$)=34\% \footnote{The large coupling of $Z_R$ with the neutralinos are due to them being triplets of $SU(2)_R$ in this case, as compared to the leptons being doublets.} and BR($\chi^0_2 \rightarrow Z +   { \chi^0_1 }$)=98\%.

\subsection{Mono-$X$ plus missing transverse energy}

Events with a single $W/Z$ boson accompanied by large missing transverse energy constitute a very clean and distinctive signature in new physics searches at the LHC. This topology has been thoroughly analyzed by both the ATLAS and CMS collaborations~\cite{Aad:2014vka,ATLAS:2014pna}, mainly in the context of DM searches. In this work, we follow these search channels to probe the heavy gauge bosons for the chosen benchmark points. We present these searches for the future collider perspective, assuming the LHC will operate at the com energies of $\sqrt{s} = 14$ and $27$ TeV  with an integrated luminosity of $3000~ {\rm fb^{-1}}$.

\subsubsection{mono-$W$ + $\slashed{E}_T$ searches}
We perform a search for the heavy charged gauge bosons $W_R$ in events where a $W$ boson is produced through one-step cascade decay (see the Fig.~\ref{fig:cascadeWZ}(a)) of the chargino $\chi^\pm_1$.
Here we only consider the leptonic decay channel of the $W$ boson ($W\rightarrow l \nu$, $l=e,\mu$). 

Signal event would be characterized by the presence of a high $p_T$ lepton (electron and muon) and a large $\slashed{E}_T$ imbalance due to the undetected escaping neutrino and lightest neutralinos.
The search strategy reported in Ref.~\cite{ATLAS:2014pna}, which focused on the DM
searches, has been followed with suitable modifications
aimed to optimize the signal significance. 
The event is selected with one isolated lepton (electron or
muon) which having transverse momentum $p_T$ larger than 400 GeV. The lepton isolation criteria is same as in the previous case. Also the candidate electron(muon) is required to satisfy the rapidity cut $|\eta| < 2.5$.
The main discriminating variable used in this search is the transverse mass defined as $M_T=\sqrt{ 2 p^l_T \slashed{E}_T (1-\cos\Delta\Phi_{l,\slashed{E}_T})}$, where $p^l_T$ is the transverse momentum of the charged lepton and $\Delta\Phi_{l,\slashed{E}_T}$ is the difference in azimuthal angle between the lepton transverse
momentum and missing transverse energy $\slashed{E}_T$.

One of the main sources of SM background is $pp\rightarrow l \nu+ jj$ production channel.
Besides, processes like $VV$ ($V=W,Z$), $t\overline{t}$, etc., also contribute to the
background. Among the $VV$ processes, the contribution comes from  $ZZ \rightarrow l l \nu \nu$,  $WZ \rightarrow l \nu \nu \nu$  and $WW \rightarrow l \nu l \nu$  channels where the additional charged leptons get misidentified or remain unreconstructed. 
%%%%%%%%%%%%%%%%%%%%%%%%%%%%%%%%%%%%%%%%%%%%%%%%%%%%%
%%%%%%%%%%%%%%%%%%%%%%%%%%%%%%%%%%%%%%%%%%%%%%%%%%%%%%
%%%%%%%%%%%%%%%%%%%%%%%%%%%%%%%%%%%%%%%%%%%%%%%%%%%%%
\begin{figure}[h!]
\centering
\subfloat[]{\includegraphics[width=3.0in,height=2.8in, angle=0]{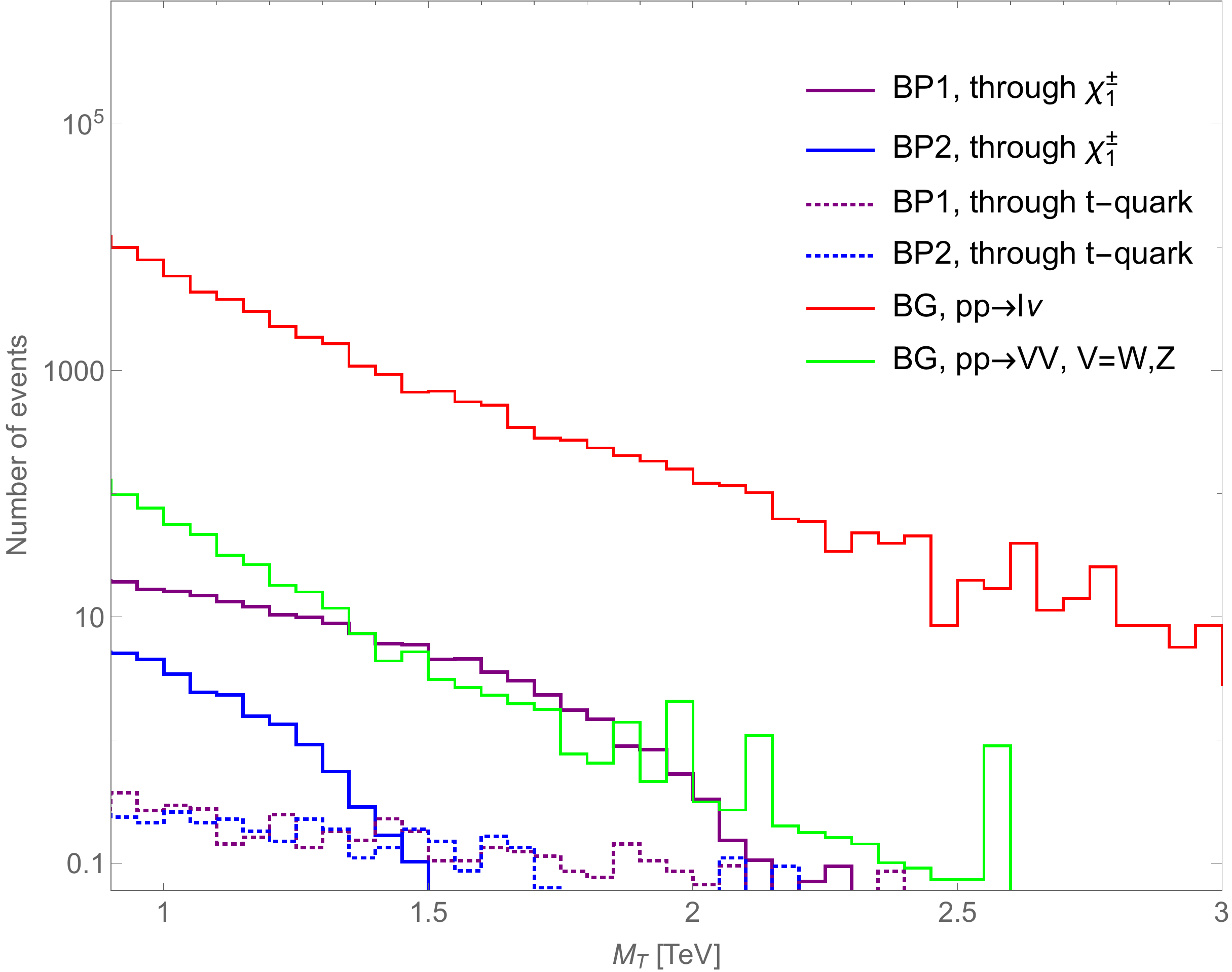}} 
\subfloat[]{\includegraphics[width=3.0in,height=2.8in, angle=0]{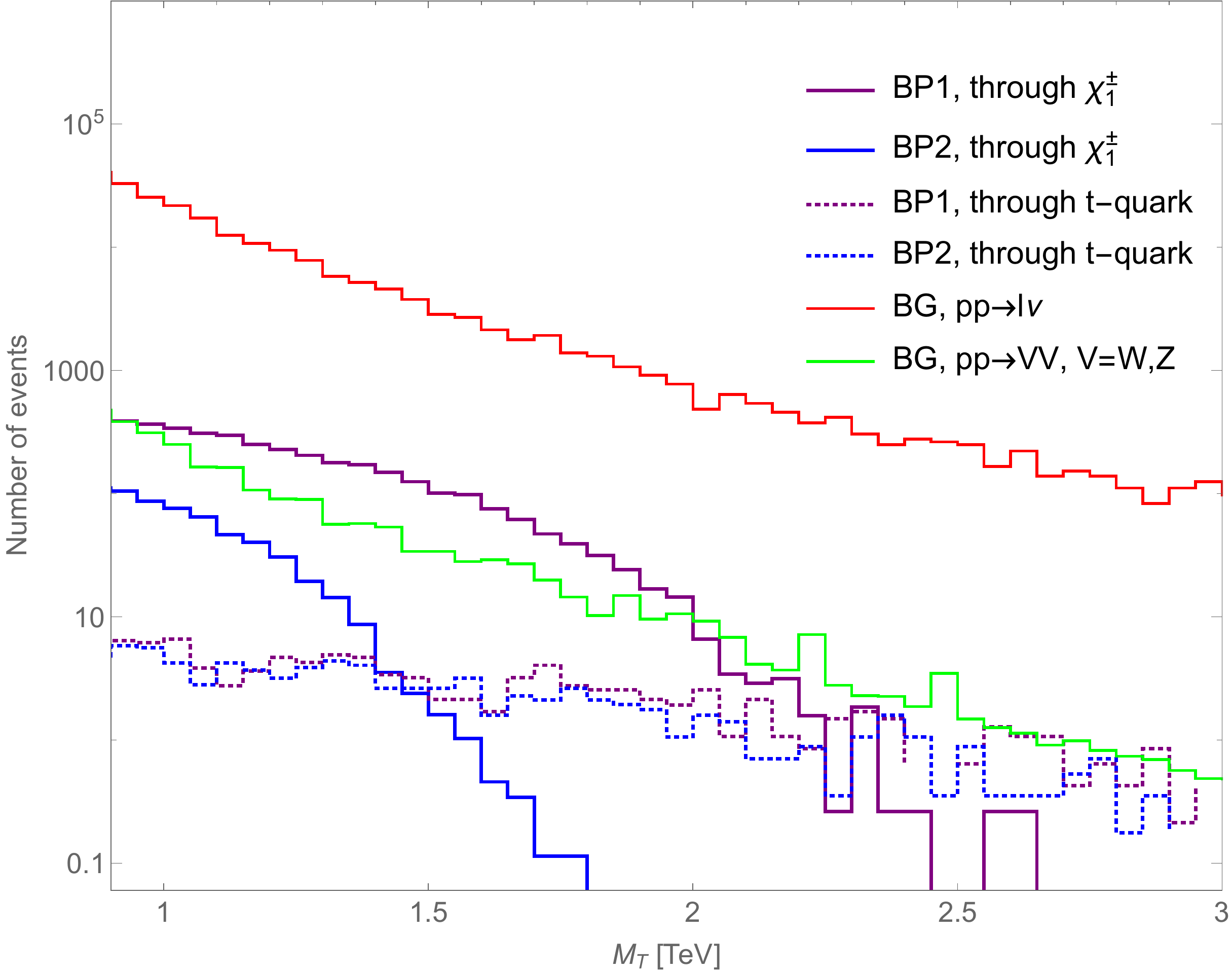}}
\caption{
The event selection criteria of the signal and background requires at least one lepton ($e$ or $\mu$) in the event, with each lepton having $p_T^l > 400$ GeV and $|\eta|<2.5$. The total number of the signal of events for the process  $pp \rightarrow 1l + \slashed{E}_T$ ($l=e,\mu$)  through the cascade (indicate by the purple and blue lines) of the $W_R$ gauge boson in the context of the LHC at (a) $14$ TeV and (b)  $27$ TeV with integrated luminosity $3000~{\rm fb^{-1}}$. The other purple and blue dotted lines stand for the additional signal events through another cascade decay process $pp\rightarrow W_R \rightarrow t \ov b, t \rightarrow W b, W\rightarrow l \nu$. The number of events with $1l + \slashed{E}_T$ final state in this channel remain less than unity as the lepton isolation cut significantly wane the signal. The dominant SM $pp\rightarrow l\nu jj$ background indicates by the red line whereas the green line stands for SM $pp\rightarrow VV (V=W,Z)$. The contribution comes from  $ZZ \rightarrow l l \nu \nu$,  $WZ \rightarrow l \nu \nu \nu$  and $WW \rightarrow l \nu l \nu$  channels where the additional charged leptons get misidentified or remain unreconstructed.} 
\label{fig:1ev} 
\end{figure} 
%%%%%%%%%%%%%%%%%%%%%%%%%%%%%%%%%%%%%%%%%%%%%%%%%%%%% 
%%%%%%%%%%%%%%%%%%%%%%%%%%%%%%%%%%%%%%%%%%%%%%%%%%%%%
%%%%%%%%%%%%%%%%%%%%%%%%%%%%%%%%%%%%%%%%%%%%%%%%%%%%%
%%%%%%%%%%%%%%%%%%%%%%%%%%%%%%%%%%%%%%

%%%%%%%%%%%%%%%%%%%%%%%%%%%%%%%%%%%%%%%%%%%%%%%%%%%%%
%%%%%%%%%%%%%%%%%%%%%%%%%%%%%%%%%%%%%%%%%%%%%%%%%%%%%%
%%%%%%%%%%%%%%%%%%%%%%%%%%%%%%%%%%%%%%%%%%%%%%%%%%%%%
\begin{figure}[h!]
\centering
\subfloat[]{\includegraphics[width=3.0in,height=2.8in, angle=0]{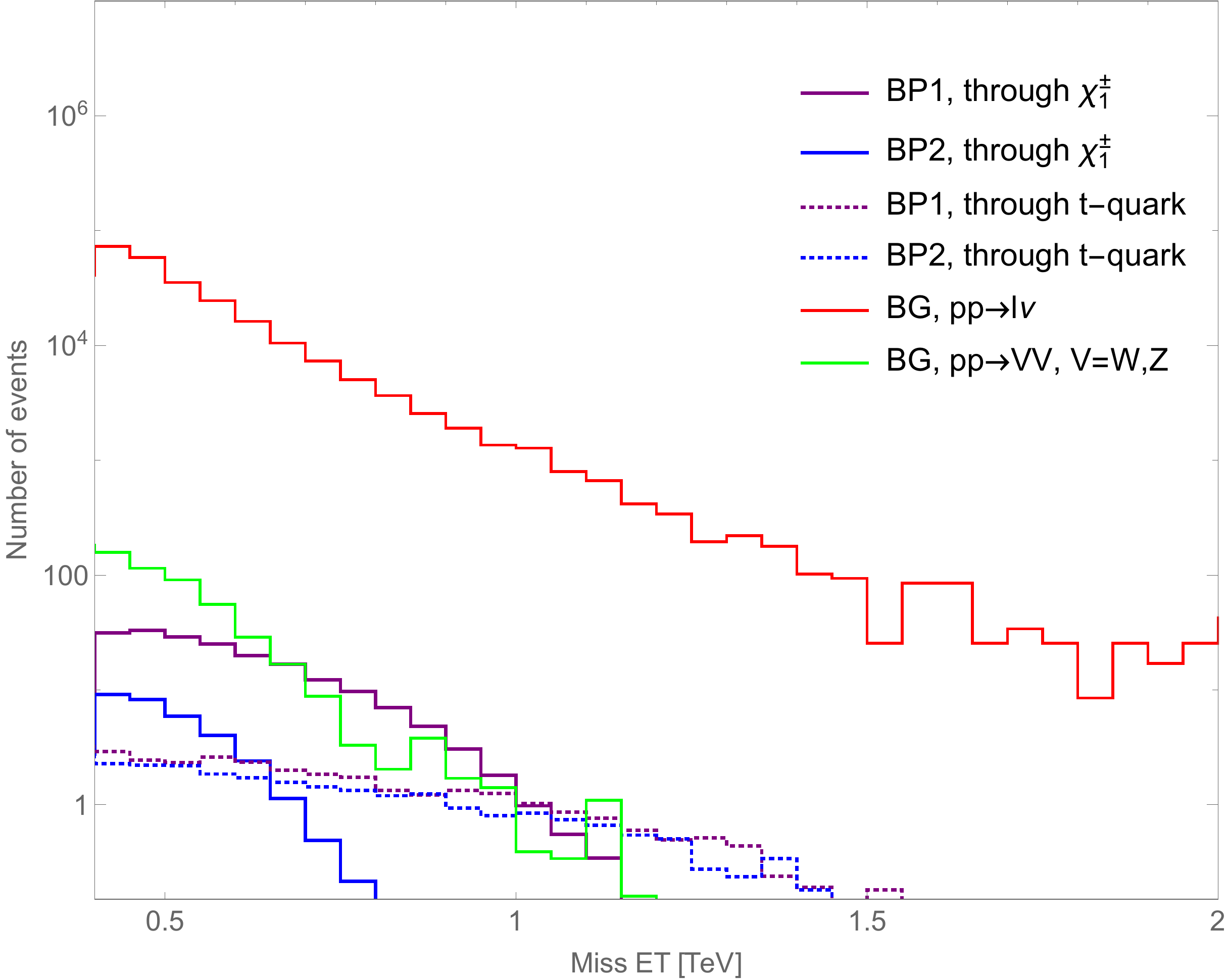}} 
\subfloat[]{\includegraphics[width=3.0in,height=2.8in, angle=0]{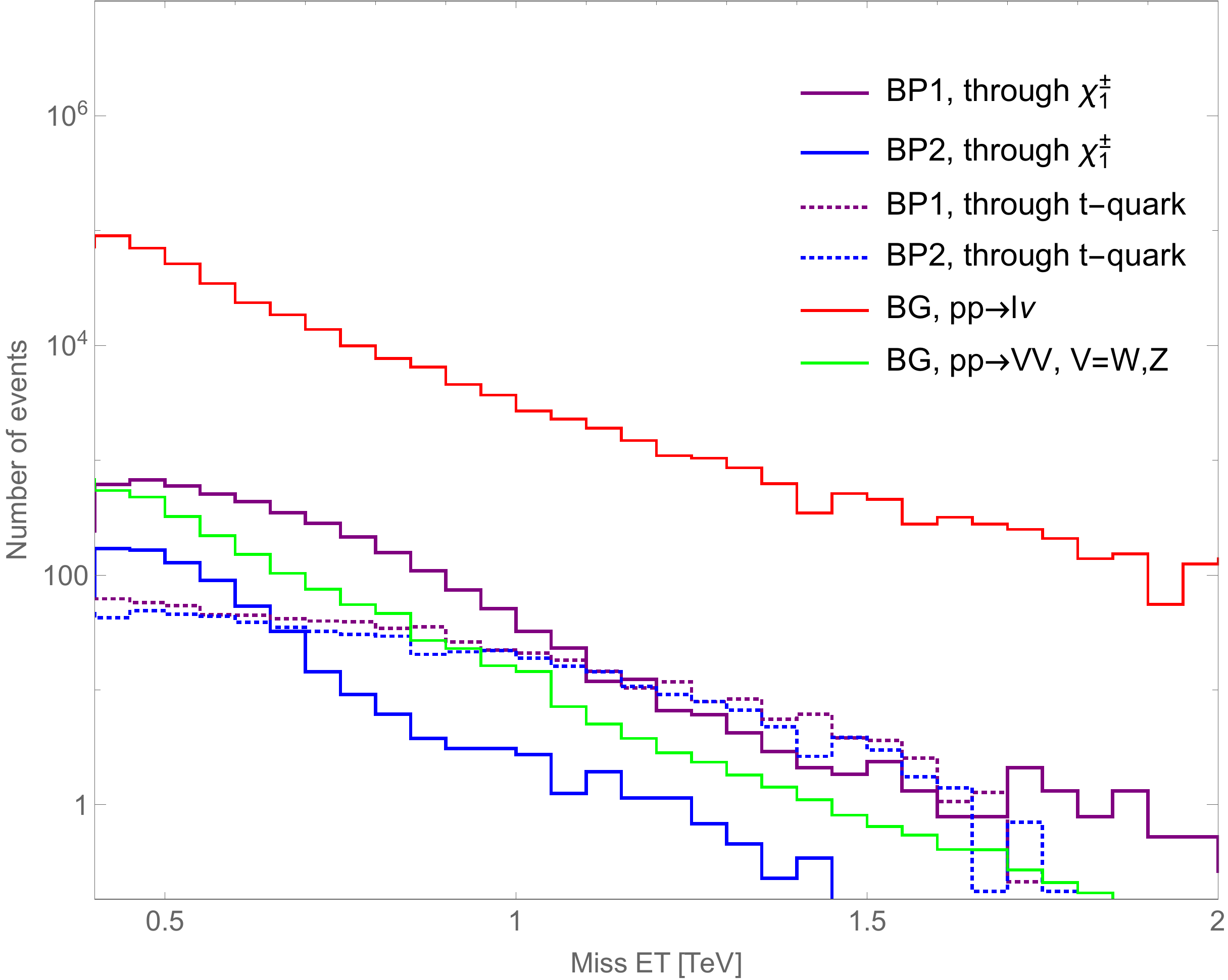}}
\caption{The transverse missing energy distribution for the signal  $pp \rightarrow 1l + \slashed{E}_T$ ($l=e,\mu$) and background events at (a) $14$ TeV and (b)  $27$ TeV with integrated luminosity $3000~{\rm fb^{-1}}$. The selection criteria of the events and color coding for the shown lines is the same as in Fig.~\ref{fig:1ev}.} 
\label{fig:1ev2} 
\end{figure} 
%%%%%%%%%%%%%%%%%%%%%%%%%%%%%%%%%%%%%%%%%%%%%%%%%%%%% 
%%%%%%%%%%%%%%%%%%%%%%%%%%%%%%%%%%%%%%%%%%%%%%%%%%%%%
%%%%%%%%%%%%%%%%%%%%%%%%%%%%%%%%%%%%%%%%%%%%%%%%%%%%%
%%%%%%%%%%%%%%%%%%%%%%%%%%%%%%%%%%%%%%

The transverse mass and missing energy distribution for the number of signal $S$ as well as  background $B$ events at $\sqrt{s}=14$ TeV (left) and  27 TeV (right) with integrated luminosity $L=3000 ~{\rm fb^{-1}}$ are shown in Figs.~\ref{fig:1ev} and ~\ref{fig:1ev2} respectively. The purple and blue solid lines indicate the signal events for BP1 and BP2 respectively.
It is to be noted that the direct decay of the heavy charged boson also can give final state with one lepton plus missing energy $W_R^\pm \rightarrow l \nu $ but this channel is absent for the chosen parameter spaces in this model. Hence, we could not see any peak in the $M_T$ distribution at $M_{W_R}\sim 4.5$ TeV. 
This is a major difference of this model compared to the other models.
In both these figures, the dominant SM $pp\rightarrow l\nu jj$ background is represented by the red line whereas the green line is for the SM $pp\rightarrow VV (V=W,Z)$ background. The $pp\rightarrow t\overline{t}$ remain almost zero for the chosen signal regions (see the Tab.~\ref{table-BGS1}) hence it is not demonstrated in these figures.
The number of events are calculated as $S,B= \epsilon A \sigma L$, where $\epsilon, A$ stand for the efficiency and acceptance for the signal or background events, $\sigma$ indicates the cross-section of the final state events and luminosity is denoted by $L$.

Several combinations of selection variables $M_T$ and $\slashed{E}_T$ are investigated. Among them,
a signal region with $0.6<\slashed{E}_T <0.9$ TeV and $1.2<\slashed{E}_T <1.6$ TeV is chosen which
yields the most efficient optimization of signal significance.
The expected number of signal and background events in this region are shown in Tab.~\ref{table-BGS1}. 
%%%%%%%%%%%%%%%%%%%%%%%%%%%%%%%%%%%%%%%%%%%%%
\begin{table}[h!]
\begin{center}\scalebox{0.7}{
\begin{tabular}{|c|c|c|c|c|c|c|c|c|c|c|c|}
\hline 
&\multicolumn{9}{c|}{$W+ \slashed{E}_T,$ $W\rightarrow l \nu_l$ final state through charginos $\chi^\pm_1$}\\
\cline{2-10}
Energy & \multicolumn{5}{c|}{ SM backgrounds}&\multicolumn{2}{c|}{Total signal events}&\multicolumn{2}{c|}{Significance}\\
\cline{2-10}
[TeV]&~~ $W\rightarrow l \nu_l$~~& ~~~~$ZZ$~~~& ~~~$WZ$~~~& ~~~$WW$~~~ &~~~ $t\overline{t}$&BP1&BP2&BP1&BP2\\
\cline{1-10}
&&&&&&&&&\\
14&6929.3&10.5&3.7&15.7&0&45.37&0.62&0.54&--\\
&&&&&&&&&\\
27&$2.5\times10^4$&28.9&28.9&28.7&0&945.2&56.5& 5.83&0.35\\
&&&&&&&&&\\
\hline
\end{tabular}}
\end{center}
\caption{ The event selection criteria of the signal and background requires at least one lepton ($\e$ or $\mu$) in the event, with each lepton having $p_T^l > 400$ GeV and $|\eta|<2.5$. Total number of events of the signal and background at the 14 and 27 TeV  run at LHC with $L=3000 ~{\rm fb^{-1}}$ are obtained after the optimization cuts $0.6<\slashed{E}_T <0.9$ TeV and $1.2<\slashed{E}_T <1.6$ TeV.  The jet faking lepton $pp\rightarrow jj$ channels also add to the background. It is found to be $22.5$ and $936.2$ at 14 and 27 TeV respectively for the similar optimization cuts $0.6<\slashed{E}_T <0.9$ TeV and $1.2<M_T^{j,\slashed{E}_T}<1.6$ TeV. }
\label{table-BGS1}
\end{table}
%%%%%%%%%%%%%%%%%%%%%%%%%%%%%%%%%%%%%%%%%%%%%%%%%%%%%%%%
The expected number of the signal events for BP1 at the 14 TeV run of the LHC with luminosity $L=3000 ~{\rm fb^{-1}}$ is 45.37. 
This number remain less than unity for the choice BP2. 
The LHC run at 27 TeV with $L=3000 ~{\rm fb^{-1}}$ gives the large number of signal events, 945.2 for BP1 whereas 56.2 for BP2 due to large production cross-section.
The SM $pp\rightarrow l\nu jj$ process background remain dominant at 14 TeV as well as 27 TeV even after the optimization cuts. The $pp\rightarrow VV$ background is negligibly smaller compared to the $pp\rightarrow l\nu jj$ process.
It is also to be noted that the jet faking lepton (assuming $0.1\%$ to electron and $0.5\%$ to muon~\cite{Khachatryan:2014fba}) $pp\rightarrow jj$ channels also give additional contribution to the background. It is found to be $22.5$ and $936.2$ events at 14 and 27 TeV respectively for similar optimization cuts with $0.6<\slashed{E}_T <0.9$ TeV and $1.2<M_T^{j,\slashed{E}_T}<1.6$ TeV.
Signal significance ($\frac{S}{\sqrt{B}},\, S,B>3$) attain a values of 0.54 for BP1 at 14 TeV. Whereas the significance become 5.83 for the LHC run at 27 TeV. The expected number or the significance for BP2 at 14 TeV as well as 27 TeV remain less than unity. 
The signal events and corresponding significance for BP1 indicate that the heavy gauge bosons could be discovered at HE-HL-LHC. Or one can exclude these parameter space if we won't get any signal at future collider. The BP2 demands that one need more high-energetic collider with high integrated luminosity to discover or exclude such region. Now in the following discussion, we will see the features of these BPs in the context of mono-$Z$ plus $\slashed{E}_T$ searches.

In this analysis, only the $W\rightarrow l \nu$, ($l=e,\mu$) decay modes have been considered. However the $W\rightarrow \tau \nu_\tau$ can also enhance the signal as tau can give one lepton (electron or muon) in the final state through its decay, e.g., $\tau \rightarrow W \nu_\tau, W\rightarrow l \nu$. It could be quantified as follows. Using the same selection and optimization cuts $p_T^l>400$ GeV, $0.6<\slashed{E}_T <0.9$ TeV and $1.2<M_T^{j,\slashed{E}_T}<1.6$ TeV, it is found that the number of events at 14 TeV run of the LHC with luminosity $L=3000 ~{\rm fb^{-1}}$ is enhanced by 5.01 (0.28) for BP1 (BP2). The significance now becomes 0.60 for the BP1. Whereas the number of events for these channels go to 106.32 (6.52) for BP1 (BP2) at 27 TeV LHC and the significance corresponding BP1 and BP2 attain values of 6.48 and 0.37 respectively.

The $l \nu $ final state can also come through another cascade decay process with $pp\rightarrow W_R \rightarrow t \ov b, t \rightarrow W b, W\rightarrow l \nu$. It can be seen from the BPs that the branching of the heavy charged gauge boson into the top-bottom quark final state is quite large. Also these particles are emitted almost back-to-back and remain boosted with $p_T^{t,\ov b}\sim 2$ TeV as $M_{W_R}=4.5$ TeV. Hence the separation $\Delta R_{lb}$ between the final state bottom quark and lepton (both coming from decay of the top quark) become very small. It can also be understood form the parton level distributions shown in the Fig.~\ref{fig:1evP}. The $\Delta R_{lb}$ distribution of the parton level bottom quark and lepton is demonstrated in the Fig .~\ref{fig:1evP}(a) while the correlation plot between final state lepton transverse momentum against $\Delta R_{lb}$ is shown in the Fig .~\ref{fig:1evP}(b). %From these plots, one can conclude that the final state lepton in these cascade decay can acquire large transverse momentum.
%%%%%%%%%%%%%%%%%%%%%%%%%%%%%%%%%%%%%%%%%%%%%%%%%%%%%
%%%%%%%%%%%%%%%%%%%%%%%%%%%%%%%%%%%%%%%%%%%%%%%%%%%%%%
%%%%%%%%%%%%%%%%%%%%%%%%%%%%%%%%%%%%%%%%%%%%%%%%%%%%%
\begin{figure}[h!]
\centering
\subfloat[]{\includegraphics[width=3.0in,height=2.8in, angle=0]{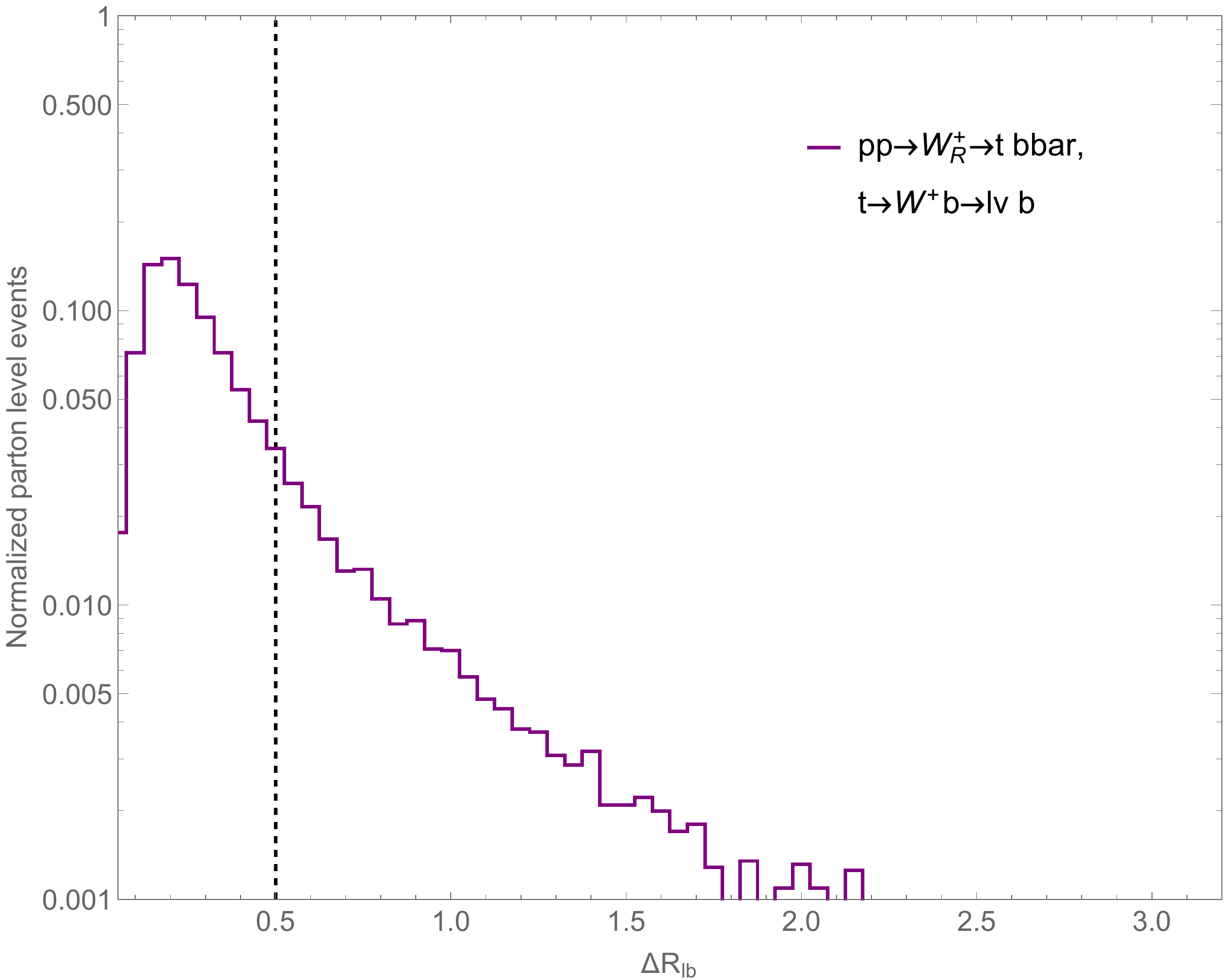}} 
\subfloat[]{\includegraphics[width=3.0in,height=2.8in, angle=0]{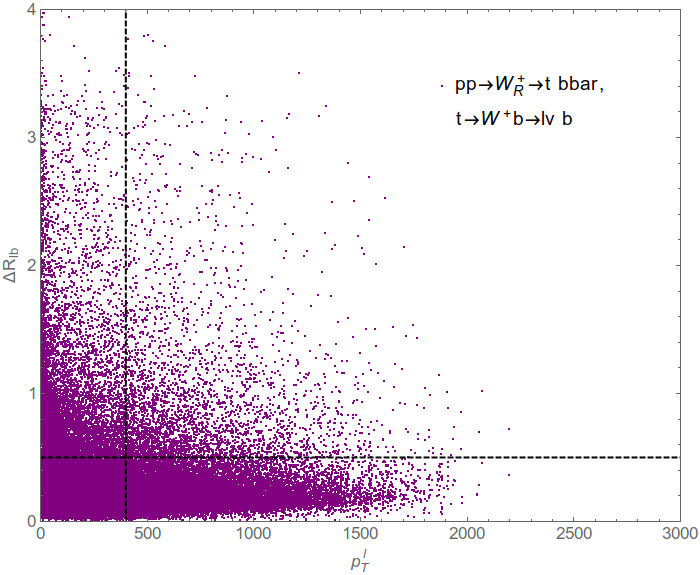}}
\caption{(a) The $\Delta R_{lb}$ parton level distribution for the signal $pp\rightarrow W_R^+ \rightarrow t \ov b, t \rightarrow W b, W\rightarrow l \nu$ events in the parton level at $\sqrt{s}=14$ TeV.} 
\label{fig:1evP} 
\end{figure} 
%%%%%%%%%%%%%%%%%%%%%%%%%%%%%%%%%%%%%%%%%%%%%%%%%%%%% 
%%%%%%%%%%%%%%%%%%%%%%%%%%%%%%%%%%%%%%%%%%%%%%%%%%%%%
%%%%%%%%%%%%%%%%%%%%%%%%%%%%%%%%%%%%%%%%%%%%%%%%%%%%%
%%%%%%%%%%%%%%%%%%%%%%%%%%%%%%%%%%%%%%
As the charged lepton isolation demands there is no other charged particle with $p_T > 0.5$ GeV within a cone of
$\Delta R < 0.5$, the number of lepton events in the final state passing this criterion is extremely small. Furthermore, the selection cut of $p_T^l>400$ GeV also significantly decrease the events at the analysis level. 
The isolation and selection criteria decrease the final state lepton events by $\sim 87\%$. The remaining number of lepton events can be identified as the purple points on the upper and right side of the black dashed lines in the Fig .~\ref{fig:1evP}(b).  
The corresponding transverse mass and missing energy distribution of these events for the BP1 (dotted purple line) and BP2 (dotted blue line) are shown in the Figs.~\ref{fig:1ev} and \ref{fig:1ev2} respectively.
These cascade decay process $pp\rightarrow W_R^+ \rightarrow t \ov b, t \rightarrow W b, W\rightarrow l \nu$ also add to these signal events of the $pp\rightarrow W_R^+ \rightarrow \chi^+_1 \chi_1^0, \chi^+_1 \rightarrow W \chi_1^0, W\rightarrow l \nu$, though the enhancement in the signals remain negligible. It is also to be noted that the b-jet veto will completely reduce these events at the analysis level.

\subsubsection{mono-$Z$ + $\slashed{E}_T$ searches}
We perform a search for the heavy neutral gauge boson $Z_R$ in events where a $Z$  boson is produced through one-step cascade decays (see the Fig.~\ref{fig:cascadeWZ}(b)) of the neutralino $\chi^0_2$.
Here we assume that the $Z$ boson decays leptonically ($Z\rightarrow ll$, $l=e,\mu$). These events also contain significant missing transverse
energy coming from the LSP $\chi^0_1$.

The events are selected with two same flavor opposite sign (SFOS) isolated electrons or muons with transverse momentum $p_T$ larger than 30 GeV satisfying $|M_{ll}- M_Z| < 15$ GeV. Here $M_{ll}$ stands for the invariant mass of the SFOS leptons pair and $M_Z=91.1876$ GeV is the SM $Z$ boson mass. The isolation criteria are the same as in the previous cases.
The charged lepton candidates  are required to be within pseudorapidity range $|\eta| < 2.5$~\cite{Aad:2014vka}. The transverse missing energy distribution $\slashed{E}_T$ can be a
useful probe to search for the $Z_R$ gauge boson.
Here processes like $pp\rightarrow ZZ ~(Z \rightarrow ll, Z\rightarrow \nu \overline{\nu})$, $pp\rightarrow ZW~(Z \rightarrow ll, W\rightarrow l \nu )$ and $pp \rightarrow WW~ (W \rightarrow l \nu)$ can add to the SM background if additional charged leptons get misidentified or remain unreconstructed. Also other reducible backgrounds like $pp\rightarrow t\overline{t}, t \rightarrow W b, W \rightarrow l \nu$ may also produce two leptons and jets in the final state.
%%%%%%%%%%%%%%%%%%%%%%%%%%%%%%%%%%%%%%%%%%%%%%%%%%%%%
%%%%%%%%%%%%%%%%%%%%%%%%%%%%%%%%%%%%%%%%%%%%%%%%%%%%%%
%%%%%%%%%%%%%%%%%%%%%%%%%%%%%%%%%%%%%%%%%%%%%%%%%%%%%
\begin{figure}[h!]
\centering
\subfloat[]{\includegraphics[width=3.0in,height=2.8in, angle=0]{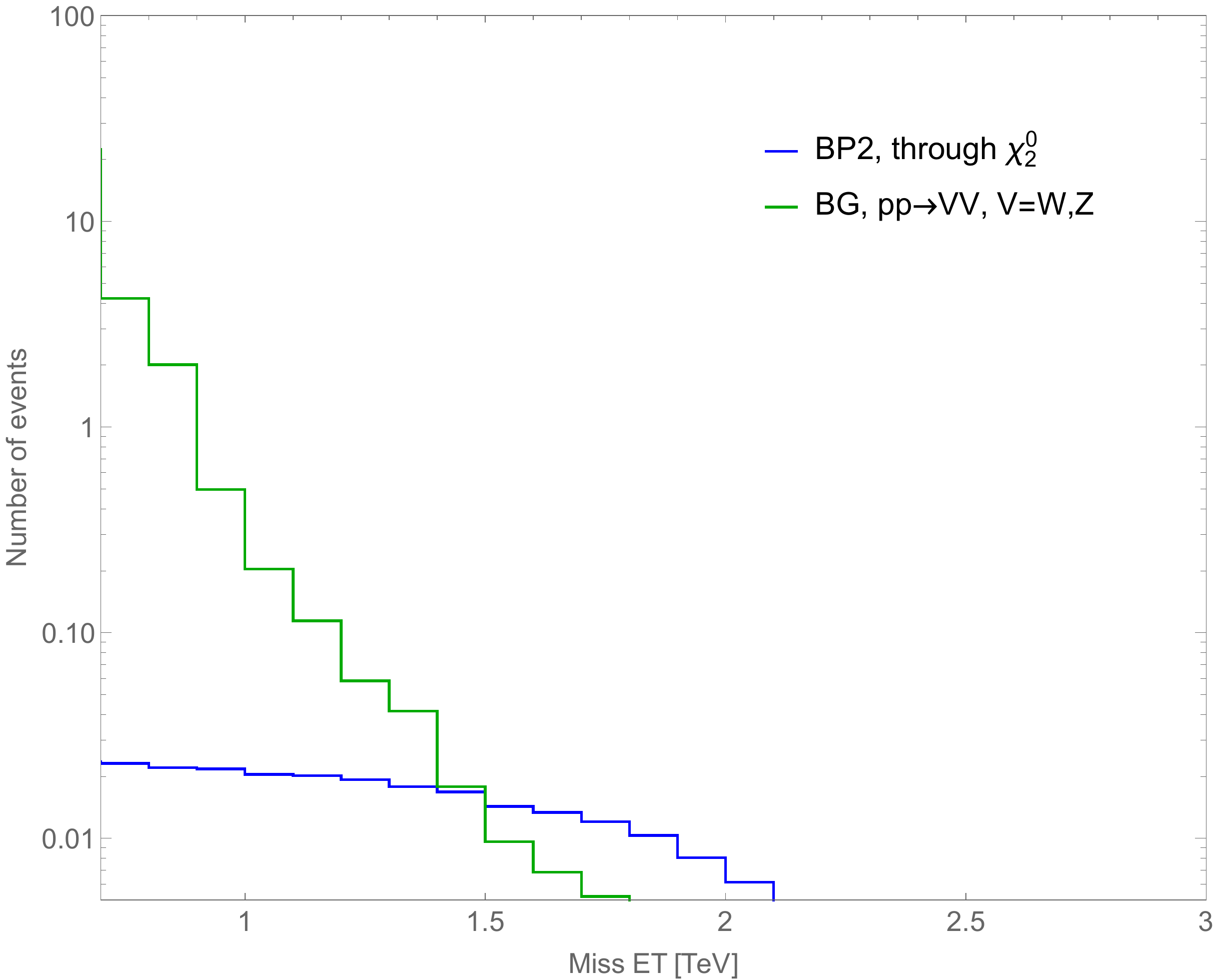}} 
\subfloat[]{\includegraphics[width=3.0in,height=2.8in, angle=0]{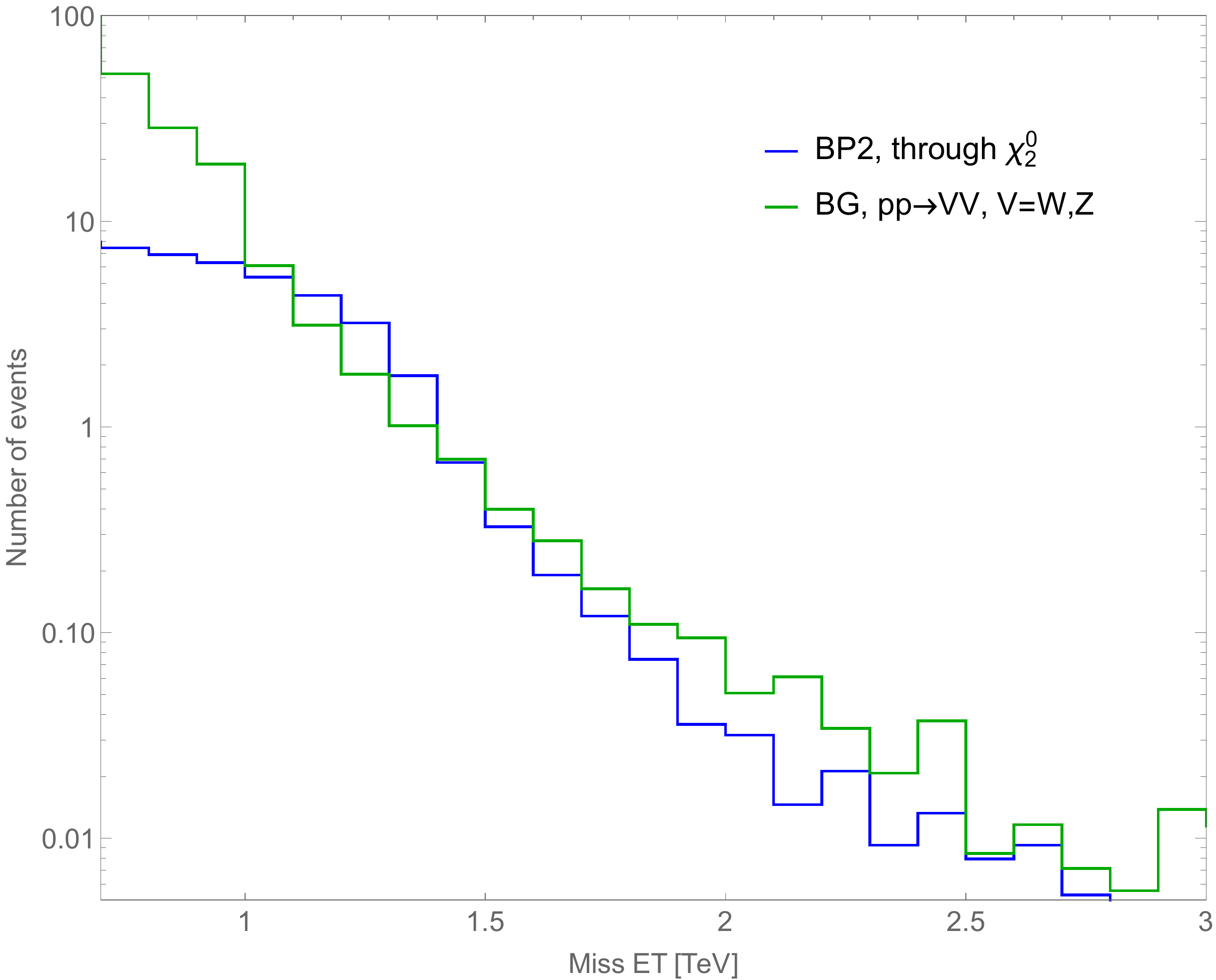}}
\caption{The event selection criteria of the signal and background requires at least two leptons ($\e$ or $\mu$) in the event, with each lepton having $p_T^l > 30$ GeV and $|\eta|<2.5$. The total number of events for the signal process $pp \rightarrow 2l$ ($l=e,\mu$) through the cascade and direct decay (indicate by the purple and blue lines) of the $Z_R$ gauge boson in the context of the LHC at (a) $14$ TeV and (b)  $27$ TeV with integrated luminosity $3000~{\rm fb^{-1}}$. The dominant SM $pp\rightarrow VV,V=W,Z$ backgrounds indicate by the green line.} 
\label{fig:2ev} 
\end{figure} 
%%%%%%%%%%%%%%%%%%%%%%%%%%%%%%%%%%%%%%%%%%%%%%%%%%%%% 
%%%%%%%%%%%%%%%%%%%%%%%%%%%%%%%%%%%%%%%%%%%%%%%%%%%%%
%%%%%%%%%%%%%%%%%%%%%%%%%%%%%%%%%%%%%%%%%%%%%%%%%%%%%
%%%%%%%%%%%%%%%%%%%%%%%%%%%%%%%%%%%%%%
The missing energy distribution for the signal as well as  background events at a center-of-mass energy of $\sqrt{s}=14$ TeV  and  $\sqrt{s}=27$ TeV (right)  with luminosity $L=3000 ~{\rm fb^{-1}}$ are shown in Fig.~\ref{fig:2ev}. Please note that the signal events for BP1 is zero as BR($Z_R \rightarrow \chi^0_1 \chi^0_2$)=0 in this case. The blue line indicates the signal events distribution for the choice of BP2. The other lines (cyan, green) stand for the SM backgrounds.
Also in this case, a signal region ${\slashed{E}_T}>0.95$ TeV is used to optimize the significance. The expected signal and background events are shown in the Tab.~\ref{table-BGS2}.
%%%%%%%%%%%%%%%%%%%%%%%%%%%%%%%%%%%%%%%%%%%%%
\begin{table}[h!]
\begin{center}\scalebox{0.7}{
\begin{tabular}{|c|c|c|c|c|c|c|c|c|c|c|c|c|}
\hline
&\multicolumn{8}{c|}{$Z+ \slashed{E}_T,$ $Z\rightarrow l l$ final state through $\chi^0_2$}\\
\cline{2-9}
Energy & \multicolumn{4}{c|}{ SM backgrounds }&\multicolumn{2}{c|}{Total signal events}&\multicolumn{2}{c|}{Significance}\\
\cline{2-9}
[TeV]& ~~$ZZ$~~& ~~$WZ$~~~& ~~~$WW$~~~ & ~~~$t\overline{t}$~~~&BP1&BP2&BP1&BP2\\
\cline{1-9}
&&&&&&&&\\
14&0.02&0.06&0&0&--&0.17&--&--\\
&&&&&&&&\\
27&3.46&3.79&0&0&--&18.27&--&6.08\\
&&&&&&&&\\
\hline
\end{tabular}}
\end{center}
\caption{The event selection criteria of the signal and background requires at least two leptons ($\e$ or $\mu$) in the event, with each lepton having $p_T^l > 30$ GeV and $|\eta|<2.5$. Total number of events of the signal and background at the 14 TeV and  27 TeV  run of LHC with $L=3000 ~{\rm fb^{-1}}$ are obtained after the optimization cut ${\slashed{E}_T}>950$. The $pp\rightarrow jj$ (where jets fake as leptons) background become negligibly small for the optimization cut ${\slashed{E}_T}>950$ along with $|M_{jj}-M_Z|< 15$ GeV. The signal events always remain $\sim zero$ for BP1 as BR$(Z_R \rightarrow \chi^0_1 \chi^0_2)\sim 0$.}
\label{table-BGS2}
\end{table}
%%%%%%%%%%%%%%%%%%%%%%%%%%%%%%%%%%%%%%%%%%%%%%%%%%%%%%%  
We find that all these SM processes including $pp\rightarrow jj$ (where jets fake as leptons) backgrounds remain small for LHC run at 14 TeV for these signal region. Also the number of signal events remain small due to the small production cross-section $\sigma(pp\rightarrow Z_R)=0.001$ fb at this energy.
However, the expected number of signal events for BP2 become quite large $18.27$ for LHC run at 27 TeV with integrated $L=3000 ~{\rm fb^{-1}}$. The corresponding dominant background $pp\rightarrow VV$ goes to 7.25. The significance attains a values of 6.08.
It is clear from this section that the expected numbers for BP1 can give an indication of heavy charged boson whereas BP2 can give hints for the heavy neutral gauge boson. Hence these cascade decay channels may lead to discovery or exclusion of these heavy gauge bosons through mono-$X$ ($X=W,Z$) plus $\slashed{E}_T$ final state searches in the context of HL-LHC and HE-LHC.

Similar to the previous cases,  only the $Z\rightarrow l l$, ($l=e,\mu$) decay modes have been considered. However the $Z\rightarrow \tau \tau$ can also enhance the signal as taus can give two SFOS lepton by decaying into electrons or muons.
Using the same selection and optimization cuts of $|M_{ll}- M_Z| < 15$ GeV and $\slashed{E}_T>0.95$ TeV, it is found that the number of events at 14 TeV run of the LHC with luminosity $L=3000 ~{\rm fb^{-1}}$ remains the same for BP2,
whereas the number of events go to 2.14 at 27 TeV LHC and the corresponding significance attains a values of 6.60. Hence it is clear that the enhancement remains relatively small even if the tau leptons in the final state are considered.

The SFOS lepton pair final state can also come through the cascade decay process with $pp\rightarrow Z_R \rightarrow t \ov t, t \rightarrow W b, W\rightarrow l \nu$. It can be seen from the BPs that the branching of the heavy neutral gauge boson into the pair of top quark final state is quite large (see the Tab.~\ref{table-NLSP}). As was already discussed in the mono-$W$ plus $\slashed{E}_T$ searches, the isolation criteria of leptons significantly wane the number of events arising from top quark decays at the detector level. Furthermore, a b-jet veto will completely remove any events in the final state and hence the enhancement in this channel remains negligible.

\section{Summary and Conclusions} \label{conc}

In this work, we have performed a detailed study of the heavy gauge boson decays and corresponding collider phenomenology in a minimal left-right supersymmetric model with automatic R-parity conservation. In our chosen scenario, the LR symmetry is broken in the SUSY limit, making the additional $W_R$ and $Z_R$ gauge bosons heavier than the SUSY particles. The heavy gauge bosons can thus decay into these SUSY states. We have studied the possible decay modes of the $W_R$ and $Z_R$ bosons into lighter electroweakinos and sfermions. In our initial analysis the sfermions were kept heavier than the right-handed gauge bosons so as to prevent their decay into sfermion final states. Our results show that the heavy gauge bosons decay into electroweakinos are strongly dependent on the composition of these states. We have thus considered all possibilities where the lightest neutralino (also the LSP in our model) is almost entirely composed of only single type of gaugino or higgsino state. The decay width of the heavy gauge bosons into these lighter electroweakinos become significant for the cases where the neutralino is either composed of the right-handed wino or is mostly composed of the higgsino superpartner of bidoublet or triplet scalar. We then looked at the cases where the LSP can be a mixture of various higgsino and gaugino states. Again, significantly large branching ratios for the heavy gauge boson decays were obtained for light neutralino states composed of a mixture of right-handed wino and higgsinos. Allowing the sfermions to be lighter than these gauge bosons opened up new decay channels with the sfermions in the final state. The BR for these channels though remained quite small except for the $Z_R$ decays into down-type squarks. These extra decay channels can suppress the heavy gauge boson direct decays into SM particles and can potentially change the measured experimental bounds on their masses. Additionally, these decays can also suggest new collider signatures hinting towards the existence of supersymmetric LR model.

We have performed detailed collider analysis of several SUSY and non-SUSY decays of the heavy gauge bosons in the context of high luminosity LHC (HL-LHC) and high energy LHC (HE-LHC) with center-of-mass energy of 14 TeV and 27 TeV respectively. We have analyzed the familiar dilepton and dijet channels arising from the direct decays of the $W_R$ and $Z_R$ bosons. We have also studied their one-step SUSY cascade decays into mono-$W$ + $\slashed{E}_T$ and mono-$Z$ +  $\slashed{E}_T$ final states. These signals have already been probed experimentally in the context of dark matter searches but not been considered as possible search channels for the heavy gauge bosons. To explore this possibility, we study the mono-$X$ +  $\slashed{E}_T$ ($X = W,Z$) final states where the $X$ particles decay leptonically. The leptonic final states produce relatively clean signals which are easy to identify in a hadron-rich environment like the LHC experiment. We have chosen two benchmark points -- BP1 which is more suited for the mono-$W$ +$\slashed{E}_T$ search and BP2 for mono-$Z$ +$\slashed{E}_T$ analysis. We have further optimized the selection cuts in order to enhance the signal significance over the SM backgrounds. 

Our study shows that the dilepton final state gives promising results for the discovery of the heavy $Z_R$ boson at the HE-LHC while the dijet channel is better suited to search for $W_R$ bosons. Even if a heavy gauge boson is seen in these channels, the SUSY nature of the model will still remain hidden. The mono-$X$ + $\slashed{E}_T$ channels, in conjunction with the dilepton and dijet channels will not only be able to tell us about the existence of these heavy gauge bosons, it can also provide significant hints towards the existence of SUSY particles.

\section{Acknowledgements}
Work of B. Bhattacherjee was supported by Department of Science and Technology, Government of India under the Grant Agreement numbers IFA13-PH-75 (INSPIRE Faculty Award). The work of Najimuddin Khan was supported by the Department of Science and Technology, Government of INDIA under the SERB-Grant PDF/2017/00372. The work of Ayon Patra was partially supported by the Department of Science and Technology, Government of INDIA under the SERB-Grant PDF/2016/000202. The authors would like to thank Sudhir K. Vempati, Rahool K. Barman and Amit Adhikary for useful discussions.

\vspace{5cm}

%\appendix

\numberwithin{equation}{section}

\begin{appendices}

\section{$W_R$ Boson-Interaction}
\label{App:WR}
 
%\subsection{Two Scalar-One Vector Boson-Interaction}

\begin{minipage}{4.5cm}
  \begin{fmffile}{FeynDia112}
\fmfframe(20,20)(20,20){ 
\begin{fmfgraph*}(75,75)  
\fmfL(-6,37.50002,r){$A_{{i}}$}%
\fmfL(71.24773,-4.68521,lt){$H^{-}_{{j}}$}%
\fmfL(71.24773,79.68521,lb){$W^{-,*}_{R,{\mu }}$}%
\end{fmfgraph*}
} 
\end{fmffile} 
\end{minipage}
:
\begin{minipage}{10cm}
{\allowdisplaybreaks
\begin{align} 
 &\frac{g_R}{2} \Big(- \sqrt{2} U^{{Hm},*}_{j 1}  Z_{{i 1}}^{A} +\sqrt{2}  U^{{Hm},*}_{j 2}  Z_{{i 2}}^{A} \nonumber \\ & + U^{{Hm},*}_{j 4}  Z_{{i 3}}^{A}  + U^{{Hm},*}_{j 3}  Z_{{i 4}}^{A}  \Big)\Big(- p^{H^{-}_{{j}}}_{\mu}  + p^{A_{{i}}}_{\mu}\Big)
 \label{coupWRAH}
 \end{align}
} 
\end{minipage}

\vskip .25cm

\begin{minipage}{4.5cm}
\begin{fmffile}{FeynDia118} 
\fmfframe(20,20)(20,20){ 
\begin{fmfgraph*}(75,75) 
\fmfL(-6,37.50002,r){$\widetilde{d}_{{i \alpha}}$}
\fmfL(71.24773,-4.68521,lt){$\widetilde{u}^*_{{j \beta}}$}
\fmfL(71.24773,79.68521,lb){$W^{-,*}_{R,{\mu}}$}
\end{fmfgraph*}} 
\end{fmffile} 

\end{minipage}
:
\begin{minipage}{10cm}

{\allowdisplaybreaks
\begin{align} 
 &\frac{ i g_R}{\sqrt{2}} \delta_{\alpha \beta} \sum_{a=1}^{3}U^{{DL},*}_{i 3 + a} U_{{j 3 + a}}^{UL} \Big(p^{\widetilde{u}^*_{{j \beta}}}_{\mu}  - p^{\widetilde{d}_{{i \alpha}}}_{\mu}\Big)
 \label{coupWRsdsu}
\end{align}
}
\end{minipage}

\vskip .25cm

\begin{minipage}{4.5cm}

\begin{fmffile}{FeynDia118} 
\fmfframe(20,20)(20,20){ 
\begin{fmfgraph*}(75,75) 
\fmfL(-6,37.50002,r){$\widetilde{e}_{{i}}$}
\fmfL(71.24773,-4.68521,lt){$\widetilde{\nu}^*_{{j}}$}
\fmfL(71.24773,79.68521,lb){$W^{-,*}_{R,{\mu}}$}
\end{fmfgraph*}} 
\end{fmffile} 

\end{minipage}
:
\begin{minipage}{10cm}
{\allowdisplaybreaks
\begin{align} 
 &\frac{i g_R}{\sqrt{2}}  \sum_{a=1}^{3}U^{{EL},*}_{i 3 + a} U_{{j 3 + a}}^{VL} \Big( p^{\widetilde{\nu}^*_{{j}}}_{\mu}  - p^{\widetilde{e}_{{i}}}_{\mu}\Big)
 \label{coupsesv}
 \end{align}
} 

\end{minipage}

\vskip .25cm

\begin{minipage}{4.5cm}

\begin{fmffile}{FeynDia112} 
\fmfframe(20,20)(20,20){ 
\begin{fmfgraph*}(75,75) 
\fmfL(-6,37.50002,r){$H_{{i}}$}
\fmfL(71.24773,-4.68521,lt){$H^{-}_{{j}}$}
\fmfL(71.24773,79.68521,lb){$W^{-,*}_{R,{\mu}}$}
\end{fmfgraph*}} 
\end{fmffile}

\end{minipage}
:
\begin{minipage}{10cm}
{\allowdisplaybreaks
\begin{align} 
&\frac{i g_R}{2} \Big(\sqrt{2}  U^{{Hm},*}_{j 1}  Z_{{i 1}}^{H} +\sqrt{2} U^{{Hm},*}_{j 2}  Z_{{i 2}}^{H} \nonumber \\ 
 & + U^{{Hm},*}_{j 4}  Z_{{i 3}}^{H} -  U^{{Hm},*}_{j 3}  Z_{{i 4}}^{H}  \Big)\Big(p^{H^{-}_{{j}}}_{\mu}  - p^{H_{{i}}}_{\mu}\Big)
 \label{coupWRHHm}
 \end{align}
} 
\end{minipage}

\vskip .25cm

\begin{minipage}{4.5cm}

\begin{fmffile}{FeynDia118} 
\fmfframe(20,20)(20,20){ 
\begin{fmfgraph*}(75,75)
\fmfL(-6,37.50002,r){$H^{-}_{{i}}$}
\fmfL(71.24773,-4.68521,lt){$H^{++}_{{j}}$}
\fmfL(71.24773,79.68521,lb){$W^-_{R,{\mu}}$}
\end{fmfgraph*}} 
\end{fmffile} 

\end{minipage}
:
\begin{minipage}{10cm}

{\allowdisplaybreaks
\begin{align} 
 &i g_R  \Big(U^{{Hm,*}}_{i 1} U_{{j 1}}^{Hmm}  + U^{{Hm,*}}_{i 2} U_{{j 2}}^{Hmm} \Big)\Big(p^{H^{++}_{{j}}}_{\mu}  - p^{H^{-}_{{i}}}_{\mu}\Big)
 \label{coupWRHPHmm}
 \end{align}
}
\end{minipage}

\vskip .25cm

\begin{minipage}{4.5cm}

\begin{fmffile}{FeynDia152} 
\fmfframe(20,20)(20,20){ 
\begin{fmfgraph*}(75,75)
\fmfL(-6,37.50002,r){$H_{{i}}$}
\fmfL(71.24773,-4.68521,lt){$W^{-,*}_{R,{\sigma}}$}
\fmfL(71.24773,79.68521,lb){$W^-_{L,{\mu}}$}
\end{fmfgraph*}} 
\end{fmffile} 

\end{minipage}
:
\begin{minipage}{10cm}

{\allowdisplaybreaks
\begin{align} 
 & \frac{i g_2 g_R}{\sqrt{2}} \Big( v_u    Z_{{i 3}}^{H} + v_d    Z_{{i 4}}^{H} \Big)g_{\sigma \mu}
 \label{coupWRHWL}
 \end{align}
} 
\end{minipage}

\vskip .25cm

\begin{minipage}{4.5cm}
\begin{fmffile}{FeynDia152} 
\fmfframe(20,20)(20,20){ 
\begin{fmfgraph*}(75,75)
\fmfL(-6,37.50002,r){$A_{{i}}$}
\fmfL(71.24773,-4.68521,lt){$W^{-,*}_{L,{\sigma}}$}
\fmfL(71.24773,79.68521,lb){$W^-_{R,{\mu}}$}
\end{fmfgraph*}} 
\end{fmffile} 

\end{minipage}
:
\begin{minipage}{10cm}
{\allowdisplaybreaks
\begin{align} 
 &\frac{g_2 g_R}{\sqrt{2}}  \Big( v_u Z_{{i 3}}^{A} - v_d Z_{{i 4}}^{A}\Big)g_{\sigma \mu}
 \label{coupWRAWL}
 \end{align}
} 

\end{minipage}

\vskip .25cm

\begin{minipage}{4.5cm}
\begin{fmffile}{FeynDia163} 
\fmfframe(20,20)(20,20){ 
\begin{fmfgraph*}(75,75)
\fmfL(-6,37.50002,r){$H^{-}_{{i}}$}
\fmfL(71.24773,-4.68521,lt){$W^{-,*}_{R,{\sigma}}$}
\fmfL(71.24773,79.68521,lb){$\gamma_{{\mu}}$}
\end{fmfgraph*}} 
\end{fmffile} 

\end{minipage}
:
\begin{minipage}{10cm}

{\allowdisplaybreaks
\begin{align} 
 &i g_R  \Big( ( 2  g_V  Z_{{1 1}}^{Z}  - g_R Z_{{3 1}}^{Z} )(\bar{v}_R U^{{Hm},*}_{i 1} + v_R U^{{Hm},*}_{i 2} )\nonumber \\ 
 &+\frac{g_2}{\sqrt{2}} Z_{{2 1}}^{Z} (v_d U^{{Hm},*}_{i 4} - v_u U^{{Hm},*}_{i 3})\Big)g_{\sigma \mu}
 \label{coupWRgammaH}
 \end{align}
} 
\end{minipage}

\vskip .25cm

\begin{minipage}{4.5cm}
\begin{fmffile}{FeynDia163} 
\fmfframe(20,20)(20,20){ 
\begin{fmfgraph*}(75,75)
\fmfL(-6,37.50002,r){$H^{-}_{{i}}$}
\fmfL(71.24773,-4.68521,lt){$W^{-,*}_{R,{\sigma}}$}
\fmfL(71.24773,79.68521,lb){$Z_{{\mu}}$}
\end{fmfgraph*}} 
\end{fmffile}

\end{minipage}
:
\begin{minipage}{10cm}
{\allowdisplaybreaks
\begin{align} 
 &i g_R  \Big( (2  g_V  Z_{{1 2}}^{Z}  - g_R Z_{{3 2}}^{Z} )(\bar{v}_R U^{{Hm},*}_{i 1}  + v_R U^{{Hm},*}_{i 2} )\nonumber \\ 
 &+\frac{g_2}{\sqrt{2}} Z_{{2 2}}^{Z} ( v_d    U^{{Hm},*}_{i 4} - v_u U^{{Hm},*}_{i 3} )\Big)g_{\sigma \mu}
 \label{coupWRZHm}
 \end{align}
}

\end{minipage}

\vskip .25cm

\begin{minipage}{4.5cm}
\begin{fmffile}{FeynDia420} 
\fmfframe(20,20)(20,20){ 
\begin{fmfgraph*}(75,75)
\fmfL(-6,37.50002,r){$\bar{u}_{{i \alpha}}$}
\fmfL(71.24773,-4.68521,lt){$d_{{j \beta}}$}
\fmfL(71.24773,79.68521,lb){$W^{-,*}_{R,{\mu}}$}
\end{fmfgraph*}} 
\end{fmffile}

\end{minipage}
:
\begin{minipage}{10cm}
{\allowdisplaybreaks
\begin{align}  
  & \, -\frac{ig_R}{\sqrt{2}}   \delta_{\alpha \beta} \sum_{a=1}^{3}U^{{Uu},*}_{i a} U_{{j a}}^{Ud}  \gamma_{\mu}P_R
\label{coupWRFud}
 \end{align}
} 

\end{minipage}

\vskip .25cm

\begin{minipage}{4.5cm}
\begin{fmffile}{FeynDia420} 
\fmfframe(20,20)(20,20){ 
\begin{fmfgraph*}(75,75)
\fmfL(-6,37.50002,r){$\nu_{{i}}$}
\fmfL(71.24773,-4.68521,lt){$e_{{j}}$}
\fmfL(71.24773,79.68521,lb){$W^{-,*}_{R,{\mu}}$}
\end{fmfgraph*}} 
\end{fmffile}

\end{minipage}
:
\begin{minipage}{10cm}
{\allowdisplaybreaks
\begin{align} 
 &\,- \frac{i g_R }{\sqrt{2}}  \sum_{a=1}^{3}U^{{PMNS},*}_{i 3 + a} U_{{j a}}^{Ue}  \gamma_{\mu}P_R
 \label{coupWRFev}
 \end{align}
} 

\end{minipage}

\vskip .25cm

\begin{minipage}{4.5cm}

\begin{fmffile}{FeynDia445} 
\fmfframe(20,20)(20,20){ 
\begin{fmfgraph*}(75,75)
\fmfL(-6,37.50002,r){${\chi}^+_{i}$}
\fmfL(71.24773,-4.68521,lt){${\chi}_{{j}}^0$}
\fmfL(71.24773,79.68521,lb){$W^-_{R,{\mu}}$}
\end{fmfgraph*}} 
\end{fmffile} 
 
\end{minipage}
:
\begin{minipage}{10cm}
{\allowdisplaybreaks
\begin{align} 
\label{coupWRFXXm}
 &-\frac{i g_R}{2} \Big(2 Z^{{fN},*}_{j 2}  U_{{i 1}}^{Lm}  -2 Z^{{fN},*}_{j 4}  U_{{i 3}}^{Lm}+\sqrt{2} Z^{{fN},*}_{j 7}  U_{{i 4}}^{Lm}  \Big)\gamma_{\mu}P_L\nonumber\\ 
  & -\frac{i g_R}{2} \Big(2 g_R U^{{Rp},*}_{i 1}  Z_{{j 2}}^{fN} -2 U^{{Rp},*}_{i 3}  Z_{{j 5}}^{fN} - \sqrt{2} U^{{Rp},*}_{i 4}  Z_{{j 6}}^{fN} \Big)\gamma_{\mu}P_R
 \end{align}
}

\end{minipage}

\vskip .25cm

\begin{minipage}{4.5cm}
\begin{fmffile}{FeynDia420} 
\fmfframe(20,20)(20,20){ 
\begin{fmfgraph*}(75,75)
\fmfL(-6,37.50002,r){${\chi}^{++}$}
\fmfL(71.24773,-4.68521,lt){${\chi}^-_{j}$}
\fmfL(71.24773,79.68521,lb){$W^-_{R,{\mu}}$}
\end{fmfgraph*}} 
\end{fmffile}

\end{minipage}
:
\begin{minipage}{10cm}
{\allowdisplaybreaks
\begin{align}
\label{coupWRFXppXm}
 &-i g_R (U^{{Lm},*}_{j 3}  \gamma_{\mu}P_L +  U_{{j 3}}^{Rp}) \gamma_{\mu}P_R 
\end{align}
}

\end{minipage}

\vskip .25cm

\section{$ {Z_R} $ Boson-Interaction}
\label{App:ZR}

%\subsection{Two Scalar-One Vector Boson-Interaction}

\begin{minipage}{4.5cm}

\begin{fmffile}{FeynDia118} 
\fmfframe(20,20)(20,20){ 
\begin{fmfgraph*}(75,75)
\fmfL(-6,37.50002,r){$A_{{i}}$}
\fmfL(71.24773,-4.68521,lt){$H_{{j}}$}
\fmfL(71.24773,79.68521,lb){$ {Z_R} _{{\mu}}$}
\end{fmfgraph*}} 
\end{fmffile}

\end{minipage}
:
\begin{minipage}{10cm}

{\allowdisplaybreaks
\begin{align} 
 & \Big((Z_{{i 2}}^{A} Z_{{j 2}}^{H}+2 Z_{{i 1}}^{A} Z_{{j 1}}^{H}) ( g_V  Z_{{1 3}}^{Z}  - g_R Z_{{3 3}}^{Z} )\nonumber \\ 
 &+\frac{1}{2}(Z_{{i 3}}^{A} Z_{{j 3}}^{H}  - Z_{{i 4}}^{A} Z_{{j 4}}^{H})(g_2 Z_{{2 3}}^{Z}  - g_R Z_{{3 3}}^{Z})\Big)\Big(p^{A_{{i}}}_{\mu} - p^{H_{{j}}}_{\mu} \Big)
 \label{coupZRAH}
 \end{align}
} 
 
\end{minipage}

\vskip .25cm

\begin{minipage}{4.5cm}

\begin{fmffile}{FeynDia118} 
\fmfframe(20,20)(20,20){ 
\begin{fmfgraph*}(75,75)
\fmfL(-6,37.50002,r){$H^{-}_{{i}}$}
\fmfL(71.24773,-4.68521,lt){$H^{+}_{{j}}$}
\fmfL(71.24773,79.68521,lb){$Z'_{{\mu}}$}
\end{fmfgraph*}} 
\end{fmffile}

\end{minipage}
:
\begin{minipage}{10cm}

{\allowdisplaybreaks
\begin{align} 
 &\frac{i}{2} \Big(2  g_V Z_{{1 3}}^{Z}  (U^{{Hm},*}_{i 1} U_{{j 1}}^{Hm}  +  U^{{Hm},*}_{i 2} U_{{j 2}}^{Hm}) \nonumber\\
 & + (U^{{Hm},*}_{i 3} U_{{j 3}}^{Hm}  + U^{{Hm},*}_{i 4} U_{{j 4}}^{Hm})(g_2 Z_{{2 3}}^{Z}  + g_R Z_{{3 3}}^{Z})\Big)\Big(p^{H^{-}_{{i}}}_{\mu}- p^{H^{+}_{{j}}}_{\mu}\Big)
  \label{coupZRHpHm}
 \end{align}
} 
 
\end{minipage}

\vskip .25cm

\begin{minipage}{4.5cm}

\begin{fmffile}{FeynDia118} 
\fmfframe(20,20)(20,20){ 
\begin{fmfgraph*}(75,75)
\fmfL(-6,37.50002,r){$H^{--}_{{i}}$}
\fmfL(71.24773,-4.68521,lt){$H^{++}_{{j}}$}
\fmfL(71.24773,79.68521,lb){$ {Z_R} _{{\mu}}$}
\end{fmfgraph*}} 
\end{fmffile}

\end{minipage}
:
\begin{minipage}{10cm}
{\allowdisplaybreaks
\begin{align} 
 &i \Big(U^{{Hmm},*}_{i 1} U_{{j 1}}^{Hmm}  + U^{{Hmm},*}_{i 2} U_{{j 2}}^{Hmm} \Big)\Big( g_V  Z_{{1 3}}^{Z}  + g_R Z_{{3 3}}^{Z} \Big)\nonumber\\
 &\times \Big(p^{H^{--}_{{i}}}_{\mu}- p^{H^{++}_{{j}}}_{\mu}\Big)
  \label{coupZRHmmHpp}
 \end{align}
} 

\end{minipage}

\vskip .25cm

\begin{minipage}{4.5cm}
\begin{fmffile}{FeynDia118} 
\fmfframe(20,20)(20,20){ 
\begin{fmfgraph*}(75,75)
\fmfL(-6,37.50002,r){$\widetilde{d}_{{i \alpha}}$}
\fmfL(71.24773,-4.68521,lt){$\widetilde{d}^*_{{j \beta}}$}
\fmfL(71.24773,79.68521,lb){$ {Z_R} _{{\mu}}$}
\end{fmfgraph*}} 
\end{fmffile}

\end{minipage}
:
\begin{minipage}{10cm}
{\allowdisplaybreaks
\begin{align} 
 &-\frac{i}{6} \delta_{\alpha \beta} \Big(\sum_{a=4}^{3}U^{{DL},*}_{ia} U_{{ja}}^{DL}  (-3 g_R Z_{{3 3}}^{Z}  +  g_V  Z_{{1 3}}^{Z}) 
 \nonumber\\
 &+ \sum_{a=1}^{3}U^{{DL},*}_{i a} U_{{j a}}^{DL} (-3 g_2 Z_{{2 3}}^{Z}  +  g_V  Z_{{1 3}}^{Z} )\Big)\Big(p^{\widetilde{d}_{{i \alpha}}}_{\mu}- p^{\widetilde{d}^*_{{j \beta}}}_{\mu}\Big)
 \label{coupZRsdsd}
 \end{align}
} 

\end{minipage}

\vskip .25cm

\begin{minipage}{4.5cm}

\begin{fmffile}{FeynDia118} 
\fmfframe(20,20)(20,20){ 
\begin{fmfgraph*}(75,75)
\fmfL(-6,37.50002,r){$\widetilde{e}_{{i}}$}
\fmfL(71.24773,-4.68521,lt){$\widetilde{e}^*_{{j}}$}
\fmfL(71.24773,79.68521,lb){$ {Z_R} _{{\mu}}$}
\end{fmfgraph*}} 
\end{fmffile}

\end{minipage}
:
\begin{minipage}{10cm}

{\allowdisplaybreaks
\begin{align} 
 &\frac{i}{2} \Big(\sum_{a=4}^{3}U^{{EL},*}_{ia} U_{{ja}}^{EL}  ( g_V  Z_{{1 3}}^{Z}  + g_R Z_{{3 3}}^{Z}) \nonumber\\
 &+ \sum_{a=1}^{3}U^{{EL},*}_{i a} U_{{j a}}^{EL}  (g_2 Z_{{2 3}}^{Z}  +  g_V  Z_{{1 3}}^{Z})\Big)\Big(p^{\widetilde{e}_{{i}}}_{\mu}- p^{\widetilde{e}^*_{{j}}}_{\mu}\Big)
 \label{coupZRsese}
 \end{align}
} 

\end{minipage}

\vskip .25cm

\begin{minipage}{4.5cm}
\begin{fmffile}{FeynDia118} 
\fmfframe(20,20)(20,20){ 
\begin{fmfgraph*}(75,75)
\fmfL(-6,37.50002,r){$\widetilde{\nu}_{{i}}$}
\fmfL(71.24773,-4.68521,lt){$\widetilde{\nu}^*_{{j}}$}
\fmfL(71.24773,79.68521,lb){$ {Z_R} _{{\mu}}$}
\end{fmfgraph*}} 
\end{fmffile}

\end{minipage}
:
\begin{minipage}{10cm}

{\allowdisplaybreaks
\begin{align} 
 &\frac{i}{2} \Big(\sum_{a=1}^{3}U_{{i 3 + a}}^{VL} U_{{j 3 + a}}^{VL} ( g_V  Z_{{1 3}}^{Z}  - g_R Z_{{3 3}}^{Z} ) \nonumber\\
 &+ \sum_{a=1}^{3}U_{{i a}}^{VL} U_{{j a}}^{VL} (- g_2 Z_{{2 3}}^{Z}  +  g_V  Z_{{1 3}}^{Z})\Big)\Big(p^{\widetilde{\nu}_{{i}}}_{\mu}- p^{\widetilde{\nu}^*_{{j}}}_{\mu}\Big)
  \label{coupZRsvsv}
 \end{align}
}
\end{minipage}

\vskip .25cm

\begin{minipage}{4.5cm}

\begin{fmffile}{FeynDia118} 
\fmfframe(20,20)(20,20){ 
\begin{fmfgraph*}(75,75)
\fmfL(-6,37.50002,r){$\widetilde{u}_{{i \alpha}}$}
\fmfL(71.24773,-4.68521,lt){$\widetilde{u}^*_{{j \beta}}$}
\fmfL(71.24773,79.68521,lb){$ {Z_R} _{{\mu}}$}
\end{fmfgraph*}} 
\end{fmffile}

\end{minipage}
:
\begin{minipage}{10cm}

{\allowdisplaybreaks
\begin{align} 
 &-\frac{i}{6} \delta_{\alpha \beta} \Big(\sum_{a=1}^{3}U^{{UL},*}_{i 3 + a} U_{{j 3 + a}}^{UL}  (3 g_R Z_{{3 3}}^{Z}  +  g_V  Z_{{1 3}}^{Z} ) \nonumber\\
 & + \sum_{a=1}^{3}U^{{UL},*}_{i a} U_{{j a}}^{UL}  (3 g_2 Z_{{2 3}}^{Z}  +  g_V  Z_{{1 3}}^{Z})\Big)\Big(p^{\widetilde{u}_{{i \alpha}}}_{\mu}- p^{\widetilde{u}^*_{{j \beta}}}_{\mu}\Big)
  \label{coupZRsusu}
 \end{align}
}
 
\end{minipage}

\vskip .25cm

%\subsection{One Scalar-Two Vector Boson-Interaction}

\begin{minipage}{4.5cm}

\begin{fmffile}{FeynDia152} 
\fmfframe(20,20)(20,20){ 
\begin{fmfgraph*}(75,75)
\fmfL(-6,37.50002,r){$H_{{i}}$}
\fmfL(71.24773,-4.68521,lt){$\gamma_\sigma$}
\fmfL(71.24773,79.68521,lb){$ {Z_R} _{{\mu}}$}
\end{fmfgraph*}} 
\end{fmffile}

\end{minipage}
:
\begin{minipage}{10cm}

{\allowdisplaybreaks
\begin{align} 
   \label{coupZRHgamma}
 &\frac{i}{\sqrt{2}} \Big(4 (\bar{v}_R Z_{{i 1}}^{H}+v_R Z_{{i 2}}^{H}) ( g_V  Z_{{1 1}}^{Z}  - g_R Z_{{3 1}}^{Z} )( g_V  Z_{{1 3}}^{Z}  - g_R Z_{{3 3}}^{Z})\nonumber \\  
 &+(v_d Z_{{i 3}}^{H}  + v_u Z_{{i 4}}^{H} )(g_2 Z_{{2 1}}^{Z}  - g_R Z_{{3 1}}^{Z} )(g_2 Z_{{2 3}}^{Z}  - g_R Z_{{3 3}}^{Z})\Big) g_{\sigma \mu}
 \end{align}
}
 
\end{minipage}

 \vskip .25cm

\begin{minipage}{4.5cm}

\begin{fmffile}{FeynDia152} 
\fmfframe(20,20)(20,20){ 
\begin{fmfgraph*}(75,75)
\fmfL(-6,37.50002,r){$H_{{i}}$}
\fmfL(71.24773,-4.68521,lt){$Z_\sigma$}
\fmfL(71.24773,79.68521,lb){$ {Z_R} _{{\mu}}$}
\end{fmfgraph*}} 
\end{fmffile}

\end{minipage}
:
\begin{minipage}{10cm}

{\allowdisplaybreaks
\begin{align} 
\label{coupZRHZ}
 &i \frac{1}{\sqrt{2}} \Big(4 (\bar{v}_R Z_{{i 1}}^{H}+v_R Z_{{i 2}}^{H})( g_V  Z_{{1 2}}^{Z}  - g_R Z_{{3 2}}^{Z} )( g_V  Z_{{1 3}}^{Z}  - g_R Z_{{3 3}}^{Z})\nonumber \\ 
 &+(v_d Z_{{i 3}}^{H}  + v_u Z_{{i 4}}^{H})(g_2 Z_{{2 2}}^{Z}  - g_R Z_{{3 2}}^{Z} )(g_2 Z_{{2 3}}^{Z}  - g_R Z_{{3 3}}^{Z})\Big) g_{\sigma \mu} 
 \end{align}
}
 
\end{minipage}

\vskip .25cm

\begin{minipage}{4.5cm}

\begin{fmffile}{FeynDia163} 
\fmfframe(20,20)(20,20){ 
\begin{fmfgraph*}(75,75)
\fmfL(-6,37.50002,r){$H^-_{{i}}$}
\fmfL(71.24773,-4.68521,lt){$W^{-*}_\sigma$}
\fmfL(71.24773,79.68521,lb){$ {Z_R} _{{\mu}}$}
\end{fmfgraph*}} 
\end{fmffile}

\end{minipage}
:
\begin{minipage}{10cm}

{\allowdisplaybreaks
\begin{align}
\label{coupZRHmWp} 
 &\frac{i g_2 g_R  }{\sqrt{2}} 
  \Big(v_u U^{{Hm},*}_{i 4} Z_{{3 2}}^{Z}- v_d U^{{Hm},*}_{i 3}  Z_{{3 2}}^{Z} \Big)g_{\sigma \mu}
 \end{align}
}
 
\end{minipage}

\vskip .25cm

\begin{minipage}{4.5cm}

\begin{fmffile}{FeynDia163} 
\fmfframe(20,20)(20,20){ 
\begin{fmfgraph*}(75,75)
\fmfL(-6,37.50002,r){$H^-_{{i}}$}
\fmfL(71.24773,-4.68521,lt){$W^{-*}_{R,\sigma}$}
\fmfL(71.24773,79.68521,lb){$ {Z_R} _{{\mu}}$}
\end{fmfgraph*}} 
\end{fmffile}

\end{minipage}
:
\begin{minipage}{10cm}

{\allowdisplaybreaks
\begin{align}
\label{coupZRHmWRp} 
&\frac{i g_R }{2}  \Big(2  ( \bar{v}_R U^{{Hm},*}_{i 1} + v_R U^{{Hm},*}_{i 2} )  (2  g_V  Z_{{1 3}}^{Z}  - g_R Z_{{3 3}}^{Z} ) \nonumber \\ 
 &+\sqrt{2} g_2 (v_d U^{{Hm},*}_{i 4}  Z_{{2 3}}^{Z}- v_u U^{{Hm},*}_{i 3}  Z_{{2 3}}^{Z})\Big) g_{\sigma \mu}
 \end{align}
}
 
\end{minipage}

\vskip .25cm

%\subsection{Two Fermion-One Vector Boson-Interaction}

\begin{minipage}{4.5cm}
\begin{fmffile}{FeynDia420} 
\fmfframe(20,20)(20,20){ 
\begin{fmfgraph*}(75,75)
\fmfL(-6,37.50002,r){$\bar{d}_{{i \alpha}}$}
\fmfL(71.24773,-4.68521,lt){$d_{{j \beta}}$}
\fmfL(71.24773,79.68521,lb){$ {Z_R} _{{\mu}}$}
\end{fmfgraph*}} 
\end{fmffile}

\end{minipage}
:
\begin{minipage}{10cm}

{\allowdisplaybreaks
\begin{align} 
\label{coupZRdd}
 &-\frac{i}{6} \delta_{\alpha \beta} \delta_{i j} \Big(-3 g_2 Z_{{2 3}}^{Z}  +  g_V  Z_{{1 3}}^{Z} \Big) \gamma_{\mu}P_L\nonumber\\ 
  & - \frac{i}{6} \delta_{\alpha \beta} \delta_{i j} \Big(-3 g_R Z_{{3 3}}^{Z}  +  g_V  Z_{{1 3}}^{Z} \Big)\gamma_{\mu}P_R  
 \end{align}
} 
\end{minipage}

\vskip .25cm 

\begin{minipage}{4.5cm}

\begin{fmffile}{FeynDia420} 
\fmfframe(20,20)(20,20){ 
\begin{fmfgraph*}(75,75)
\fmfL(-6,37.50002,r){$\bar{e}_{{i}}$}
\fmfL(71.24773,-4.68521,lt){$e_{{j}}$}
\fmfL(71.24773,79.68521,lb){$ {Z_R} _{{\mu}}$}
\end{fmfgraph*}} 
\end{fmffile} 
 
\end{minipage}
:
\begin{minipage}{10cm}

{\allowdisplaybreaks
\begin{align} 
 &\frac{i}{2} \delta_{i j} \Big(g_2 Z_{{2 3}}^{Z}  +  g_V  Z_{{1 3}}^{Z} \Big)\gamma_{\mu}P_L+ \,\frac{i}{2} \delta_{i j} \Big( g_V  Z_{{1 3}}^{Z}  + g_R Z_{{3 3}}^{Z} \Big)\gamma_{\mu}P_R
    \label{coupZRee}
 \end{align}
}
  
\end{minipage}

\vskip .25cm

\begin{minipage}{4.5cm}
\begin{fmffile}{FeynDia420} 
\fmfframe(20,20)(20,20){ 
\begin{fmfgraph*}(75,75)
\fmfL(-6,37.50002,r){$\bar{u}_{{i \alpha}}$}
\fmfL(71.24773,-4.68521,lt){$u_{{j \beta}}$}
\fmfL(71.24773,79.68521,lb){$ {Z_R} _{{\mu}}$}
\end{fmfgraph*}} 
\end{fmffile}

\end{minipage}
:
\begin{minipage}{10cm}

{\allowdisplaybreaks
\begin{align} 
 &-\frac{i}{6} \delta_{\alpha \beta} \delta_{i j} \Big(3 g_2 Z_{{2 3}}^{Z}  +  g_V  Z_{{1 3}}^{Z} \Big)\gamma_{\mu}P_L\nonumber\\ 
  & + \,-\frac{i}{6} \delta_{\alpha \beta} \delta_{i j} \Big(3 g_R Z_{{3 3}}^{Z}  +  g_V  Z_{{1 3}}^{Z} \Big)\gamma_{\mu}P_R
    \label{coupZRuu}
 \end{align}
}

\end{minipage}

\vskip .25cm

\begin{minipage}{4.5cm}

\begin{fmffile}{FeynDia420} 
\fmfframe(20,20)(20,20){ 
\begin{fmfgraph*}(75,75)
\fmfL(-6,37.50002,r){$\nu_{{i}}$}
\fmfL(71.24773,-4.68521,lt){$\nu_{{j}}$}
\fmfL(71.24773,79.68521,lb){$ {Z_R} _{{\mu}}$}
\end{fmfgraph*}} 
\end{fmffile}

\end{minipage}
:
\begin{minipage}{10cm}

{\allowdisplaybreaks
\begin{align} 
 &\frac{i}{2} \Big(\sum_{a=4}^{3}U^{{PMNS},*}_{j a} U_{{i a}}^{PMNS}  \Big(-  g_V  Z_{{1 3}}^{Z}  + g_R Z_{{3 3}}^{Z} \Big) \nonumber\\
 &+ \sum_{a=1}^{3}U^{{PMNS},*}_{j a} U_{{i a}}^{PMNS}  \Big(- g_2 Z_{{2 3}}^{Z}  +  g_V  Z_{{1 3}}^{Z} \Big)\Big)\gamma_{\mu}P_L\nonumber\\ 
  & -\frac{i}{2} \Big(\sum_{a=4}^{3}U^{{PMNS},*}_{i a} U_{{j a}}^{PMNS}  \Big(-  g_V  Z_{{1 3}}^{Z}  + g_R Z_{{3 3}}^{Z} \Big) \nonumber\\
 &+ \sum_{a=1}^{3}U^{{PMNS},*}_{i a} U_{{j a}}^{PMNS}  \Big(- g_2 Z_{{2 3}}^{Z}  +  g_V  Z_{{1 3}}^{Z} \Big)\Big)\gamma_{\mu}P_R
   \label{coupZRvv}
 \end{align}
} 

\end{minipage}

\vskip .25cm

\begin{minipage}{4.5cm}

\begin{fmffile}{FeynDia420} 
\fmfframe(20,20)(20,20){ 
\begin{fmfgraph*}(75,75)
\fmfL(-6,37.50002,r){${\chi}_{{i}}^0$}
\fmfL(71.24773,-4.68521,lt){${\chi}_{{j}}^0$}
\fmfL(71.24773,79.68521,lb){$ {Z_R} _{{\mu}}$}
\end{fmfgraph*}} 
\end{fmffile}

\end{minipage}
:
\begin{minipage}{10cm}

{\allowdisplaybreaks
\begin{align} 
 &\frac{i}{2} \Big(2 \Big(Z^{{fN},*}_{j 4} Z_{{i 4}}^{fN}-Z^{{fN},*}_{j 5} Z_{{i 5}}^{fN} \Big) \Big( g_V  Z_{{1 3}}^{Z}  - g_R Z_{{3 3}}^{Z} \Big) \nonumber \\ 
 &- \Big(Z^{{fN},*}_{j 6} Z_{{i 6}}^{fN}  - Z^{{fN},*}_{j 7} Z_{{i 7}}^{fN} \Big)\Big(g_2 Z_{{2 3}}^{Z} - g_R Z_{{3 3}}^{Z} \Big) \Big)\gamma_{\mu}P_L \nonumber\\ 
  & - \frac{i}{2} \Big(2 \Big(Z^{{fN},*}_{i 4} Z_{{j 4}}^{fN}-Z^{{fN},*}_{i 5} Z_{{j 5}}^{fN}\Big) \Big( g_V  Z_{{1 3}}^{Z}  - g_R Z_{{3 3}}^{Z} \Big) \nonumber \\ 
 &- \Big(Z^{{fN},*}_{i 6} Z_{{j 6}}^{fN}  - Z^{{fN},*}_{i 7} Z_{{j 7}}^{fN} \Big)\Big(g_2 Z_{{2 3}}^{Z} - g_R Z_{{3 3}}^{Z} \Big) \Big)\gamma_{\mu}P_R
   \label{coupZRXX}
 \end{align}
} 
 
\end{minipage}

\vskip .25cm

\begin{minipage}{4.5cm}

\begin{fmffile}{FeynDia420} 
\fmfframe(20,20)(20,20){ 
\begin{fmfgraph*}(75,75)
\fmfL(-6,37.50002,r){${\chi}^+_{i}$}
\fmfL(71.24773,-4.68521,lt){${\chi}^-_{j}$}
\fmfL(71.24773,79.68521,lb){$ {Z_R} _{{\mu}}$}
\end{fmfgraph*}} 
\end{fmffile}

\end{minipage}
:
\begin{minipage}{10cm}

{\allowdisplaybreaks
\begin{align} 
 &\frac{i}{2} \Big(2  g_V  U^{{Lm},*}_{j 3} U_{{i 3}}^{Lm} Z_{{1 3}}^{Z} +2 g_2 U^{{Lm},*}_{j 2} U_{{i 2}}^{Lm} Z_{{2 3}}^{Z} +g_2 U^{{Lm},*}_{j 4} U_{{i 4}}^{Lm} Z_{{2 3}}^{Z} \nonumber \\ 
 &+2 g_R U^{{Lm},*}_{j 1} U_{{i 1}}^{Lm} Z_{{3 3}}^{Z} +g_R U^{{Lm},*}_{j 4} U_{{i 4}}^{Lm} Z_{{3 3}}^{Z} \Big)\gamma_{\mu}P_L\nonumber\\ 
  & + \,\frac{i}{2} \Big(2  g_V  U^{{Rp},*}_{i 3} U_{{j 3}}^{Rp} Z_{{1 3}}^{Z} +2 g_2 U^{{Rp},*}_{i 2} U_{{j 2}}^{Rp} Z_{{2 3}}^{Z} +g_2 U^{{Rp},*}_{i 4} U_{{j 4}}^{Rp} Z_{{2 3}}^{Z} \nonumber \\ 
 &+2 g_R U^{{Rp},*}_{i 1} U_{{j 1}}^{Rp} Z_{{3 3}}^{Z} +g_R U^{{Rp},*}_{i 4} U_{{j 4}}^{Rp} Z_{{3 3}}^{Z} \Big)\gamma_{\mu}P_R
  \label{coupZRXmXp}
 \end{align}
}

\end{minipage}

\vskip .25cm

\begin{minipage}{4.5cm}
\begin{fmffile}{FeynDia420} 
\fmfframe(20,20)(20,20){ 
\begin{fmfgraph*}(75,75)
\fmfL(-6,37.50002,r){${\chi}^{++}$}
\fmfL(71.24773,-4.68521,lt){${\chi}^{--}$}
\fmfL(71.24773,79.68521,lb){$ {Z_R} _{{\mu}}$}
\end{fmfgraph*}} 
\end{fmffile}

\end{minipage}
:
\begin{minipage}{10cm}

{\allowdisplaybreaks
\begin{align} 
 &i \Big( g_V  Z_{{1 3}}^{Z}  + g_R Z_{{3 3}}^{Z} \Big)
    \label{coupZRXmmXpp}
 \end{align}
}
  
\end{minipage}

\vskip .25cm

\begin{minipage}{4.5cm}
\begin{fmffile}{FeynDia354} 
\fmfframe(20,20)(20,20){ 
\begin{fmfgraph*}(75,75)
\fmfL(-6,37.50002,r){$W^{-,*}_{L,{\rho}}$}
\fmfL(71.24773,-4.68521,lt){$W^-_{L,{\sigma}}$}
\fmfL(71.24773,79.68521,lb){$ {Z_R} _{{\mu}}$}
\end{fmfgraph*}} 
\end{fmffile}

\end{minipage}
:
\begin{minipage}{10cm}
{\allowdisplaybreaks
\begin{align}
\label{coupZRWmWp}
 &-\frac{i}{2} \Big( g_2 Z_{{2 3}}^{Z}  + g_R Z_{{3 3}}^{Z} \Big)\Big(g_{\rho \mu} (p^{W^{-,*}_{L,{\rho}}}_{\sigma} - p^{ {Z_R} _{{\mu}}}_{\sigma}) \nonumber\\
 &+ g_{\rho \sigma} ( p^{W^-_{L,{\sigma}}}_{\mu}- p^{W^{-,*}_{L,{\rho}}}_{\mu}) + g_{\sigma \mu} ( p^{ {Z_R}_{{\mu}}}_{\rho}- p^{W^-_{L,{\sigma}}}_{\rho})\Big)
\end{align}
}
\end{minipage}
\vskip .25cm

\end{appendices}

\end{document}